\documentclass[aps,prd,reprint,groupedaddress,preprintnumbers,nofootinbib,amsmath,amssymb,longbibliography,floatfix]{revtex4-2}
\usepackage[T1]{fontenc}
\usepackage{newtxtext,newtxmath}
\usepackage[utf8]{inputenc}
\usepackage{graphicx}
\usepackage{xcolor}
\usepackage{slashed}
\usepackage{bbm}
\usepackage{mathrsfs}
\usepackage{multirow}
\usepackage{comment}
\usepackage{mathtools}
\usepackage{booktabs}
\usepackage{array}
\usepackage{siunitx}
\usepackage[normalem]{ulem}

\newcommand{\GeV}{\mathrm{GeV}}
\definecolor{mycyan}{RGB}{0,130,200}
\usepackage[colorlinks=true,linkcolor=mycyan,citecolor=mycyan,urlcolor=mycyan]{hyperref}
\usepackage[nameinlink,capitalize]{cleveref}

\usepackage{orcidlink}

\newcommand{\be}{\begin{equation}}
\newcommand{\ee}{\end{equation}}
\newcommand{\ba}{\begin{eqnarray}}
\newcommand{\ea}{\end{eqnarray}}

\definecolor{darkermagenta}{rgb}{0.7, 0, 0.7}
\definecolor{darkgreen}{RGB}{0,175,10}
\definecolor{AK}{rgb}{0,0,1.0}

\allowdisplaybreaks[1]
\hypersetup{pdftitle={Baryon-Number Violating Lambda-c Decays from Light-Cone Sum Rules}}
\begin{document}
\raggedbottom
\preprint{SI-HEP-2026-25}
\preprint{P3H-26-090}
\title{Baryon-Number Violating \texorpdfstring{$\Lambda_c$}{Lambda-c} Decays from Light-Cone Sum Rules}
\author{Ali Mohamed\,\orcidlink{0009-0006-3466-2709}}
\email{ali.mohamed@uni-siegen.de}
\affiliation{Theoretische Physik 1, Center for Particle Physics Siegen,\\
and Cluster of Excellence ``Color meets Flavor,''\\
Universit\"at Siegen, 57068 Siegen, Germany}
\begin{abstract}
Baryon-number-violating decays with missing energy probe new
interactions connecting Standard-Model quarks to light
invisible fermions. Such signatures can arise from
dimension-six operators coupling three quarks to an
SM-singlet fermion $\chi$, which escapes detection. With
a suitable baryon-number assignment for $\chi$, total
baryon number is conserved while its violation is apparent
in the visible sector. We study the exclusive decays
$\Lambda_c\to\pi\bar\chi$ and $\Lambda_c\to K\bar\chi$,
determining the required form factors using light-cone
sum rules with pion and kaon distribution amplitudes.
The calculation includes leading-order two-particle
contributions up to twist four and factorisable
four-particle terms. Two $\Lambda_c$ interpolating currents
are employed to assess the current dependence.
Using single-operator benchmarks based on collider bounds,
we obtain branching fractions of order $10^{-7}$ over
sizeable parts of the physical region, with uncertainty
envelopes reaching $10^{-6}$ in some scenarios.
These are indicative estimates within the retained
light-cone expansion. We compare with existing
chiral-perturbation-theory and QCD-factorisation estimates
and discuss experimental prospects. Future data from STCF and
Belle II could probe part of the predicted range,
while FCC-ee could reach smaller branching fractions.
\end{abstract}
\maketitle

\tableofcontents

\section{Introduction}
\label{sec:introduction}

Baryon number is an accidental global symmetry of the
Standard Model (SM), and its possible violation is a generic feature of
many extensions of the SM.
Searches for baryon-number violation (BNV) therefore provide a
particularly sensitive probe of physics beyond the SM. They are also
closely connected to one of the central open questions in particle
physics and cosmology, namely the origin of the observed
matter--antimatter asymmetry of the Universe. Indeed, baryon-number
violation constitutes one of the Sakharov conditions required for the
dynamical generation of a baryon asymmetry~\cite{Sakharov:1967dj}.
At the same time, BNV arises naturally in a variety of ultraviolet
extensions of the SM, most notably in grand-unified theories
(GUTs). These considerations have motivated
extensive experimental searches for BNV, traditionally focused on
processes such as proton decay and neutron--antineutron
oscillations, see e.g.~\cite{Dev:2022jbf}.

An interesting possibility is that baryon number is shared
between the visible sector and a light, feebly interacting
SM-singlet fermion $\chi$. We assign $\chi$ a baryon number
such that dimension-six interactions of the form $qqq\chi$
conserve total baryon number. These interactions can induce
apparent baryon-number violation in the visible sector, with
baryon number carried away by an undetected $\chi$ or $\bar\chi$.

A prominent realisation of this idea in the heavy-flavour sector is
$B$-mesogenesis~\cite{Elor:2018twp,Alonso-Alvarez:2021qfd,Nelson:2019fln,Elahi:2021jia,Alonso-Alvarez:2021oaj,BurgosMarcos:2026lwv}, in which the baryon asymmetry of the Universe is
generated through the production and decay of $B$ mesons, with baryon
number shared between visible baryons and dark-sector states. A characteristic prediction of this framework is
the decay of a $B$ meson into a visible baryon and a dark antibaryon,
giving rise experimentally to baryon-plus-missing-energy signatures.
These decays have been studied extensively at the inclusive and
exclusive levels
\cite{Khodjamirian:2022vta,Elor:2022jxy,Boushmelev:2023huu,Lenz:2024rwi,Mohamed:2025zgx,Biswas:2026oxq}
and have motivated dedicated experimental searches
\cite{Belle-II:2026tyb,BaBar:2023dtq,BaBar:2023rer,BaBar:2024qqx,Belle:2021gmc}.
More generally, such studies demonstrate that heavy-flavour decays
provide access to BNV interactions and flavour structures that are
complementary to conventional searches in the light-quark sector.

This motivates extending the study of BNV decays
with missing energy to the charm sector. Operators containing a charm
quark can mediate the two-body decays
$\Lambda_c\to\pi(K)\bar\chi$. Predicting their branching fractions
requires the nonperturbative matrix elements of the corresponding
three-quark operators between the $\Lambda_c$ baryon and the
light meson. These matrix elements are parametrised by transition
form factors, whose momentum dependence determines the decay
rates as functions of the invisible-fermion mass $m_\chi$.

The decays $\Lambda_c\to\pi(K)\bar\chi$ have previously been
studied using different descriptions of the hadronic matrix
elements. Reference~\cite{Hiller:2026osz} estimates the form
factors within QCD factorisation (QCDF), using heavy-quark and
large-recoil expansions near $q^2\simeq0$. Their application
at the charm scale entails sizeable hadronic uncertainties.
A complementary analysis employs baryon chiral perturbation
theory (BChPT), matching the relevant operators onto a hadronic
description with pole and non-pole contributions
\cite{Wang:2026vxh}. This treatment requires an extrapolation
of light-baryon matrix elements to the charmed-baryon sector,
introducing additional uncertainty. An independent determination
of the $\Lambda_c\to\pi(K)$ transition form factors is therefore
valuable for assessing the hadronic uncertainties in the
predicted decay rates.

In this work, we determine the complete set of
$\Lambda_c\to\pi(K)$ form factors for the dimension-six operators
considered here using QCD light-cone sum rules (LCSRs) with
light-meson distribution amplitudes (DAs). We evaluate the
light-cone OPE at leading order in $\alpha_s$, retaining
two-particle DAs up to twist four and factorised four-particle
contributions expressed through the quark condensate and
two-particle DAs. The $\Lambda_c$ baryon is interpolated by
local three-quark currents. To assess the dependence on the
choice of interpolating current, we use both pseudoscalar
$(\mathcal P)$ and axial-vector $(\mathcal A)$ currents and compare
the resulting form factors. Following the heavy-baryon LCSR
framework in~\cite{Khodjamirian:2011jp}, we include the lowest
negative-parity $\Lambda_c^\ast$ state alongside the
positive-parity ground state in the hadronic representation
and eliminate its contribution through suitable combinations
of invariant amplitudes.

Combining our form factors with the single-operator bounds from
LHC missing-energy searches~\cite{Hiller:2026osz}, we estimate
the $\Lambda_c\to\pi(K)\bar\chi$ branching fractions as functions
of the invisible-fermion mass. We compare our predictions with
existing ChPT and QCDF results using consistent operator and
Wilson-coefficient conventions, and discuss the experimental
prospects at BESIII, Belle II, STCF, and FCC-ee.

The paper is organised as follows.
\Cref{sec:eff_L} introduces the effective interactions and form
factors, while \cref{sec:LCSR,sec:OPE} present the sum rules and
their QCD representation. The numerical analysis and continuation
to the physical region are discussed in
\cref{sec:Numerical,sec:zexpansion}, followed by branching
fractions and experimental prospects in
\cref{sec:branching_fractions}. Comparisons with ChPT and QCDF
and the explicit OPE coefficient functions are collected in
\cref{app:comparison_previous,app:Pis}, respectively.

\section{Effective interactions}
\label{sec:eff_L}

Following the conventions of~\cite{Hiller:2026osz}, with the
invisible fermion denoted by $\chi$, we consider dimension-six
quark-portal interactions of the form $qqq\chi$. With
$B(\chi)=-1$, these interactions conserve total baryon number.
The effective Lagrangian reads
\begin{equation}
    \mathcal{L}_{\rm eff}
    =
    \frac{C^{qqd\chi}_{ijk}}{\Lambda^2}\,
    \mathcal{O}^{qqd\chi}_{ijk}
    +
    \frac{C^{udd\chi}_{ijk}}{\Lambda^2}\,
    \mathcal{O}^{udd\chi}_{ijk}
    + \text{h.c.},
    \label{eq:Leff}
\end{equation}
where
\begin{align}
    \mathcal{O}^{qqd\chi}_{ijk}
    &=
    \epsilon^{ab}\epsilon^{\alpha\beta\gamma}
    \big(
        \overline{q}^{\,C}_{L,a,\alpha,i}\,
        q_{L,b,\beta,j}
    \big)
    \big(
        \overline{d}^{\,C}_{R,\gamma,k}\,
        \chi_R
    \big),
    \label{eq:op_qqdN}
    \\
    \mathcal{O}^{udd\chi}_{ijk}
    &=
    \epsilon^{\alpha\beta\gamma}
    \big(
        \overline{u}^{\,C}_{R,\alpha,i}\,
        d_{R,\beta,j}
    \big)
    \big(
        \overline{d}^{\,C}_{R,\gamma,k}\,
        \chi_R
    \big).
    \label{eq:op_uddN}
\end{align}
Here, $a,b$ are $SU(2)_L$ indices,
$\alpha,\beta,\gamma$ are colour indices, and $i,j,k$ label
quark generations. Since $\mathcal O^{qqd\chi}_{ijk}$ is
symmetric under $i\leftrightarrow j$, we take
\begin{equation}
    C^{qqd\chi}_{ijk}
    =
    C^{qqd\chi}_{jik}
    \equiv
    C^{qqd\chi}_{\{ij\}k}.
\end{equation}
The flavour structures relevant for $\Lambda_c$ decays contain
one charm quark. With the generation assignments $1=(u,d)$
and $2=(c,s)$, the coefficients entering the pion and kaon
channels are
\begin{align}
\Lambda_c\to\pi\bar\chi:\quad&
C^{qqd\chi}_{\{12\}1},\quad
C^{udd\chi}_{211},
\\
\Lambda_c\to K\bar\chi:\quad&
C^{qqd\chi}_{\{12\}2},\quad
C^{qqd\chi}_{\{22\}1},\quad
C^{udd\chi}_{212},\quad
C^{udd\chi}_{221}.
\end{align}
We factor out the invisible-fermion spinor and define the
three-quark operators
\begin{align}
    \widetilde{\mathcal O}_{1,X}
    &=
    i\epsilon^{\alpha\beta\gamma}
    P_R s^\gamma
    \big(\overline{c}^{\,C,\alpha} P_X d^\beta\big),
    \label{eq:effop1}
    \\
    \widetilde{\mathcal O}_{2,X}
    &=
    i\epsilon^{\alpha\beta\gamma}
    P_R d^\gamma
    \big(\overline{c}^{\,C,\alpha} P_X s^\beta\big),
    \label{eq:effop2}
    \\
    \widetilde{\mathcal O}_{3,X}
    &=
    i\epsilon^{\alpha\beta\gamma}
    P_R d^\gamma
    \big(\overline{c}^{\,C,\alpha} P_X d^\beta\big),
    \label{eq:effop3}
\end{align}
where $X=L,R$ and $P_{L,R}=(1\mp\gamma_5)/2$. The decay amplitudes can then be expressed directly in
terms of the corresponding hadronic matrix elements as
\begin{align}
i\mathcal M(\Lambda_c\to\pi\overline\chi)
&=
-i\,\frac{2C^{qqd\chi}_{\{12\}1}}{\Lambda^2}\,
\bar v_\chi^C(q)
\langle\pi(p)|
\widetilde{\mathcal O}_{3,L}
|\Lambda_c(p+q)\rangle
\nonumber\\*
&{}
-i\,\frac{C^{udd\chi}_{211}}{\Lambda^2}\,
\bar v_\chi^C(q)
\langle\pi(p)|
\widetilde{\mathcal O}_{3,R}
|\Lambda_c(p+q)\rangle ,
\label{eq:amplitude_pion}
\\
i\mathcal M(\Lambda_c\to K\overline\chi)
&=
-i\,\frac{2C^{qqd\chi}_{\{12\}2}}{\Lambda^2}\,
\bar v_\chi^C(q)
\langle K(p)|
\widetilde{\mathcal O}_{1,L}
|\Lambda_c(p+q)\rangle
\nonumber\\*
&{}
-i\,\frac{C^{udd\chi}_{212}}{\Lambda^2}\,
\bar v_\chi^C(q)
\langle K(p)|
\widetilde{\mathcal O}_{1,R}
|\Lambda_c(p+q)\rangle
\nonumber\\*
&{}
-i\,\frac{2C^{qqd\chi}_{\{22\}1}}{\Lambda^2}\,
\bar v_\chi^C(q)
\langle K(p)|
\widetilde{\mathcal O}_{2,L}
|\Lambda_c(p+q)\rangle
\nonumber\\*
&{}
-i\,\frac{C^{udd\chi}_{221}}{\Lambda^2}\,
\bar v_\chi^C(q)
\langle K(p)|
\widetilde{\mathcal O}_{2,R}
|\Lambda_c(p+q)\rangle .
\label{eq:amplitude_kaon}
\end{align}

The factors of two multiplying the $C^{qqd\chi}$ contributions arise
from the symmetry under interchange of the first two flavour indices.

The hadronic matrix elements of the three-quark operators
$\widetilde{\mathcal O}_{I,X}$ between a pseudoscalar meson
$M=\pi,K$ and the $\Lambda_c$ baryon can be parametrised in terms of
two independent form factors:
\begin{equation}
\begin{aligned}
&    \langle M(p)|
    \widetilde{\mathcal O}_{I,X}
    |\Lambda_c(p+q)\rangle
\\
&=
    P_R
    \left[
        F_{I,X}(q^2)
        +
        \frac{\slashed q}{m_{\Lambda_c}}\,
        \widetilde F_{I,X}(q^2)
    \right]
    u_{\Lambda^+_c}(p+q),
    \label{eq:FFdecomposition}
\end{aligned}
\end{equation}
where $p$ and $q$ denote the momenta of the final-state meson and
invisible fermion, respectively. Here $I=1,2,3$
labels the flavour structure of the three-quark operator.

\section{Light-Cone Sum Rules}
\label{sec:LCSR}

To determine the form factors in \cref{eq:FFdecomposition},
we consider the correlation function
\begin{equation}
\begin{aligned}
\Pi_{I,X}^{(\mathcal J)}(P,q)
&=
-i\int d^4x\,e^{-iP\cdot x}\\
&\quad\times\langle M(p)|
T\left\{
\bar j_{\Lambda_c}^{(\mathcal J)}(x),
\widetilde{\mathcal O}_{I,X}(0)
\right\}
|0\rangle,
\\
&\hspace{2em}\mathcal J=\mathcal P,\mathcal A,
\label{eq:correlator}
\end{aligned}
\end{equation}
where $j_{\Lambda_c}^{(\mathcal J)}$ is the $\Lambda_c$
interpolating current. The momentum $q$ flows through the
effective operator, and $P=p+q$ is the momentum in the
$\Lambda_c$ channel.

The interpolating current carrying the quantum numbers of the
$\Lambda_c$ baryon is not unique. To assess the resulting
dependence within the truncated sum-rule calculation, we use
the pseudoscalar ($\mathcal P$) and axial-vector ($\mathcal A$)
currents~\cite{Khodjamirian:2011jp},
\begin{equation}
\begin{aligned}
j_{\Lambda_c}^{(\mathcal P)}
&=
\epsilon^{abc}
\left(u^{aT}C\gamma_5d^b\right)c^c,
\\
j_{\Lambda_c}^{(\mathcal A)}
&=
\epsilon^{abc}
\left(u^{aT}C\gamma_5\gamma_\mu d^b\right)
\gamma^\mu c^c,
\label{eq:LambdaCcurrent}
\end{aligned}
\end{equation}
where $a,b,c$ are colour indices and $C$ is the
charge-conjugation matrix.

The correlation function is decomposed into independent
Dirac structures as
\begin{flalign}
&\Pi^{(\mathcal J)}_{I,X}(P,q) = &&
\nonumber\\[-2pt]
&\Big[
\Pi^{(1,\mathcal J)}_{I,X}
+\slashed P\,\Pi^{(2,\mathcal J)}_{I,X}
+\slashed q\,\Pi^{(3,\mathcal J)}_{I,X}
+\slashed q\slashed P\,\Pi^{(4,\mathcal J)}_{I,X}
\Big]P_R
&&\nonumber\\
&+\Big[
\Pi^{(5,\mathcal J)}_{I,X}
+\slashed P\,\Pi^{(6,\mathcal J)}_{I,X}
+\slashed q\,\Pi^{(7,\mathcal J)}_{I,X}
+\slashed q\slashed P\,\Pi^{(8,\mathcal J)}_{I,X}
\Big]P_L\,,
&&\label{eq:corr_decomp}
\end{flalign}
where the dependence of the invariant
amplitudes on $P^2$ and $q^2$ is suppressed for brevity,
$\Pi^{(n,\mathcal J)}_{I,X}
\equiv\Pi^{(n,\mathcal J)}_{I,X}(P^2,q^2)$,
with $n=1,\ldots,8$.

The correlation function is evaluated by inserting hadronic
states with the $\Lambda_c$ quantum numbers and, at sufficiently
large spacelike virtualities,
$-P^2\gg\Lambda_{\rm QCD}^2$, through a light-cone OPE
in meson DAs. These representations are matched using a
dispersion relation in $P^2$ and quark--hadron duality.
The QCD calculation is presented in \cref{sec:OPE}.

Both currents couple to the positive-parity ground state
$\Lambda_c$ and to negative-parity states. We explicitly
retain the lowest negative-parity state,
$\Lambda_c^\ast$, following~\cite{Khodjamirian:2011jp}.
The corresponding current couplings are defined by
\begin{equation}
\begin{aligned}
\langle\Lambda_c(P,s)|
\bar j_{\Lambda_c}^{(\mathcal J)}|0\rangle
&=
m_{\Lambda_c}\lambda_{\Lambda_c}^{(\mathcal J)}
\bar u_{\Lambda_c}(P,s),
\\
\langle\Lambda_c^\ast(P,s)|
\bar j_{\Lambda_c}^{(\mathcal J)}|0\rangle
&=
m_{\Lambda_c^\ast}\lambda_{\Lambda_c^\ast}^{(\mathcal J)}
\bar u_{\Lambda_c^\ast}(P,s)\gamma_5.
\label{eq:lambda_dconstants}
\end{aligned}
\end{equation}
In analogy with \cref{eq:FFdecomposition}, the negative-parity
transition form factors are parametrised as
\begin{equation}
\begin{aligned}
&\langle M(p)|\widetilde{\mathcal O}_{I,X}
|\Lambda_c^\ast(P,s)\rangle
\\
&=
P_R
\left[
G_{I,X}(q^2)
+\frac{\slashed q}{m_{\Lambda_c^\ast}}
\widetilde G_{I,X}(q^2)
\right]
\gamma_5u_{\Lambda_c^\ast}(P,s).
\label{eq:GGdecomp_lambdastar}
\end{aligned}
\end{equation}

Inserting the intermediate states and performing the spin
sums gives the hadronic representation
\begin{equation}
\begin{aligned}
\Pi^{(\mathcal J)}_{I,X}&(P,q)=\\
&\hspace{-1.5em}
\frac{m_{\Lambda_c}\lambda_{\Lambda_c}^{(\mathcal J)}}
{m_{\Lambda_c}^2-P^2}\,P_R
\left[
F_{I,X}(q^2)
+\frac{\slashed q}{m_{\Lambda_c}}
\widetilde F_{I,X}(q^2)
\right]
(\slashed P+m_{\Lambda_c})
\\
&\hspace{-1.5em}
-\frac{m_{\Lambda_c^\ast}\lambda_{\Lambda_c^\ast}^{(\mathcal J)}}
{m_{\Lambda_c^\ast}^2-P^2}\,P_R
\left[
G_{I,X}(q^2)
+\frac{\slashed q}{m_{\Lambda_c^\ast}}
\widetilde G_{I,X}(q^2)
\right]
(\slashed P-m_{\Lambda_c^\ast})
\\
&\hspace{-1.5em}
+\int_{s_h}^{\infty}ds\,
\frac{\rho^{h,(\mathcal J)}_{I,X}(s,q^2)}
{s-P^2}.
\end{aligned}
\label{eq:hadronic_representation}
\end{equation}
Here $\rho^{h,(\mathcal J)}_{I,X}$ contains higher resonances
and continuum states and is decomposed in the Dirac basis
of \cref{eq:corr_decomp}. The physical continuum threshold is
$s_h=(m_{\Lambda_c}+2m_\pi)^2$, corresponding to the lowest baryon-two-meson continuum state.
Subtraction terms are omitted throughout, since they vanish
under the Borel transformation.

On the QCD side, each invariant amplitude admits the
dispersion representation
\begin{equation}
\Pi^{(n,\mathcal J)}_{I,X}(P^2,q^2)
=
\frac{1}{\pi}
\int_{m_c^2}^{\infty} ds\,
\frac{\operatorname{Im}\Pi^{(n,\mathcal J)}_{I,X}(s,q^2)}
{s-P^2}.
\label{eq:disp_invariants}
\end{equation}
The hadronic representation contains the isolated
$\Lambda_c$ and $\Lambda_c^\ast$ poles and a continuum
above $s_h$, whereas the QCD spectral density starts at
the partonic threshold $m_c^2$.

Quark--hadron duality approximates the higher-state
contribution by the QCD spectral density above an effective
continuum threshold. Following~\cite{Khodjamirian:2011jp},
we adopt a common threshold $s_0^{(\mathcal J)}$ for the
invariant amplitudes of each current:
\begin{equation}
\begin{aligned}
&\int_{s_h}^{\infty}ds\,
\frac{\rho^{h,(n,\mathcal J)}_{I,X}(s,q^2)}{s-P^2}
\simeq
\frac{1}{\pi}
\int_{s_0^{(\mathcal J)}}^{\infty}ds\,
\frac{\operatorname{Im}
\Pi^{(n,\mathcal J)}_{I,X}(s,q^2)}{s-P^2}.
\label{eq:duality_invariants}
\end{aligned}
\end{equation}
Unlike the physical threshold $s_h$, $s_0^{(\mathcal J)}$
is an effective parameter of the duality approximation.
\begin{figure}[!t]
    \centering
    \includegraphics[width=0.48\columnwidth]{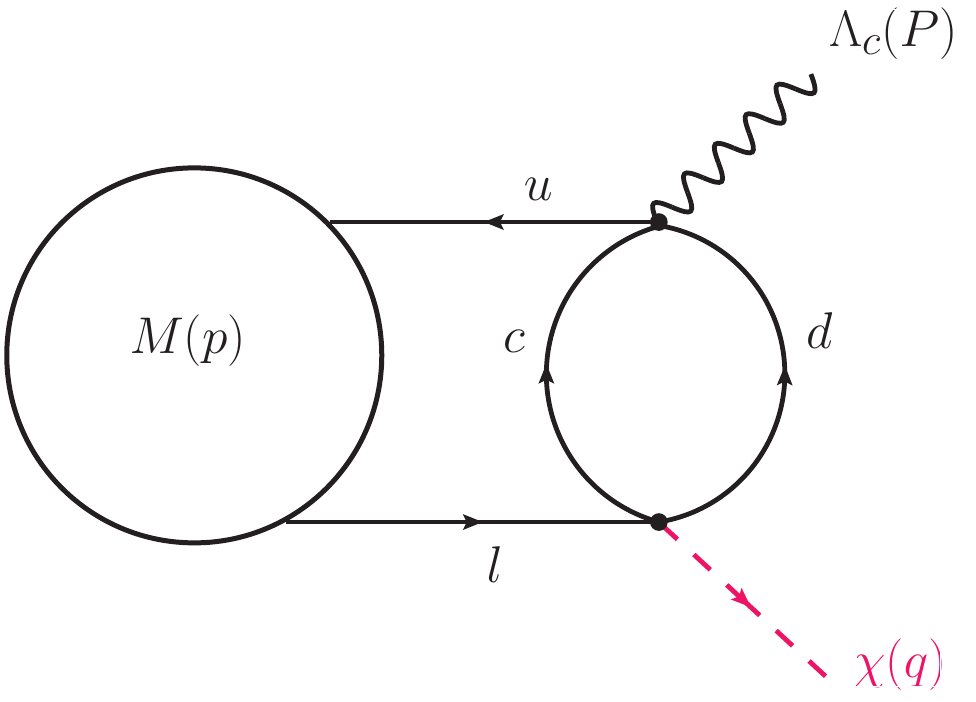}
    \hfill
    \includegraphics[width=0.48\columnwidth]{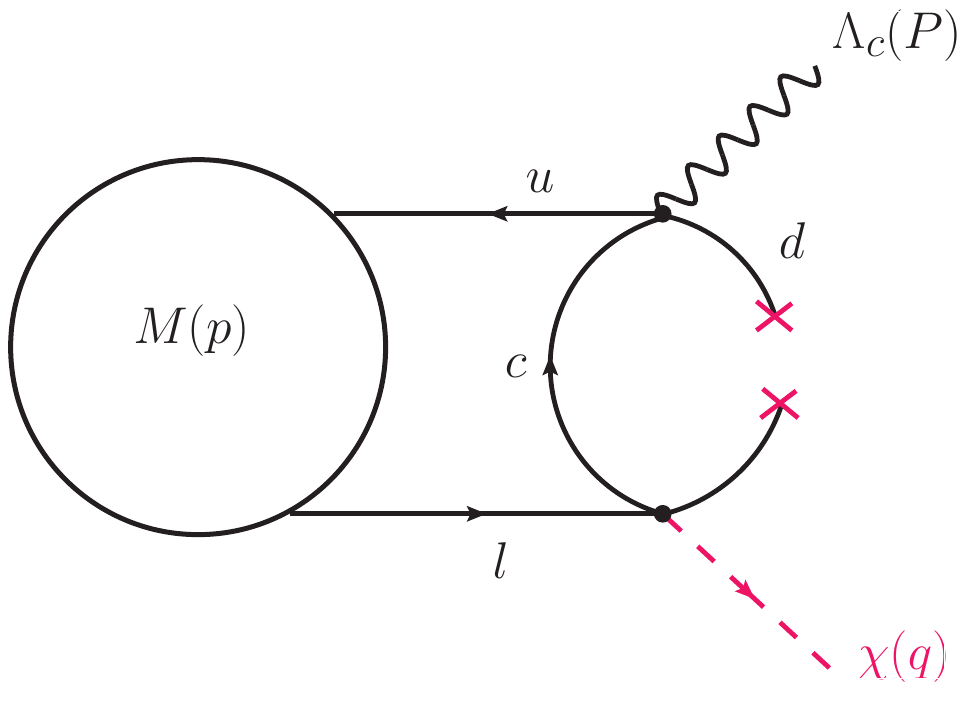}
    \caption{The two contributions retained in the light-cone OPE:
LO two-particle terms (left) and factorised four-particle
terms (right). Here $M=\pi^+,K^+$ and $\ell=d,s$,
respectively.}
    \label{fig:OPEdiagrams}
\end{figure}

Matching the Dirac structures in
\cref{eq:hadronic_representation,eq:corr_decomp}, the
negative-parity pole cancels in the combinations
\begin{equation}
\begin{gathered}
\Pi^{(1,\mathcal J)}_{I,X}(P^2,q^2)
+m_{\Lambda_c^\ast}\Pi^{(6,\mathcal J)}_{I,X}(P^2,q^2),
\\
m_{\Lambda_c^\ast}\Pi^{(4,\mathcal J)}_{I,X}(P^2,q^2)
+\Pi^{(7,\mathcal J)}_{I,X}(P^2,q^2).
\label{eq:parity_projection}
\end{gathered}
\end{equation}
The common-threshold prescription allows continuum
subtraction to be applied to these combinations.
A Borel transformation in $P^2$, with Borel parameter
$M^2$, then yields
\begin{widetext}
\begin{align}
F^{(\mathcal J)}_{I,X}(q^2)
&=\frac{e^{m_{\Lambda_c}^2/M^2}}{\pi m_{\Lambda_c}\lambda_{\Lambda_c}^{(\mathcal J)}(m_{\Lambda_c}+m_{\Lambda_c^\ast})}
\int_{m_c^2}^{s_0^{(\mathcal J)}}\!ds\,e^{-s/M^2}
\Big[\operatorname{Im}\Pi^{(1,\mathcal J)}_{I,X}(s,q^2)+m_{\Lambda_c^\ast}\operatorname{Im}\Pi^{(6,\mathcal J)}_{I,X}(s,q^2)\Big],
\label{eq:sumrule_F}
\\[4pt]
\widetilde F^{(\mathcal J)}_{I,X}(q^2)
&=\frac{e^{m_{\Lambda_c}^2/M^2}}{\pi\lambda_{\Lambda_c}^{(\mathcal J)}(m_{\Lambda_c}+m_{\Lambda_c^\ast})}
\int_{m_c^2}^{s_0^{(\mathcal J)}}\!ds\,e^{-s/M^2}
\Big[m_{\Lambda_c^\ast}\operatorname{Im}\Pi^{(4,\mathcal J)}_{I,X}(s,q^2)+\operatorname{Im}\Pi^{(7,\mathcal J)}_{I,X}(s,q^2)\Big].
\label{eq:sumrule_Ftilde}
\end{align}
\end{widetext}
Further details of the construction of baryonic hadronic
representations, including contributions from states of
both parities, can be found in~\cite{Biswas:2026oxq}.
\section{Light-cone OPE}
\label{sec:OPE}

We now turn to the QCD calculation of the correlation function in
\cref{eq:correlator}. For sufficiently large spacelike virtualities,
$-P^2\gg\Lambda_{\rm QCD}^2$, the correlation function is dominated by
light-cone distances, $x^2\simeq 0$, and can be expanded near the light
cone in terms of the DAs of the external
pseudoscalar meson.

For definiteness, in the following we illustrate the light-cone OPE using the operator
$\widetilde{\mathcal O}_{3,R}$ in the pion channel and the pseudoscalar
interpolating current $j_{\Lambda_c}^{(\mathcal P)}$. For this channel,
we neglect $m_\pi^2$ in the kinematics and in the hard-scattering
kernels, while retaining the nonperturbative parameters entering the
pion DAs. 
At the accuracy considered in this work, we include two classes of
contributions to the light-cone OPE, shown in
\cref{fig:OPEdiagrams}: the LO two-particle terms and the factorised
four-particle terms involving the quark condensate and
two-particle DAs. We discuss these contributions
separately in the following.
\subsection{LO two-particle contribution}
\label{sec:loop_contribution}

At leading order in $\alpha_s$, the charm- and $d$-quark
fields in the interpolating current are contracted with the
corresponding fields in the effective operator. The resulting bilocal
matrix element of the remaining light-quark fields is then expressed
in terms of the light-cone DAs of the pion. The calculation for the
axial-vector current, as well as for the remaining operators and the
kaon channel, proceeds analogously.

Using
\begin{equation}
\bar j_{\Lambda_c}^{(\mathcal P)}(x)
=
-\epsilon^{abc}\,
\bar c^c(x)
\left[
\bar d^b(x)\gamma_5 C\,\bar u^{aT}(x)
\right],
\end{equation}
together with
\begin{equation}
\widetilde{\mathcal O}_{3,R}(0)
=
i\epsilon^{\alpha\beta\gamma}
P_R d^\gamma(0)
\left[
\bar c^{C,\alpha}(0)P_R d^\beta(0)
\right],
\end{equation}
after contracting the charm-quark fields, $\bar d^b(x)$ can
be contracted with either $d^\gamma(0)$ outside the diquark
bilinear or $d^\beta(0)$ inside it. We refer to these as the
outer-$d$ and diquark-$d$ contractions, respectively.
In the following, we illustrate the calculation explicitly for the
outer-$d$ contraction; the diquark-$d$ contraction is treated
analogously.

For the outer-$d$ contraction, the remaining light-quark fields form
the bilocal pion matrix element
\begin{equation}
\langle \pi^+(p)|
\bar u_\omega^a(x)d_\xi^\beta(0)
|0\rangle
=
\delta^{a\beta}
[\mathcal M_\pi(x)]_{\xi\omega},
\label{eq:pion_matrix_element}
\end{equation}
where $\mathcal M_\pi(x)$ denotes the colour-stripped bilocal pion
matrix element in Dirac space. Performing the colour contractions, the outer-$d$ contribution
to the correlation function becomes
\begin{align}
\Pi_{3,R}^{(\mathcal P,\mathrm{out})}(P,q)
&=
-6
\int d^4x\,e^{-iP\cdot x}\, P_R
\big[iS_d(-x)\big]
\gamma_5 \nonumber\\*
&\quad\times C\,
\mathcal M_\pi^T(x)\, C^{-1} P_R
\big[iS_c(-x)\big].
\label{eq:pion_contraction_outer_color}
\end{align}
The nonlocal pion matrix element is expanded near the light
cone using the two-particle DA decomposition
of~\cite{Duplancic:2008ix}, with $x_1=x$ and $x_2=0$.
We retain the twist-two DA $\varphi_\pi$, the twist-three
DAs $\phi^p_{3\pi}$ and $\phi^\sigma_{3\pi}$, and the
twist-four DAs $\phi_{4\pi}$ and $\psi_{4\pi}$.

In the massless-light-quark limit, the Dirac structure in
\cref{eq:pion_contraction_outer_color} selects the combination
\begin{equation}
P_L\,C\mathcal M_\pi^T(x)C^{-1}P_R .
\label{eq:selection_P}
\end{equation}
Only structures containing an odd number of ordinary gamma
matrices survive. Thus, the twist-2 DA $\varphi_\pi$ and
twist-4 DAs $\phi_{4\pi}$ and $\psi_{4\pi}$ contribute,
whereas the twist-3 structures $\gamma_5$ and
$\sigma_{\rho\tau}\gamma_5$ are projected out.

For the axial-vector current, the outer-$d$ contraction gives
\begin{align}
\Pi_{3,R}^{(\mathcal A,\mathrm{out})}(P,q)
&=
-6
\int d^4x\,e^{-iP\cdot x}\, P_R
\big[iS_d(-x)\big]
\gamma_\mu\gamma_5 \nonumber\\*
&\qquad\times C\,
\mathcal M_\pi^T(x)\,
C^{-1}P_R
\big[iS_c(-x)\big]
\gamma^\mu .
\label{eq:pion_contraction_outer_color_A}
\end{align}
The additional $\gamma_\mu$ changes the adjacent chiral
projector, yielding
\begin{equation}
P_R\,C\mathcal M_\pi^T(x)C^{-1}P_R .
\label{eq:selection_A}
\end{equation}
The selection is therefore reversed: the twist-3 structures
$\gamma_5$ and $\sigma_{\rho\tau}\gamma_5$ survive, while
the twist-2 and twist-4 structures are projected out. The two
currents thus probe complementary subsets of pion DAs for
this operator at leading order.

The selection pattern also depends on the chirality of the
effective operator. Replacing the right-handed projector in
the quark bilinear by a left-handed one changes the projector
adjacent to the pion matrix element and reverses the selection
rule. The contributing DAs are therefore determined by both
the operator chirality and the
interpolating current.

Importantly, this selection occurs entirely within the light-quark part
of the correlation function. The charm-quark propagator appears only
after the chiral projector and does not participate in selecting the
pion DAs.

We now evaluate the light-cone expansion for the outer-$d$
contraction of $\widetilde{\mathcal O}_{3,R}$ with the
pseudoscalar current. Using the two-particle pion DA
decomposition~\cite{Duplancic:2008ix}, we extract the invariant amplitudes defined
in \cref{eq:corr_decomp}. The nonvanishing amplitudes are
$\Pi^{(1,\mathcal P)}_{3,R}$,
$\Pi^{(4,\mathcal P)}_{3,R}$,
$\Pi^{(6,\mathcal P)}_{3,R}$, and
$\Pi^{(7,\mathcal P)}_{3,R}$.

The same procedure applies to the remaining operators,
chiralities, and interpolating currents. After Fourier and
loop integration, the LO two-particle contribution to each
invariant amplitude takes the form\footnote{In \cref{eq:pion_OPE_generic}, only terms generating the discontinuity across the physical
charm cut in $P^2$ are displayed. Polynomial subtraction terms and
nonlogarithmic terms cancelling apparent kinematic singularities do not
contribute independently to this discontinuity.}
\begin{equation}
\begin{aligned}
\Pi^{(i,\mathcal J)}_{I,X}(P^2,q^2)
&=
\int_0^1 du\,
\mathcal F_{i;I,X}^{(\mathcal J)}
(u,P^2,q^2,m_c,m_M)\,
\\*
&\quad\times\ln\!\left[
\frac{m_c^2}
{\Delta_M(u,P^2,q^2)}
\right],
\label{eq:pion_OPE_generic}
\end{aligned}
\end{equation}
where
\begin{equation}
\Delta_M(u,P^2,q^2)
=
m_c^2
-(1-u)P^2
-u q^2
+u(1-u)m_M^2 ,
\label{eq:Delta_M}
\end{equation}
with $M=\pi,K$. The function
$\mathcal F_{i;I,X}^{(\mathcal J)}$ can be decomposed in terms of the
two-particle light-meson DAs as
\begin{align}
\mathcal F_{i;I,X}^{(\mathcal J)}
={}&
\mathcal F_{i;I,X}^{(2,\mathcal J)}
\,\phi_{2M}(u)+
\mathcal F_{i;I,X}^{(3,p,\mathcal J)}
\,\phi_{3M}^{p}(u)\nonumber\\*
&+
\mathcal F_{i;I,X}^{(3,\sigma,\mathcal J)}
\,\phi_{3M}^{\sigma}(u)
+
\mathcal F_{i;I,X}^{(4,\phi,\mathcal J)}
\,\phi_{4M}(u)\nonumber\\*
&+
\mathcal F_{i;I,X}^{(4,\psi,\mathcal J)}
\,\Psi_{4M}(u).
\label{eq:pion_OPE_DA_structure}
\end{align}
Here $\phi_{2\pi}\equiv\varphi_\pi$, and
$\Psi_{4M}(u)=\int_0^u dv\,\psi_{4M}(v)$. The superscripts
identify the twist and, where necessary, distinguish DAs of
the same twist. Only a subset contributes for each operator
and interpolating current; the nonvanishing coefficient
functions are collected in \cref{app:Pis}.

We next extract the spectral densities entering the sum rules.
In the pion channel, setting $m_\pi^2=0$, the logarithm becomes
\begin{equation}
L_\pi(u,P^2,q^2)
=
\ln\!\left[
\frac{m_c^2}
{m_c^2-(1-u)P^2-uq^2}
\right].
\label{eq:L_pi}
\end{equation}
With $P^2=s+i\epsilon$, its imaginary part is
\begin{equation}
\frac{1}{\pi}\operatorname{Im}L_\pi(u,s,q^2)
=
\theta\!\left((1-u)s+uq^2-m_c^2\right).
\label{eq:imag_log}
\end{equation}
For $s\geq m_c^2$ and $q^2<m_c^2$, the cut restricts
the momentum fraction to
\begin{equation}
0\leq u\leq u_0(s,q^2),
\qquad
u_0(s,q^2)
=
\frac{s-m_c^2}{s-q^2}.
\label{eq:u0_pi}
\end{equation}
The invariant spectral densities are obtained by integrating
the corresponding spectral integrands over this interval,
\begin{equation}
\operatorname{Im}\Pi^{(i,\mathcal J)}_{I,X}(s,q^2)
=
\int_0^{u_0(s,q^2)}du\,
\operatorname{Im}\Pi^{(i,\mathcal J)}_{I,X}(s,q^2,u).
\label{eq:pion_imaginary_u}
\end{equation}
The remaining integration over $s$ is then Borel weighted and
continuum subtracted according to the sum rules in
\cref{eq:sumrule_F,eq:sumrule_Ftilde}.

The remaining operators and the axial-vector current are
treated analogously. In the kaon channel, the pion DAs are
replaced by kaon DAs and the finite kaon mass is retained.
The strange-quark field enters the bilocal kaon matrix element
rather than a hard propagator, so its mass enters through
the DA parameters rather than explicitly in the hard-scattering
kernels. The retained $SU(3)$-flavour breaking therefore arises
from the kaon mass and DAs, including their nonperturbative
parameters. The nonvanishing coefficient functions for both
channels are collected in \cref{app:Pis}.

\subsection{Factorised four-particle contribution}

The DA selection discussed above follows from the Dirac
structure of the free massless light-quark propagator,
$S_d(x)\propto\slashed{x}$.
A second contribution to the light-cone OPE arises when the
corresponding $d\bar d$ pair is left uncontracted. This leaves
four light-quark fields between the external meson state and
the vacuum. In the factorisation approximation, their matrix
element is expressed as the product of the bilocal matrix
element defining the two-particle meson DAs and the vacuum
matrix element of the remaining $d\bar d$
pair~\cite{Braun:1999uj,Khodjamirian:2002pk}. At leading order in the local
expansion, the latter is parametrised as
\begin{equation}
\langle 0|
d_\alpha^a(0)\,\bar d_\beta^b(0)
|0\rangle
=
-\frac{\delta^{ab}\delta_{\alpha\beta}}{12}
\langle\bar d d\rangle ,
\label{eq:d_quark_condensate}
\end{equation}
where $a,b$ and $\alpha,\beta$ are colour and Dirac indices,
respectively.

These terms represent the factorisable four-particle
contributions to the light-cone OPE. Their total twist is
five, six, or seven, depending on whether the accompanying
two-particle DA has twist two, three, or four,
respectively~\cite{Rusov:2017chr}. Below, the coefficient functions are labelled by the twist
of the accompanying two-particle DA, while the numerical
decomposition distinguishes the total twists.
Non-factorisable four-particle contributions are not included,
as their systematic treatment requires additional
nonperturbative input. The retained terms involve no loop
integration and can therefore be numerically enhanced relative
to the LO two-particle contribution.

The local vacuum matrix element is proportional to the identity
in Dirac space and therefore changes the DA selection pattern.
Consequently, the factorisable four-particle terms can involve
DAs projected out in the LO two-particle contribution.

For illustration, we again consider the outer-$d$ contraction
of $\widetilde{\mathcal O}_{3,R}$ with the pseudoscalar current.
Applying \cref{eq:d_quark_condensate} to
\cref{eq:pion_contraction_outer_color} gives
\begin{align}
\Pi_{3,R}^{(\mathcal P,\mathrm{out}),\langle\bar d d\rangle}(P,q)
&=
\frac{\langle\bar d d\rangle}{2}
\int d^4x\,e^{-iP\cdot x}\,\nonumber\\*
&\quad\times P_R C\,
\mathcal M_\pi^T(x)\,
C^{-1}P_R
\left[iS_c(-x)\right].
\label{eq:pion_contraction_outer_condensate}
\end{align}
The absence of $\slashed x$ changes the chiral projection to
$P_R C\mathcal M_\pi^T(x)C^{-1}P_R$, reversing the DA
selection. Only the twist-3 DAs $\phi^p_{3\pi}$ and
$\phi^\sigma_{3\pi}$ survive, yielding a factorisable
twist-six contribution. The LO two-particle and factorisable
four-particle terms thus probe complementary subsets of the
pion DAs and together involve all two-particle pion DAs
retained in our analysis.

For $M=\pi,K$, the factorisable four-particle contributions
 to the invariant amplitudes can be written as
\begin{equation}
\begin{aligned}
\Pi^{(i,\mathcal J),\langle\bar d d\rangle}_{I,X}(P^2,q^2)
&= f_M N_c\langle\bar d d\rangle \int_0^1 du\,\Bigl[
\mathcal C^{(2,\mathcal J)}_{i;I,X}\,\phi_{2M}(u)
\\
&\hspace{-2.5em}
+\mu_M\mathcal C^{(3,p,\mathcal J)}_{i;I,X}\,\phi^p_{3M}(u)
+\mu_M\mathcal C^{(3,\sigma,\mathcal J)}_{i;I,X}\,\phi^\sigma_{3M}(u)
\\
&\hspace{-2.5em}
+\mathcal C^{(4,\phi,\mathcal J)}_{i;I,X}\,\phi_{4M}(u)
+\mathcal C^{(4,\psi,\mathcal J)}_{i;I,X}\,\Psi_{4M}(u)
\Bigr].
\end{aligned}
\label{eq:condensate_generic}
\end{equation}
Here $N_c=3$ is the number of colours, and $\mu_M$ denotes
the chirally enhanced parameter of the meson $M$. 
For brevity, we write
$\mathcal C^{(t,\mathcal J)}_{i;I,X}
\equiv
\mathcal C^{(t,\mathcal J)}_{i;I,X}(u,P^2,q^2,m_c,m_M)$. Their superscripts
refer to the twist of the accompanying two-particle DA,
rather than the total twist of the four-particle contribution,
which is five, six, or seven, respectively. The additional
labels $p$, $\sigma$, $\phi$, and $\psi$ distinguish DAs
of the same twist. The coefficients contain powers of
$\Delta_M^{-1}$ up to $\Delta_M^{-3}$, with $\Delta_M$
defined in \cref{eq:Delta_M}. Only a subset is nonvanishing
for each operator and interpolating current; the explicit
coefficients are collected in \cref{app:Pis}.

Unlike the LO two-particle terms, whose physical discontinuity
arises from the logarithm, the factorisable four-particle terms
contain poles in $\Delta_M$. Their imaginary parts generate
delta functions and their derivatives:
\begin{equation}
\begin{gathered}
\frac{1}{\pi}\operatorname{Im}
\frac{1}{(\Delta_M-i0)^n}
=
\frac{(-1)^{n-1}}{(n-1)!}\,
\delta^{(n-1)}(\Delta_M),
\\ n=1,2,\ldots,
\label{eq:condensate_imaginary}
\end{gathered}
\end{equation}
where $\delta^{(n-1)}$ denotes the $(n-1)$th derivative
with respect to its argument. 
This relation determines the spectral densities from
\cref{eq:condensate_generic}. They are added to the LO
two-particle spectral densities before insertion into
\cref{eq:sumrule_F,eq:sumrule_Ftilde}.

\section{Numerical analysis}
\label{sec:Numerical}

We now turn to the numerical evaluation of the sum rules in
\cref{eq:sumrule_F,eq:sumrule_Ftilde}. The required inputs include the
meson-DA parameters, the quark and hadron masses, the couplings
$\lambda_{\Lambda_c}^{(\mathcal J)}$ of the $\Lambda_c$ interpolating
currents, and the effective continuum thresholds
$s_0^{(\mathcal J)}$.
The input parameters used in our analysis are summarised in
\cref{tab:input}.

We adopt the pion and kaon DA parametrisations and inputs
used in~\cite{Mohamed:2026hab}, including two-particle DAs
up to twist four. The pion parametrisations follow
\cite{Duplancic:2008ix}, and the kaon parametrisations follow
\cite{Khodjamirian:2009ys,Ball:2006wn}.

The twist-two Gegenbauer expansions are truncated at
$a_4^\pi$ and $a_2^K$, respectively; odd pion moments vanish
in the isospin limit. The moments $a_2^\pi$ and $a_4^\pi$
are taken from~\cite{Cheng:2020vwr}, and $a_1^K$ and $a_2^K$
from~\cite{Chetyrkin:2007vm,Ball:2006wn}. The parameters
$\mu_\pi$, $\mu_K$, and the remaining twist-three and
twist-four inputs follow~\cite{Mohamed:2026hab}.
All scale-dependent nonperturbative parameters are evolved
to the factorisation scale using leading-order
renormalisation-group equations.

The meson decay constants are taken from FLAG averages and
the quark and hadron masses from the PDG. In the sum rules,
we neglect $m_\pi^2$ but retain the kaon mass in the kinematics.
As discussed in \cref{sec:OPE}, $m_s$ enters through the kaon
DA parameters rather than explicitly in the hard-scattering
kernels.

The factorisable four-particle terms involve
$\langle\bar d d\rangle$, which we express through the
Gell-Mann--Oakes--Renner relation in the isospin
limit~\cite{Gell-Mann:1968hlm},
\begin{equation}
\langle\bar d d\rangle
=
-\frac{1}{2}f_\pi^2\mu_\pi .
\label{eq:QC-dd-relation}
\end{equation}
The current couplings $\lambda_{\Lambda_c}^{(\mathcal J)}$,
defined in \cref{eq:lambda_dconstants}, are taken from the
two-point QCD sum-rule analysis for the pseudoscalar and
axial-vector currents~\cite{Khodjamirian:2011jp}.

We next specify the effective continuum threshold and the Borel window
entering the sum rules. For the continuum threshold, we adopt the value
used in~\cite{Khodjamirian:2011jp} and use a common threshold for
the two interpolating currents,
\begin{equation}
s_0^{(\mathcal P)}
=
s_0^{(\mathcal A)}
\equiv s_0
=
(10.0\pm0.5)\,\mathrm{GeV}^2 .
\label{eq:s0_numerical}
\end{equation}
The Borel window is chosen by examining the suppression of
higher-state contributions, the relative sizes of the retained
OPE terms, and the sensitivity of the form factors to $M^2$.
We adopt the working interval
\begin{equation}
3\,\GeV^2 \leq M^2 \leq 5\,\GeV^2 .
\label{eq:Borel_window}
\end{equation}
\renewcommand{\arraystretch}{1.2}
\begin{table*}[!t]
\centering
\caption{
Inputs used in the numerical analysis. The DA parameters
are quoted at $\mu=1\,\GeV$, except $\mu_\pi$ and $\mu_K$,
which are quoted at $2\,\GeV$. The quark masses are specified
at the scales indicated in the table. All scale-dependent
inputs are evolved to the working scale.
}
\label{tab:input}

\setlength{\tabcolsep}{4pt}
\begingroup
\setlength{\arrayrulewidth}{0.6pt}
\setlength{\lightrulewidth}{0.6pt}
\begin{tabular}{ccl|ccl}
\hline\hline
\multicolumn{6}{c}{Pion and kaon twist-two and twist-three DA parameters} \\
\midrule

$\mu_\pi$
& $(2.50 \pm 0.30)\,\GeV$
& \cite{Khodjamirian:2017fxg}
&
$\mu_K$
& $(2.49 \pm 0.26)\,\GeV$
& \cite{Khodjamirian:2017fxg}
\\

$a_2^\pi$
& $0.275 \pm 0.055$
& \cite{Cheng:2020vwr}
&
$a_1^K$
& $0.10 \pm 0.04$
& \cite{Chetyrkin:2007vm}
\\

$a_4^\pi$
& $0.185 \pm 0.065$
& \cite{Cheng:2020vwr}
&
$a_2^K$
& $0.25 \pm 0.15$
& \cite{Ball:2006wn}
\\

$f_3^\pi$
& $(0.0045 \pm 0.0015)\,\GeV^2$
& \cite{Ball:2006wn}
&
$f_3^K$
& $(0.0045 \pm 0.0020)\,\GeV^2$
& \cite{Ball:2006wn}
\\

$\omega_3^\pi$
& $-1.5 \pm 0.7$
& \cite{Ball:2006wn}
&
$\omega_3^K$
& $-1.2 \pm 0.7$
& \cite{Ball:2006wn}
\\

$\lambda_3^\pi$
& $0$
& ---
&
$\lambda_3^K$
& $1.6 \pm 0.4$
& \cite{Ball:2006wn}
\\
\midrule
\multicolumn{6}{c}{Pion and kaon twist-four DA parameters} \\
\midrule

$\delta_\pi^2$
& $(0.18 \pm 0.06)\,\GeV^2$
& \cite{Ball:2006wn}
&
$\delta_K^2$
& $(0.20 \pm 0.06)\,\GeV^2$
& \cite{Ball:2006wn}
\\

$\omega_4^\pi$
& $0.20 \pm 0.10$
& \cite{Ball:2006wn}
&
$\omega_4^K$
& $0.20 \pm 0.10$
& \cite{Ball:2006wn}
\\

$\kappa_{4\pi}$
& $0$
& ---
&
$\kappa_{4K}$
& $-0.12 \pm 0.01$
& \cite{Ball:2006wn}
\\

\midrule
\multicolumn{6}{c}{$\Lambda_c$ sum-rule parameters} \\
\midrule

$s_0$
& $(10.0 \pm 0.5)\,\GeV^2$
& \cite{Khodjamirian:2011jp}
&
$M^2$
& $[3.0,5.0]\,\GeV^2$
& ---
\\

$\lambda_{\Lambda_c}^{(\mathcal P)}$
& $\left(1.19^{+0.19}_{-0.28}\right)\times10^{-2}\,\GeV^2$
& \cite{Khodjamirian:2011jp}
&
$\lambda_{\Lambda_c}^{(\mathcal A)}$
& $\left(1.51^{+0.37}_{-0.39}\right)\times10^{-2}\,\GeV^2$
& \cite{Khodjamirian:2011jp}
\\

\midrule
\multicolumn{6}{c}{Other parameters} \\
\midrule

$m_{\pi^\pm}$ & $0.139570\,\GeV$
& \cite{ParticleDataGroup:2026aaa}
&
$m_{K^\pm}$
& $0.493677\,\GeV$
& \cite{ParticleDataGroup:2026aaa}
\\

$f_\pi$
& $(0.1302 \pm 0.0008)\,\GeV$
& \cite{FLAG:2024oxs}
&
$f_K$
& $(0.1557 \pm 0.0003)\,\GeV$
& \cite{FLAG:2024oxs}
\\

$\overline m_c(\overline m_c)$
& $(1.27 \pm 0.02)\,\GeV$
& \cite{ParticleDataGroup:2026aaa}
&
$\overline m_s(2\,\GeV)$
& $0.0934\,\GeV$
& \cite{ParticleDataGroup:2026aaa}
\\

$m_{\Lambda_c}$
& $2.28646\,\GeV$
& \cite{ParticleDataGroup:2026aaa}
&
$m_{\Lambda_c^\ast}$
& $2.59225\,\GeV$
& \cite{ParticleDataGroup:2026aaa}
\\

$m_{\Xi_c}$
& $2.47050\,\GeV$
& \cite{ParticleDataGroup:2026aaa}
&
$\tau_{\Lambda_c^+}$
& $2.026\times10^{-13}\,\mathrm{s}$
& \cite{ParticleDataGroup:2026aaa}
\\
\hline\hline
\end{tabular}
\endgroup

\end{table*}

\begin{figure*}[!t]
    \centering

    \includegraphics[width=0.45\textwidth]{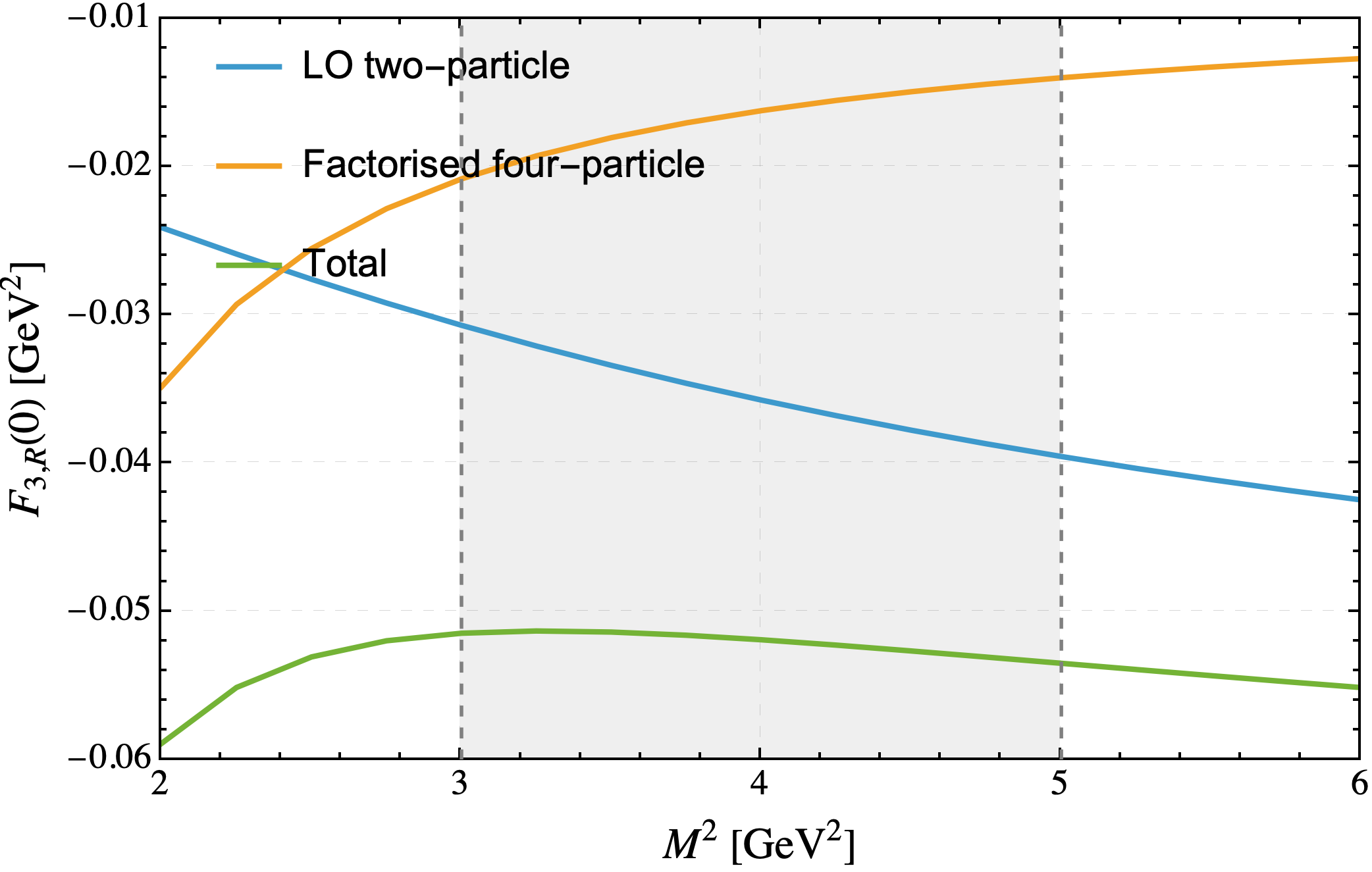}
    \hfill
    \includegraphics[width=0.45\textwidth]{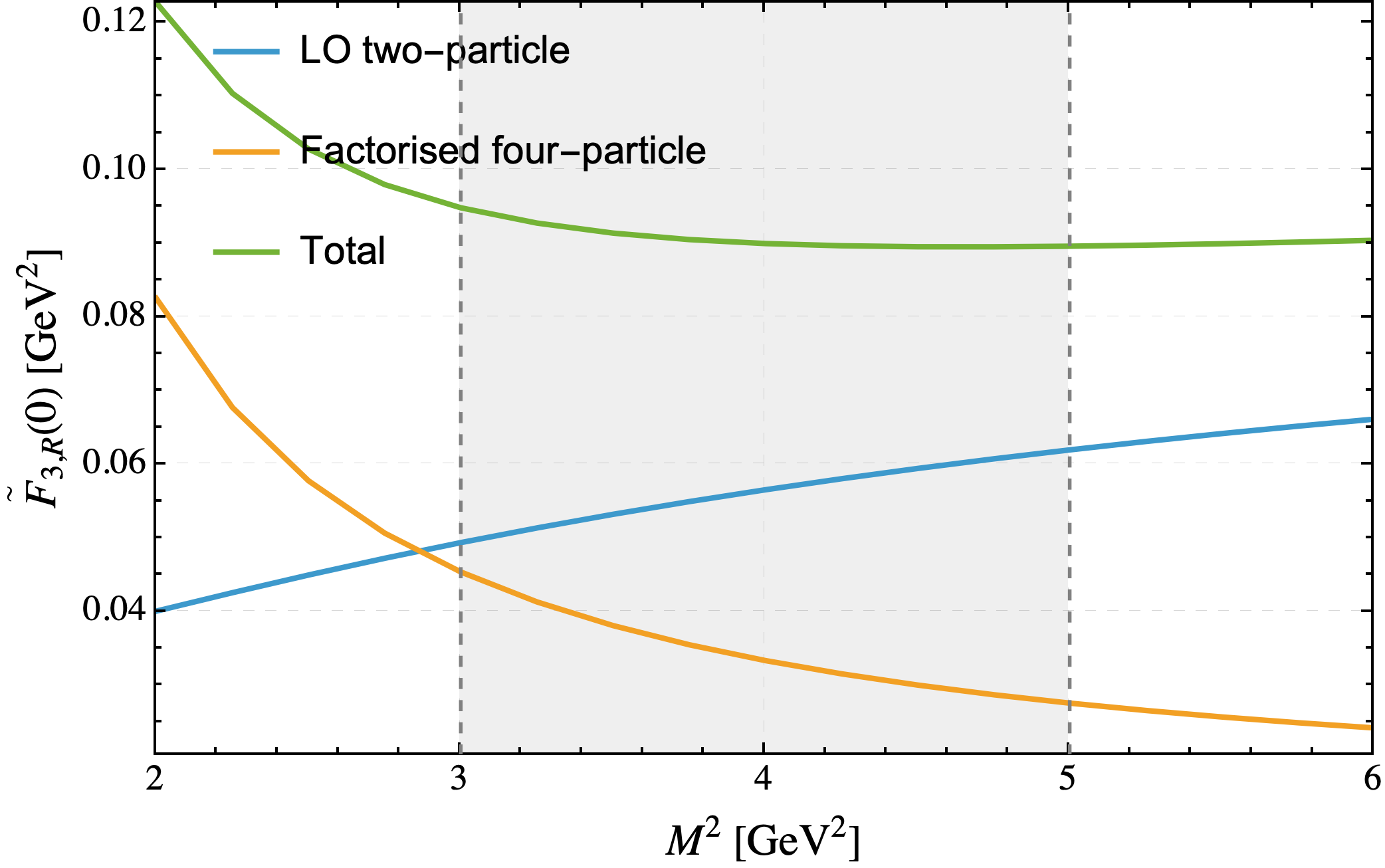}

    \vspace{0.25cm}

    \includegraphics[width=0.45\textwidth]{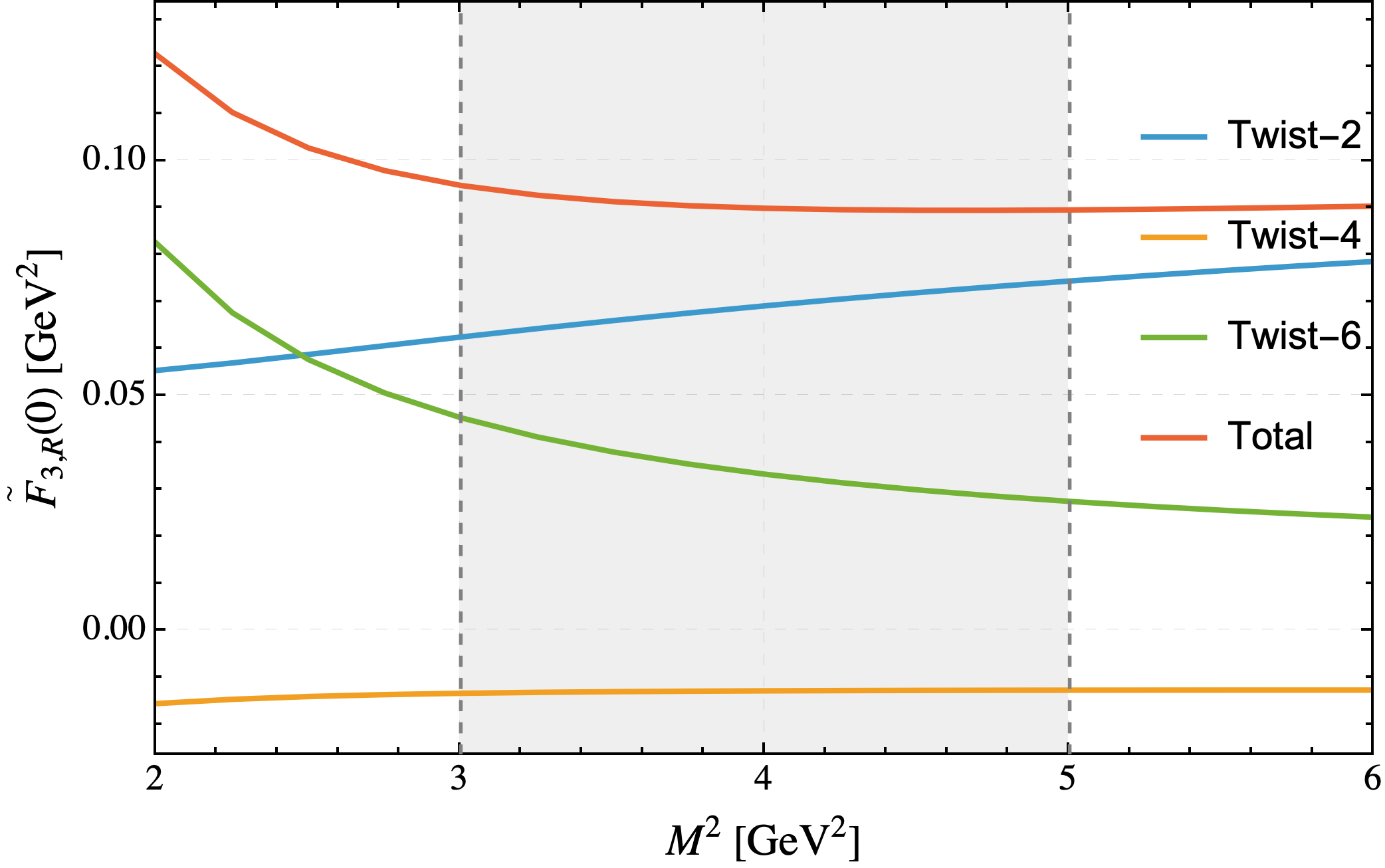}
    \hfill
    \includegraphics[width=0.45\textwidth]{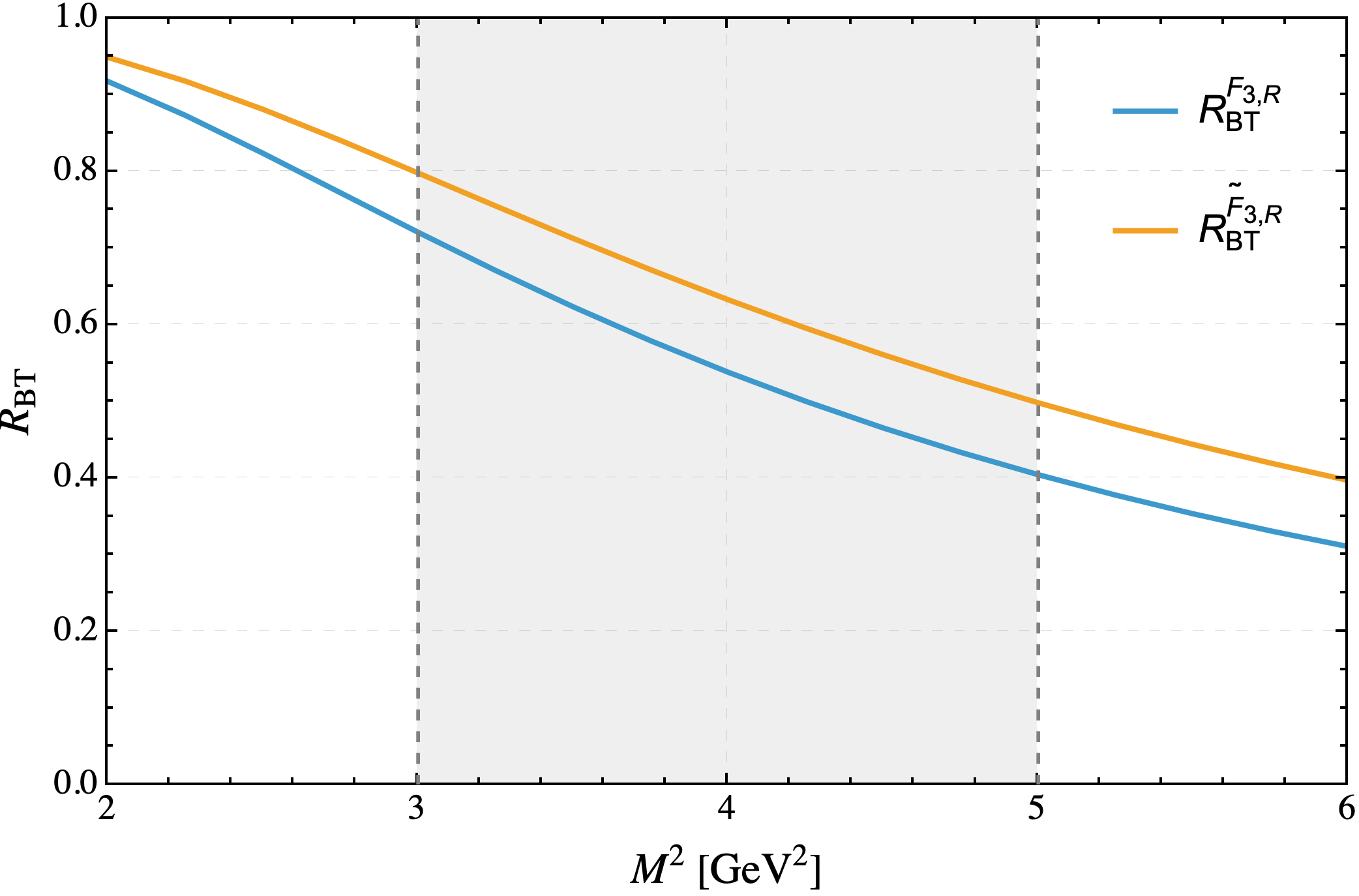}

\caption{
Borel dependence of the $\widetilde{\mathcal O}_{3,R}$
form factors obtained with the $\mathcal P$ current at
$q^2=0$ and $s_0=10\,\GeV^2$.
The upper panels show the LO two-particle, factorisable
four-particle, and total contributions to $F_{3,R}(0)$
(left) and $\widetilde F_{3,R}(0)$ (right).
The lower-left panel separates the twist-two, twist-four,
and factorisable twist-six contributions to
$\widetilde F_{3,R}(0)$. The lower-right panel shows the
below-threshold ratios for both form factors, including
both two- and four-particle contributions.
The shaded region marks the adopted Borel window
$3\,\GeV^2\leq M^2\leq5\,\GeV^2$.
}
    \label{fig:Q3R-sumrule-stability}
\end{figure*}

As a representative example, \cref{fig:Q3R-sumrule-stability}
shows $F_{3,R}$ and $\widetilde F_{3,R}$ obtained with the
$\mathcal P$ current at $q^2=0$ and $s_0=10\,\GeV^2$.
The upper panels display their Borel dependence, separating
the LO two-particle and factorisable four-particle contributions.
While the individual terms exhibit a more pronounced dependence
on $M^2$, their sum shows reduced variation throughout the
adopted Borel window.

The lower-left panel of \cref{fig:Q3R-sumrule-stability}
shows the twist decomposition of $\widetilde F_{3,R}$.
For the $\mathcal P$ current, the LO two-particle terms
have twist two and four, while the factorisable four-particle
term has twist six. Towards smaller $M^2$, the twist-six
contribution becomes comparable in magnitude to the
leading-twist term, indicating increased sensitivity to
higher-twist effects and motivating the exclusion of smaller
Borel parameters.

To quantify the relative contribution from the region below the
effective continuum threshold, we define the below-threshold fractions
\begin{equation}
\begin{gathered}
R_{\rm BT}^{F}(q^2,M^2,s_0)
=
\frac{F(q^2,M^2,s_0)}
     {F(q^2,M^2,\infty)},
\\
R_{\rm BT}^{\widetilde F}(q^2,M^2,s_0)
=
\frac{\widetilde F(q^2,M^2,s_0)}
     {\widetilde F(q^2,M^2,\infty)} .
\label{eq:RBT}
\end{gathered}
\end{equation}
These ratios are shown in the lower-right panel of
\cref{fig:Q3R-sumrule-stability}. They decrease with increasing
$M^2$, reflecting weaker continuum suppression, and remain
approximately $40\%$ and $50\%$ for $F_{3,R}$ and
$\widetilde F_{3,R}$, respectively, at $M^2=5\,\GeV^2$.
Together with the Borel dependence and relative sizes of the
retained OPE terms, they motivate the adopted working window.

For the $\mathcal A$ current, the LO two-particle contribution
involves twist-three DAs, while the factorisable four-particle
terms involve twist-two and twist-four DAs and have total
twist five and seven, respectively. The twist-three contribution
dominates throughout the adopted Borel window, and the
twist-seven term remains small for both $F_{3,R}$ and
$\widetilde F_{3,R}$.

The uncertainty of the form factors is obtained by varying the input parameters
within the ranges in \cref{tab:input}, including the meson-DA
parameters, decay constants, quark masses, and current couplings.
The dependence on the sum-rule parameters is assessed separately
by scanning the full adopted $(M^2,s_0)$ region and taking
the minimum and maximum of each form factor as its uncertainty
envelope.

Finally, the residual factorisation-scale dependence is
estimated by varying $\mu$ between $1$ and $2\,\GeV$ around
the central value $\mu=1.5\,\GeV$, a characteristic range
for charm LCSRs~\cite{Khodjamirian:2011jp}. At the present
order, this dependence enters through the running quark
masses and scale-dependent meson-DA parameters.
The parametric, sum-rule, and scale uncertainties are
combined in quadrature to obtain the quoted uncertainties below.

\begin{figure*}[!t]
    \centering
    \includegraphics[width=0.49\textwidth]{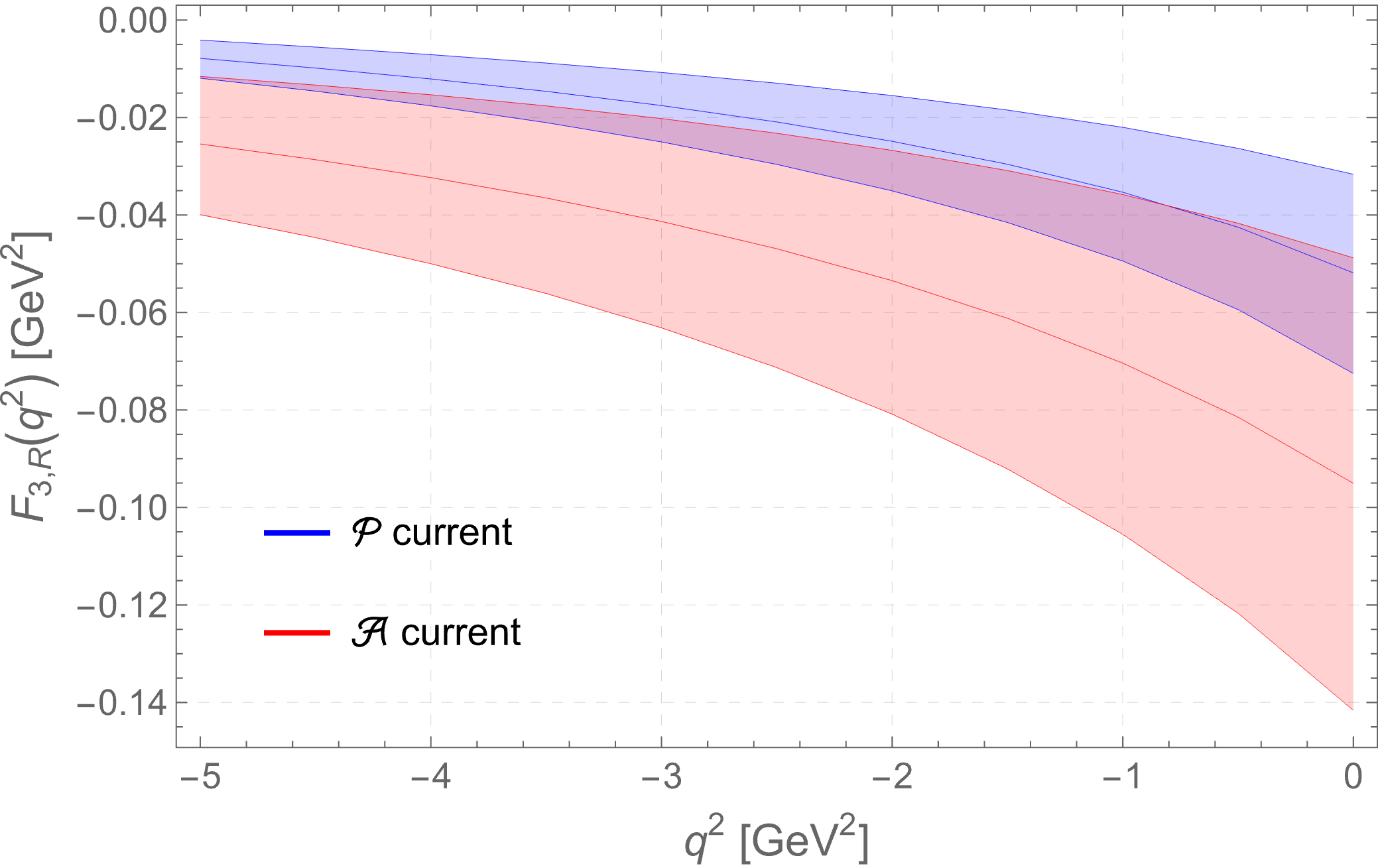}
    \hfill
    \includegraphics[width=0.49\textwidth]{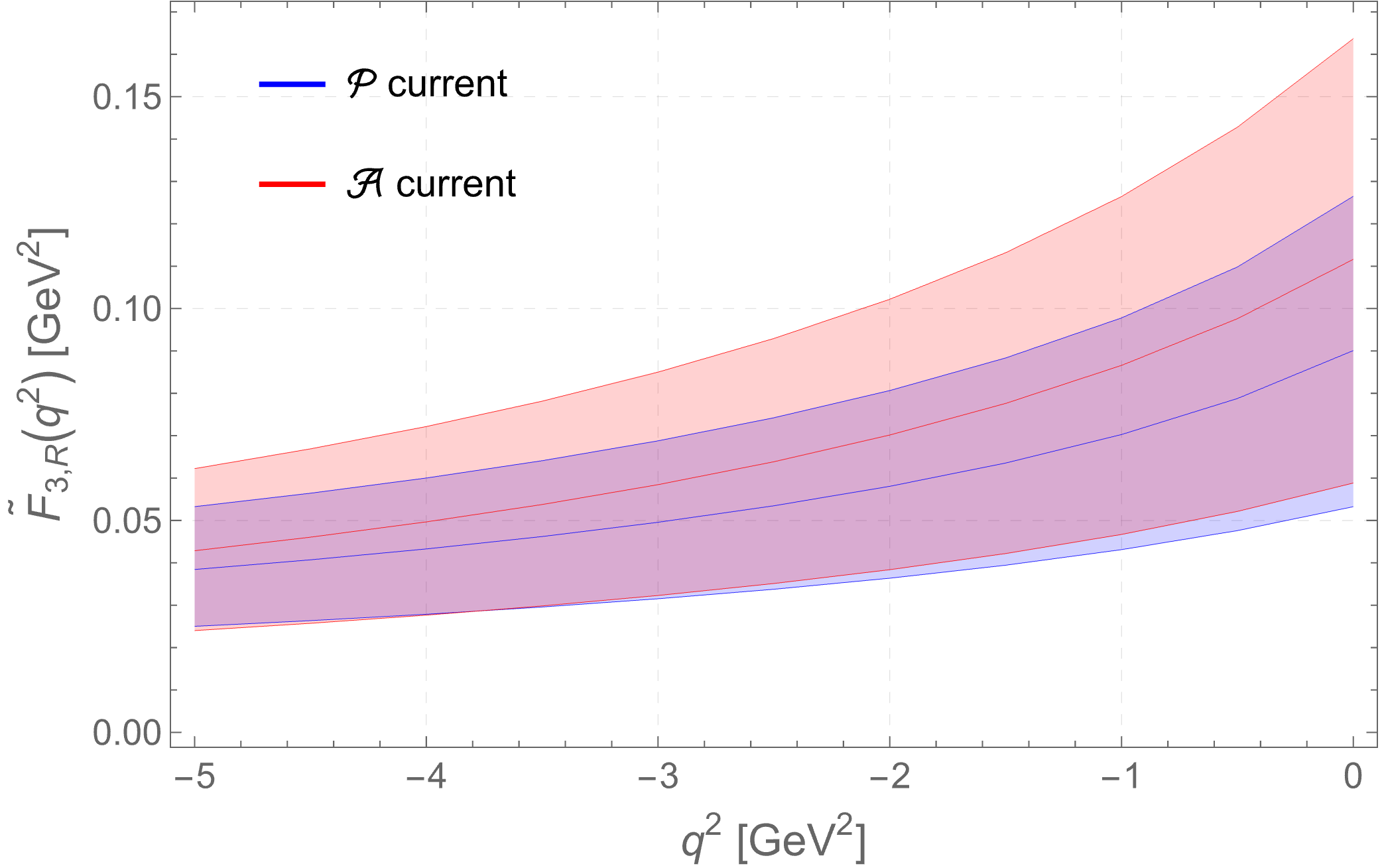}
    \caption{
    Form factors $F_{3,R}(q^2)$ (left) and
    $\widetilde F_{3,R}(q^2)$ (right) obtained with the
    $\mathcal P$ and $\mathcal A$ interpolating currents in the
    spacelike region $-5\,\GeV^2\leq q^2\leq0$.
    }
    \label{fig:Q3R-form-factors}
\end{figure*}

We first quote the form factors at $q^2=0$ as a convenient reference
point for comparing the two interpolating currents:
\begin{align}
F_{3,R}^{(\mathcal P)}(0)
&=
\left(
-5.19^{+2.02}_{-2.07}
\right)\times10^{-2}\,\GeV^2,\nonumber\\*
F_{3,R}^{(\mathcal A)}(0)
&=
\left(
-9.50^{+4.63}_{-4.66}
\right)\times10^{-2}\,\GeV^2,
\\
\widetilde F_{3,R}^{(\mathcal P)}(0)
&=
\left(
9.01^{+3.64}_{-3.68}
\right)\times10^{-2}\,\GeV^2,\nonumber\\*
\widetilde F_{3,R}^{(\mathcal A)}(0)
&=
\left(
11.16^{+5.21}_{-5.28}
\right)\times10^{-2}\,\GeV^2.
\end{align}
The two interpolating currents lead to different central values, with
the difference being more pronounced for $F_{3,R}$ than for
$\widetilde F_{3,R}$. Nevertheless, the two determinations are
compatible within their respective LCSR uncertainties. This is
illustrated in \cref{fig:Q3R-form-factors}, where the form factors are
shown over the spacelike region used in the LCSR analysis. For
$F_{3,R}$, the overlap of the uncertainty bands decreases towards
larger spacelike momentum transfer, while for
$\widetilde F_{3,R}$ a substantial overlap persists throughout the
entire region.

The results for $\widetilde{\mathcal O}_{3,R}$ illustrate
the dependence on the interpolating current. For the remaining
form factors, \cref{tab:form-factors-q20} summarises their
values at $q^2=0$, while their full $q^2$ dependence is
retained in the phenomenological analysis.

As seen in \cref{tab:form-factors-q20}, for most form factors the two
current determinations agree within the quoted uncertainties.
The exceptions are $F_{2,R}(0)$ and $F_{1,L}(0)$, whose
uncertainty intervals do not overlap. We examine these cases
below to assess the relative quality of the corresponding
sum rules. 

We first consider $F_{2,R}$, for which the two currents yield
opposite signs at $q^2=0$. For the $\mathcal P$ current,
the contributions at the central sum-rule point are
\begin{equation}
\begin{gathered}
\bigl(
F_{2,R}^{(\mathcal P),{\rm tw2}},
F_{2,R}^{(\mathcal P),{\rm tw4}},
F_{2,R}^{(\mathcal P),{\rm tw6}}
\bigr)\\
=
(-0.440,\;0.345,\;0.597)\times10^{-2}\,\GeV^2 .
\end{gathered}
\end{equation}
The twist-six term is the factorisable four-particle
contribution. It exceeds either LO two-particle term, while
the twist-four term is comparable to the leading-twist term.
The resulting cancellations and lack of suppression of
higher-twist contributions limit the reliability of this
determination.

For the $\mathcal A$ current, the corresponding contributions are
\begin{equation}
\begin{gathered}
\bigl(
F_{2,R}^{(\mathcal A),{\rm tw3}},
F_{2,R}^{(\mathcal A),{\rm tw5}},
F_{2,R}^{(\mathcal A),{\rm tw7}}
\bigr)\\
=
(-1.749,\;0.234,\;0.159)\times10^{-2}\,\GeV^2 .
\end{gathered}
\end{equation}
The LO twist-three contribution dominates, while the
factorisable twist-five and twist-seven terms are smaller.
The retained contributions therefore show a clearer hierarchy
than for the $\mathcal P$ current.

The $\mathcal A$-current result is more sensitive to the
continuum contribution than its $\mathcal P$-current
counterpart. Nevertheless, its smaller factorisable
four-particle contribution and clearer hierarchy among the
retained OPE terms motivate using the $\mathcal A$-current
determination of $F_{2,R}$ in the phenomenological analysis.

We now turn to $F_{1,L}$. At $q^2=0$, the two determinations
have the same sign and differ less markedly than for $F_{2,R}$.
For the
$\mathcal P$ current, the retained contributions at the
central sum-rule point are
\begin{equation}
\begin{gathered}
\bigl(
F_{1,L}^{(\mathcal P),{\rm tw3}},
F_{1,L}^{(\mathcal P),{\rm tw5}},
F_{1,L}^{(\mathcal P),{\rm tw7}}
\bigr)\\
=
(5.93,\;1.81,\;-0.10)\times10^{-2}\,\GeV^2 .
\end{gathered}
\end{equation}
The LO twist-three term dominates, while the factorisable
twist-five contribution is appreciable and the twist-seven
term remains small throughout the adopted Borel window.
The full form factor also exhibits only a mild dependence
on $M^2$.

For the $\mathcal A$ current, the contributions at the same
central sum-rule point are
\begin{equation}
\begin{gathered}
\bigl(
F_{1,L}^{(\mathcal A),{\rm tw2}},
F_{1,L}^{(\mathcal A),{\rm tw4}},
F_{1,L}^{(\mathcal A),{\rm tw6}}
\bigr)\\
=
(4.94,\;-2.08,\;-0.14)\times10^{-2}\,\GeV^2 .
\end{gathered}
\end{equation}
The LO twist-four term has the opposite sign to the
leading-twist contribution and amounts to approximately
$42\%$ of its magnitude. This cancellation reduces
$F_{1,L}^{(\mathcal A)}$, while the full result exhibits
a stronger Borel-parameter dependence.
The stronger suppression of the retained twist-four DA
contribution and milder Borel dependence favour the
$\mathcal P$-current determination of $F_{1,L}$. 

Together with the preference for the
$\mathcal A$ current for $F_{2,R}$, this illustrates that
neither current is uniformly preferred. Their relative
performance depends on the operator structure and the
resulting DA selection and cancellations, as discussed
in \cref{sec:OPE}.
\begin{table}[!t]
\centering
\caption{Form factors $F_{I,X}^{(\mathcal J)}(0)$ and
$\widetilde F_{I,X}^{(\mathcal J)}(0)$ for
$\mathcal J=\mathcal P,\mathcal A$, in units of
$10^{-2}\,\GeV^2$. The uncertainties include input-parameter,
sum-rule, and renormalisation-scale variations.}
\label{tab:form-factors-q20}

\begingroup
\setlength{\arrayrulewidth}{0.6pt}
\setlength{\lightrulewidth}{0.6pt}
\renewcommand{\arraystretch}{1.38}
\begin{tabular}{c|cc|cc}
\hline\hline
& \multicolumn{2}{c|}{$F_{I,X}(0)$}
& \multicolumn{2}{c}{$\widetilde F_{I,X}(0)$}
\\
$(I,X)$
& $\mathcal P$ & $\mathcal A$
& $\mathcal P$ & $\mathcal A$
\\
\midrule
$(1,R)$
& $-4.98^{+2.15}_{-2.08}$ & $-8.23^{+4.06}_{-4.01}$
& $3.44^{+1.43}_{-1.54}$ & $6.13^{+2.92}_{-3.01}$
\\
$(1,L)$
& $7.64^{+3.17}_{-3.26}$ & $2.72^{+1.44}_{-1.48}$
& $-5.72^{+2.54}_{-2.41}$ & $-5.21^{+2.73}_{-2.62}$
\\
$(2,R)$
& $0.50^{+0.27}_{-0.22}$ & $-1.36^{+0.93}_{-0.91}$
& $4.58^{+2.06}_{-2.21}$ & $6.20^{+2.90}_{-3.00}$
\\
$(2,L)$
& $7.64^{+3.17}_{-3.26}$ & $4.07^{+1.82}_{-1.84}$
& $-5.72^{+2.54}_{-2.41}$ & $-3.57^{+1.72}_{-1.69}$
\\
$(3,R)$
& $-5.19^{+2.02}_{-2.07}$ & $-9.50^{+4.63}_{-4.66}$
& $9.01^{+3.64}_{-3.68}$ & $11.16^{+5.21}_{-5.28}$
\\
$(3,L)$
& $14.15^{+5.78}_{-5.69}$ & $8.08^{+3.49}_{-3.35}$
& $-10.89^{+4.55}_{-4.48}$ & $-9.33^{+4.33}_{-4.34}$
\\
\hline\hline
\end{tabular}
\endgroup

\end{table}

For the remaining form factors, analogous checks are performed
on the retained OPE contributions, Borel dependence, and
below-threshold ratios. The relative size of the factorisable
four-particle terms depends on both the operator and the
interpolating current. For the kaon 
$\widetilde{\mathcal O}_{2,L}$ operator, these
terms exceed the LO two-particle contribution in
$\widetilde F_{2,L}^{(\mathcal A)}$ throughout the adopted
Borel window, whereas they remain subleading for both form
factors obtained with the $\mathcal P$ current. We therefore
prefer the $\mathcal P$-current determination for this operator.
For some other determinations, comparable contributions occur
towards the lower edge of the Borel window, highlighting the
usefulness of comparing the two currents.

These checks guide the choice of current and identify the
sensitivity to the retained higher-twist terms. The resulting
form factors provide estimates within the present partial
OPE; the quoted uncertainties do not include the effects of
omitted three-particle and non-factorisable four-particle
contributions.

\section{Extrapolation to the physical region}
\label{sec:zexpansion}

The LCSR results are obtained in the spacelike region, whereas the
decay amplitudes require the form factors in the physical timelike
region\footnote{While the pion mass is neglected in the kinematics and
hard-scattering kernels of the LCSR calculation, we retain its physical
value when specifying the analytic structure and kinematic boundaries
of the form factors.},
\begin{equation}
\begin{gathered}
0\leq q^2=m_\chi^2\leq t_-^{(M)},
\\
t_-^{(M)}
=
\left(m_{\Lambda_c}-m_M\right)^2,
\\ M=\pi,K .
\label{eq:physical-q2-region}
\end{gathered}
\end{equation}
To extrapolate the LCSR results to the physical region, we introduce
the conformal variable
\begin{equation}
z_M(q^2)
=
\frac{
\sqrt{t_+^{(M)}-q^2}-\sqrt{t_+^{(M)}-t_0^{(M)}}
}{
\sqrt{t_+^{(M)}-q^2}+\sqrt{t_+^{(M)}-t_0^{(M)}}
},
\qquad M=\pi,K ,
\label{eq:z_definition}
\end{equation}
where $t_+^{(M)}$ denotes the lowest branch point of the form factor in
the corresponding crossed channel. We choose
\begin{equation}
t_0^{(M)}
=
t_+^{(M)}
\left(
1-\sqrt{1-\frac{t_-^{(M)}}{t_+^{(M)}}}
\right),
\label{eq:t0_definition}
\end{equation}
which minimises the maximum value of $|z_M|$ over the physical region.
For later convenience, we define
\begin{equation}
\Delta z_M(q^2)
=
z_M(q^2)-z_M(0),
\label{eq:delta_z}
\end{equation}
such that $\Delta z_M(0)=0$.

For the $\Lambda_c\to\pi$ transition, the singularities in the
$q^2$ channel are associated with charmed-baryon states carrying the
quantum numbers of the corresponding three-quark operator. The lowest
hadronic continuum in this channel is $\Lambda_c\pi$, giving the
branch point
\begin{equation}
t_+^{(\pi)}
=
\left(m_{\Lambda_c}+m_\pi\right)^2 .
\label{eq:tplus_pion}
\end{equation}
The lowest charmed-baryon resonance with the relevant quantum numbers
is the $\Sigma_c$, which lies above the $\Lambda_c\pi$ threshold.
Hence, no isolated subthreshold baryon pole needs to be factored out
explicitly from the $\Lambda_c\to\pi$ form factors.

For the pion channel, we therefore employ a quadratic expansion in the
shifted conformal variable,
\begin{align}
F_{I,X}^{(\mathcal J,\pi)}(q^2)
&=
F_{I,X}^{(\mathcal J,\pi)}(0)
\nonumber\\
&\hspace{-2.5em}\times
\Bigl[
1+b_{I,X}^{(\mathcal J,\pi)}\,\Delta z_\pi(q^2)
+c_{I,X}^{(\mathcal J,\pi)}\,\Delta z_\pi(q^2)^2
\Bigr],
\label{eq:zexp_F_pi}
\\
\widetilde F_{I,X}^{(\mathcal J,\pi)}(q^2)
&=
\widetilde F_{I,X}^{(\mathcal J,\pi)}(0)
\nonumber\\
&\hspace{-2.5em}\times
\Bigl[
1+\widetilde b_{I,X}^{(\mathcal J,\pi)}\,\Delta z_\pi(q^2)
+\widetilde c_{I,X}^{(\mathcal J,\pi)}\,\Delta z_\pi(q^2)^2
\Bigr].
\label{eq:zexp_Ftilde_pi}
\end{align}
We have
compared the quadratic parametrisation with a linear expansion and
with more constrained forms inspired by the
Bourrely--Caprini--Lellouch (BCL)
parametrisation~\cite{Bourrely:2008za}. The quadratic form provides a
better description of the curvature of the spacelike LCSR results and
is therefore adopted in the following analysis.

The analytic structure is different for the $\Lambda_c\to K$
transition. In the crossed channel, the relevant three-quark operators
carry the flavour quantum numbers of a $csd$ baryon. The lowest hadronic
continuum is the $\Xi_c\pi$ state, giving
\begin{equation}
t_+^{(K)}
=
\left(m_{\Xi_c}+m_\pi\right)^2 .
\label{eq:tplus_kaon}
\end{equation}
Both the ground-state $\Xi_c$ and the $\Xi_c'$ lie below this threshold
and may therefore appear as isolated subthreshold poles.

For the kaon channel, we adopt a single-pole parametrisation inspired
by the BCL form. We explicitly factor
out the nearest subthreshold pole, corresponding to the ground-state
$\Xi_c$, and parametrise the remaining $q^2$ dependence in terms of
the conformal variable,
\begin{align}
F_{I,X}^{(\mathcal J,K)}(q^2)
&=
\frac{F_{I,X}^{(\mathcal J,K)}(0)}
{1-q^2/m_{\Xi_c}^2}
\nonumber\\
&\hspace{-2.5em}\times
\Bigl[1+\beta_{I,X}^{(\mathcal J,K)}
\bigl(\Delta z_K(q^2)
+\tfrac12[z_K^2(q^2)-z_K^2(0)]\bigr)\Bigr],
\label{eq:zexp_F_K}
\\
\widetilde F_{I,X}^{(\mathcal J,K)}(q^2)
&=
\frac{\widetilde F_{I,X}^{(\mathcal J,K)}(0)}
{1-q^2/m_{\Xi_c}^2}
\nonumber\\
&\hspace{-2.5em}\times
\Bigl[1+\widetilde\beta_{I,X}^{(\mathcal J,K)}
\bigl(\Delta z_K(q^2)
+\tfrac12[z_K^2(q^2)-z_K^2(0)]\bigr)\Bigr].
\label{eq:zexp_Ftilde_K}
\end{align}
Since the terms multiplying the shape parameters vanish at
$q^2=0$, the normalisation of each parametrisation is fixed by the
corresponding LCSR result at $q^2=0$, while
$\beta_{I,X}^{(\mathcal J,K)}$ and
$\widetilde\beta_{I,X}^{(\mathcal J,K)}$ determine the momentum
dependence.

In \cref{eq:zexp_F_K,eq:zexp_Ftilde_K}, we do not factor out the $\Xi_c'$  pole explicitly, but treat the single-pole form above as a
truncated parametrisation of the form factors over the spacelike fit
region and the physical decay region. To assess the sensitivity to
this choice, we repeat the fit for
$F_{2,R}^{(\mathcal A,K)}$ using $m_{\Xi_c'}$ instead of
$m_{\Xi_c}$ in the pole factor. The resulting form factor at the
physical endpoint changes by approximately $9\%$. The fit with the
ground-state $\Xi_c$ pole also gives the smaller unweighted
least-squares residual in the spacelike region. We therefore use the
nearest subthreshold pole, $m_{\Xi_c}$, in our central parametrisation.
\begin{table}[!t]
\centering
\caption{
Shape parameters for the form-factor determinations retained
in the phenomenological analysis. Each row specifies the
operator index $I$, chirality $X$, and interpolating current
$\mathcal J$.
Central values follow from fits to the central LCSR predictions;
asymmetric ranges encompass the central and two boundary-fit
values and are not statistical uncertainties.
The kaon $\mathcal P$-current sum rules coincide for the two
left-handed operators, giving identical fitted parameters.
}
\label{tab:zfit_parameters}

\begingroup
\small
\setlength{\tabcolsep}{2.5pt}
\renewcommand{\arraystretch}{1.18}
\begingroup
\setlength{\arrayrulewidth}{0.6pt}
\setlength{\lightrulewidth}{0.6pt}
\begin{tabular}{ccccc}
\hline\hline
\rule[-1.2ex]{0pt}{4.2ex}$(I,X,\mathcal J)$
& $b_{I,X}^{(\mathcal J,\pi)}$
& $\widetilde b_{I,X}^{(\mathcal J,\pi)}$
& $c_{I,X}^{(\mathcal J,\pi)}$
& $\widetilde c_{I,X}^{(\mathcal J,\pi)}$
\\
\midrule
$(3,R,\mathcal P)$
& $-9.05^{+0.47}_{-0.00}$
& $-6.23^{+0.83}_{-0.18}$
& $22.24^{+0.58}_{-4.40}$
& $15.89^{+1.02}_{-3.84}$
\\
$(3,R,\mathcal A)$
& $-7.43^{+0.13}_{-0.19}$
& $-6.37^{+0.52}_{-0.10}$
& $16.33^{+0.00}_{-0.23}$
& $14.65^{+0.53}_{-2.54}$
\\
$(3,L,\mathcal P)$
& $-5.60^{+0.40}_{-0.05}$
& $-6.13^{+0.73}_{-0.16}$
& $11.20^{+0.33}_{-2.14}$
& $13.76^{+0.87}_{-3.63}$
\\
$(3,L,\mathcal A)$
& $-5.78^{+0.28}_{-0.00}$
& $-7.33^{+1.38}_{-0.35}$
& $14.45^{+0.08}_{-1.30}$
& $21.88^{+2.00}_{-6.97}$
\\
\midrule
$(I,X,\mathcal J)$
& $\beta_{I,X}^{(\mathcal J,K)}$
& $\widetilde\beta_{I,X}^{(\mathcal J,K)}$
& \multicolumn{2}{c}{}
\\
\midrule
$(1,R,\mathcal P)$
& $-1.60^{+0.52}_{-0.02}$
& $-1.33^{+0.43}_{-0.00}$
& \multicolumn{2}{c}{}
\\
$(1,R,\mathcal A)$
& $-1.70^{+0.14}_{-0.00}$
& $-2.54^{+0.09}_{-0.00}$
& \multicolumn{2}{c}{}
\\
$(1,L,\mathcal P)$
& $-2.07^{+0.11}_{-0.00}$
& $-2.27^{+0.41}_{-0.07}$
& \multicolumn{2}{c}{}
\\
$(2,R,\mathcal A)$
& $-13.97^{+3.93}_{-21.51}$
& $-2.45^{+0.29}_{-0.05}$
& \multicolumn{2}{c}{}
\\
$(2,L,\mathcal P)$
& $-2.07^{+0.11}_{-0.00}$
& $-2.27^{+0.41}_{-0.07}$
& \multicolumn{2}{c}{}
\\
\hline\hline
\end{tabular}
\endgroup
\endgroup

\end{table}

The comparison of alternative parametrisations also indicates that the
extrapolation towards the upper end of the physical $q^2$ region can
be more sensitive to the assumed functional form than is apparent from
the quality of the spacelike fit alone. We therefore regard the
parametrisations above as phenomenological continuations of the LCSR
results, rather than as model-independent dispersive determinations of
the form factors.

For each form factor, we evaluate the LCSR prediction at eleven
spacelike points,
\begin{equation}
q^2=-5,-4.5,\ldots,-0.5,0\,\GeV^2 .
\end{equation}
We construct three datasets corresponding to the central values and
the pointwise lower and upper boundaries of the LCSR uncertainty band.
Each dataset is fitted independently by unweighted least squares using
the parametrisation appropriate to the corresponding channel,
\cref{eq:zexp_F_pi,eq:zexp_Ftilde_pi} for the pion and
\cref{eq:zexp_F_K,eq:zexp_Ftilde_K} for the kaon. In each case, the
normalisation is fixed to the corresponding LCSR value at $q^2=0$,
while the remaining shape parameters are determined by the fit. The
three fitted functions are then extrapolated to the physical timelike
region. We take the fit to the central LCSR values as the central
prediction and use the extrapolated lower and upper boundary fits to
estimate the corresponding uncertainty.

The fitted shape parameters are collected in
\cref{tab:zfit_parameters}. The quoted ranges should not be interpreted as statistical
uncertainties on the fitted shape parameters. Rather, they reflect the
spread obtained from independent fits to the lower and upper boundaries
of the LCSR uncertainty band. Since the shape parameters are nonlinear
characterisations of the full $q^2$ dependence, their values need not
preserve the ordering of the corresponding form-factor curves.
Consequently, the central value of a shape parameter need not lie
between the values obtained from the two boundary fits, even when the
form-factor curves themselves remain ordered throughout the physical
region. This explains the one-sided ranges in
\cref{tab:zfit_parameters}, for which both boundary fits shift the
corresponding coefficient to the same side of its central value.

A particularly pronounced variation is found for
$\beta_{2,R}^{(\mathcal A,K)}$. This is associated with the
comparatively small normalisation of
$F_{2,R}^{(\mathcal A,K)}(0)$, especially for the upper boundary of
the LCSR uncertainty band. Since the normalisation of each boundary
fit is fixed independently to the corresponding LCSR value at
$q^2=0$, a larger shape coefficient is required to reproduce the
$q^2$ dependence of this boundary curve. The large variation of
$\beta_{2,R}^{(\mathcal A,K)}$ should therefore not, by itself, be
interpreted as an instability of the extrapolated form factor. We
have explicitly verified that the central, lower, and upper fitted
form-factor curves remain ordered throughout the physical region.

\par\vspace{0.8\baselineskip}
\section{Branching fractions and phenomenology}
\label{sec:branching_fractions}

Using the form factors obtained above, we determine the
branching fractions for
\begin{equation}
\Lambda_c(P)\to M(p)+\bar\chi(q),
\qquad P=p+q,\qquad M=\pi,K,
\end{equation}
with
\begin{equation}
q^2=m_\chi^2,
\qquad
0\leq m_\chi\leq m_{\Lambda_c}-m_M.
\end{equation}
Physical meson masses are used in the phase-space factors.
For convenience, we introduce the dimensionful Wilson coefficients
\begin{equation}
\begin{gathered}
\mathcal C^{qqd\chi}_{\{ij\}k}
\equiv
\frac{C^{qqd\chi}_{\{ij\}k}}{\Lambda^2},
\qquad
\mathcal C^{udd\chi}_{ijk}
\equiv
\frac{C^{udd\chi}_{ijk}}{\Lambda^2}.
\label{eq:dimensionful_WCs}
\end{gathered}
\end{equation}
The Wilson coefficients and hadronic
matrix elements are understood to be evaluated at a common
renormalisation scale and in consistent conventions.
We then define the coefficient-weighted combinations entering the
decay amplitudes,
\begin{align}
\mathcal F_\pi(q^2)
&=
2\mathcal C^{qqd\chi}_{\{12\}1}F_{3,L}(q^2)+
\mathcal C^{udd\chi}_{211}F_{3,R}(q^2),
\label{eq:Fpi_eff}
\\
\mathcal F_K(q^2)
&=
2\mathcal C^{qqd\chi}_{\{12\}2}F_{1,L}(q^2)+
\mathcal C^{udd\chi}_{212}F_{1,R}(q^2)
\nonumber\\*
&\quad+
2\mathcal C^{qqd\chi}_{\{22\}1}F_{2,L}(q^2)+
\mathcal C^{udd\chi}_{221}F_{2,R}(q^2),
\label{eq:FK_eff}
\end{align}
with $\widetilde{\mathcal F}_M$ defined by replacing
each $F_{I,X}$ with $\widetilde F_{I,X}$.
Using the on-shell equation for the invisible fermion,
the decay amplitudes become
\begin{equation}
\begin{aligned}
i\mathcal M_M
&=
-i\,\overline{v_\chi^C}(q)\left[
\mathcal F_M(q^2)P_R
+
\frac{m_\chi}{m_{\Lambda_c}}
\widetilde{\mathcal F}_M(q^2)P_L
\right]
u_{\Lambda_c}(P),
\\
&\hspace{2em}M=\pi,K.
\label{eq:amplitude_universal}
\end{aligned}
\end{equation}
After averaging over the initial spin and summing over the
final spin, the branching fraction is
\begin{widetext}
\begin{equation}
\mathrm{BR}(\Lambda_c\to M\bar\chi)
= \tau_{\Lambda_c}\,\mathcal N_M
\left[
(m_{\Lambda_c}^2+m_\chi^2-m_M^2)
\left( |\mathcal F_M|^2
+\frac{m_\chi^2}{m_{\Lambda_c}^2}|\widetilde{\mathcal F}_M|^2 \right)
+4m_\chi^2\operatorname{Re}\!\left(\mathcal F_M\widetilde{\mathcal F}_M^\ast\right)
\right]_{q^2=m_\chi^2},
\label{eq:BR_general}
\end{equation}
\end{widetext}
where $\tau_{\Lambda_c}$ is the $\Lambda_c$ lifetime and
$\mathcal N_M
=
\frac{
\lambda^{1/2}(m_{\Lambda_c}^2,m_\chi^2,m_M^2)
}{
32\pi m_{\Lambda_c}^3
},
\\
\lambda(x,y,z)
=
x^2+y^2+z^2-2xy-2xz-2yz.$
These expressions retain interference for arbitrary complex
Wilson coefficients. For the numerical analysis, we consider one independent Wilson
coefficient at a time, setting all others to zero. As benchmarks,
we use the one-parameter upper bounds on these coefficients
derived from LHC missing-energy searches~\cite{Hiller:2026osz}. These bounds cannot, in general, be saturated
simultaneously.

\subsection{Pion channel}
For $\Lambda_c\to\pi\bar\chi$, we consider the two
single-operator scenarios associated with
$\widetilde{\mathcal O}_{3,R}$ and
$\widetilde{\mathcal O}_{3,L}$, using the benchmark values
\begin{equation}
\begin{gathered}
\mathcal C^{udd\chi}_{211}
=
0.065\times10^{-6}\,\GeV^{-2},
\\
\mathcal C^{qqd\chi}_{\{12\}1}
=
0.023\times10^{-6}\,\GeV^{-2}.
\label{eq:pion_WC_benchmark}
\end{gathered}
\end{equation}

\begin{figure*}[!t]
\centering
\begin{minipage}[t]{0.48\textwidth}
\centering
\includegraphics[width=\linewidth]{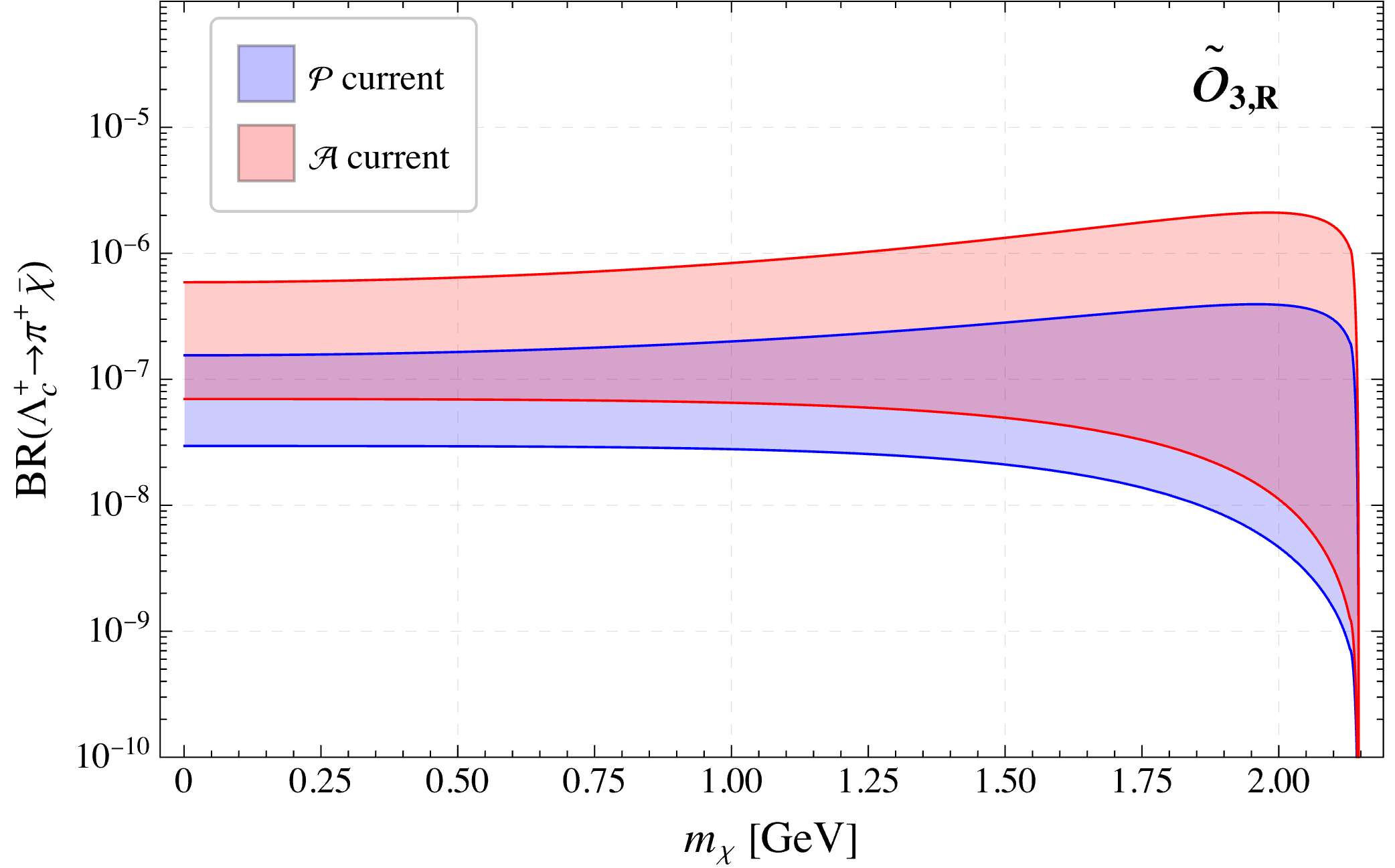}
\end{minipage}
\hfill
\begin{minipage}[t]{0.48\textwidth}
\centering
\includegraphics[width=\linewidth]{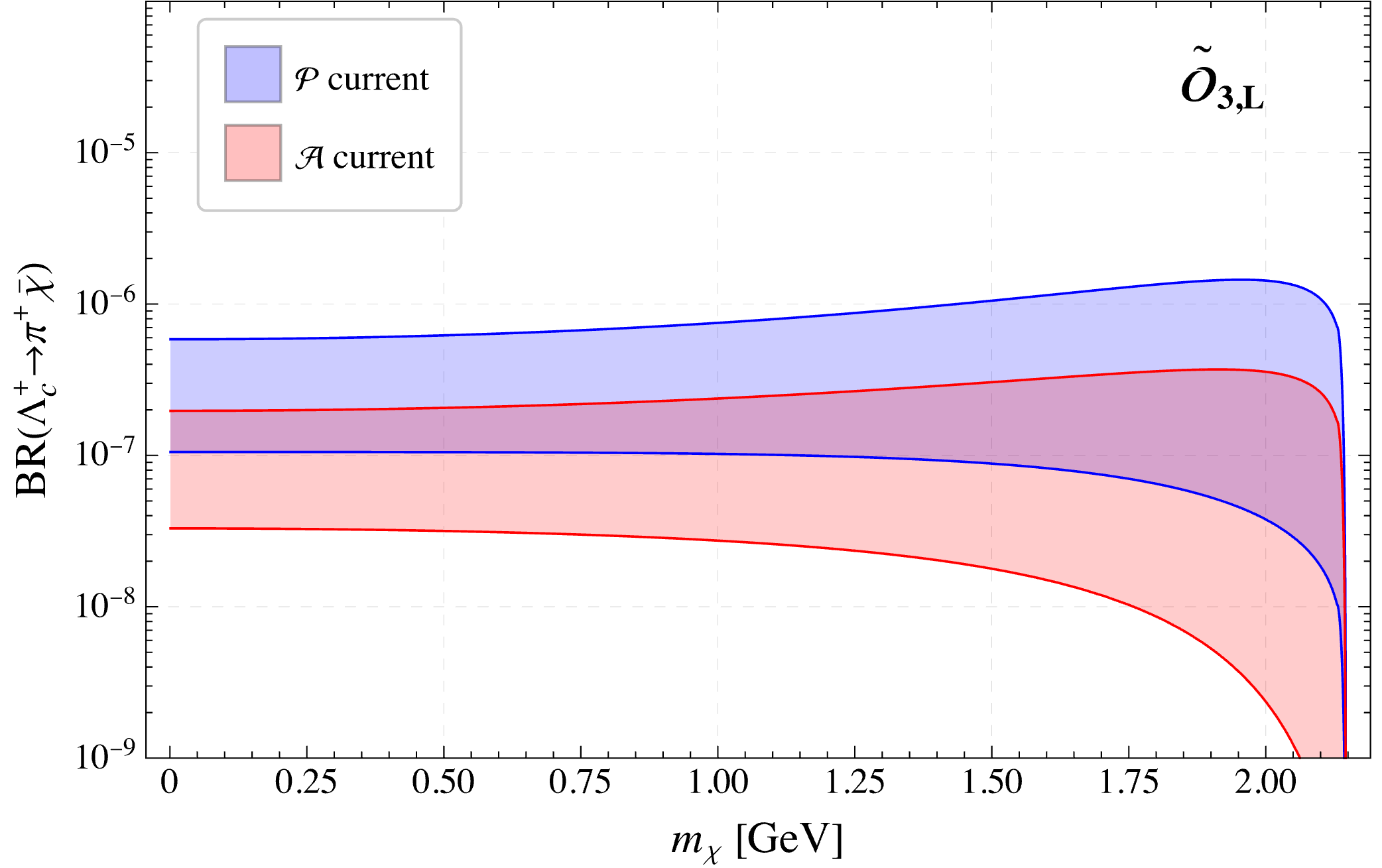}
\end{minipage}

\caption{Branching fraction for
$\Lambda_c\to\pi\bar\chi$ as a function of $m_\chi$ in the
single-operator scenarios indicated in each panel, using
the respective benchmark coefficients in
\cref{eq:pion_WC_benchmark}.
The blue and red bands show the form-factor uncertainty
envelopes from the $\mathcal P$- and $\mathcal A$-current
LCSR determinations, respectively.}
\label{fig:BR_pion}
\end{figure*}

The branching fractions are shown in \cref{fig:BR_pion},
with blue and red bands corresponding to the $\mathcal P$-
and $\mathcal A$-current form factors, respectively.
To propagate their uncertainties, each parameter of the
$z$ parametrisations is independently assigned its central,
lower, and upper values. At each $m_\chi$, the branching
fraction is evaluated for all combinations, and the minimum
and maximum define the uncertainty envelope. This scan
neglects parameter correlations and provides an estimate
of the form-factor uncertainty rather than a statistical
confidence interval.

In both single-operator scenarios, the two current determinations
yield substantially overlapping uncertainty bands over most
of the physical region. The benchmark branching fractions
are of order $10^{-7}$ over sizeable regions, with uncertainty
envelopes reaching $10^{-6}$ in some regions.
\subsection{Kaon channel}

For $\Lambda_c\to K\bar\chi$, the single-operator
benchmark values are
\begin{align}
\mathcal C^{udd\chi}_{212}
&=
0.081\times10^{-6}\,\GeV^{-2},\nonumber\\*
\mathcal C^{qqd\chi}_{\{12\}2}
&=
0.033\times10^{-6}\,\GeV^{-2},
\nonumber\\*
\mathcal C^{udd\chi}_{221}
&=
0.11\times10^{-6}\,\GeV^{-2},\nonumber\\*
\mathcal C^{qqd\chi}_{\{22\}1}
&=
0.050\times10^{-6}\,\GeV^{-2}.
\label{eq:kaon_WC_benchmark}
\end{align}

Following the selection discussed in \cref{sec:Numerical},
we retain both current determinations for
$\widetilde{\mathcal O}_{1,R}$, the $\mathcal P$-current
results for $\widetilde{\mathcal O}_{1,L}$ and
$\widetilde{\mathcal O}_{2,L}$, and the $\mathcal A$-current
result for $\widetilde{\mathcal O}_{2,R}$.

\begin{figure*}[!t]
\centering

\begin{minipage}[t]{0.48\textwidth}
\centering
\includegraphics[width=\linewidth]{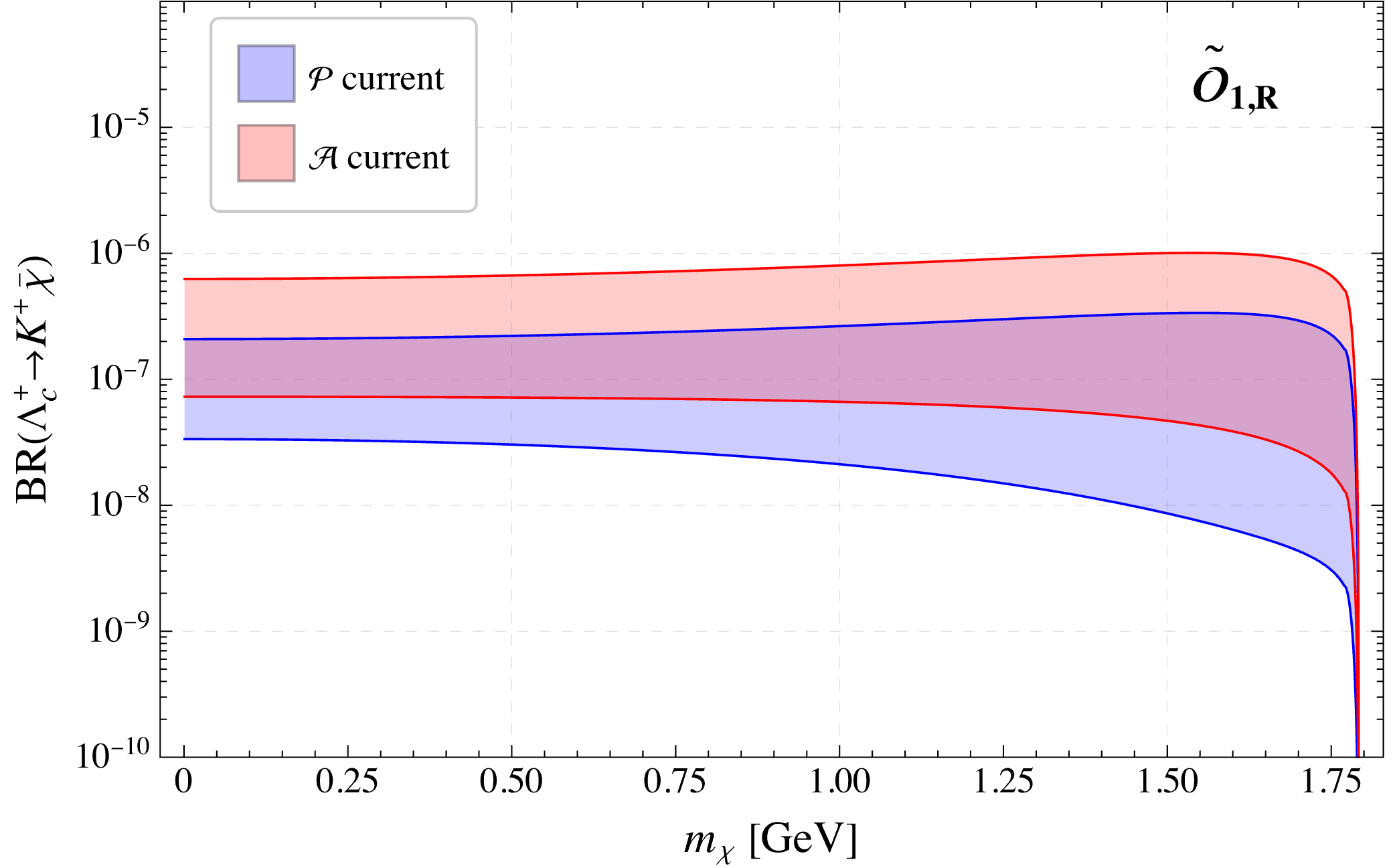}
\end{minipage}
\hfill
\begin{minipage}[t]{0.48\textwidth}
\centering
\includegraphics[width=\linewidth]{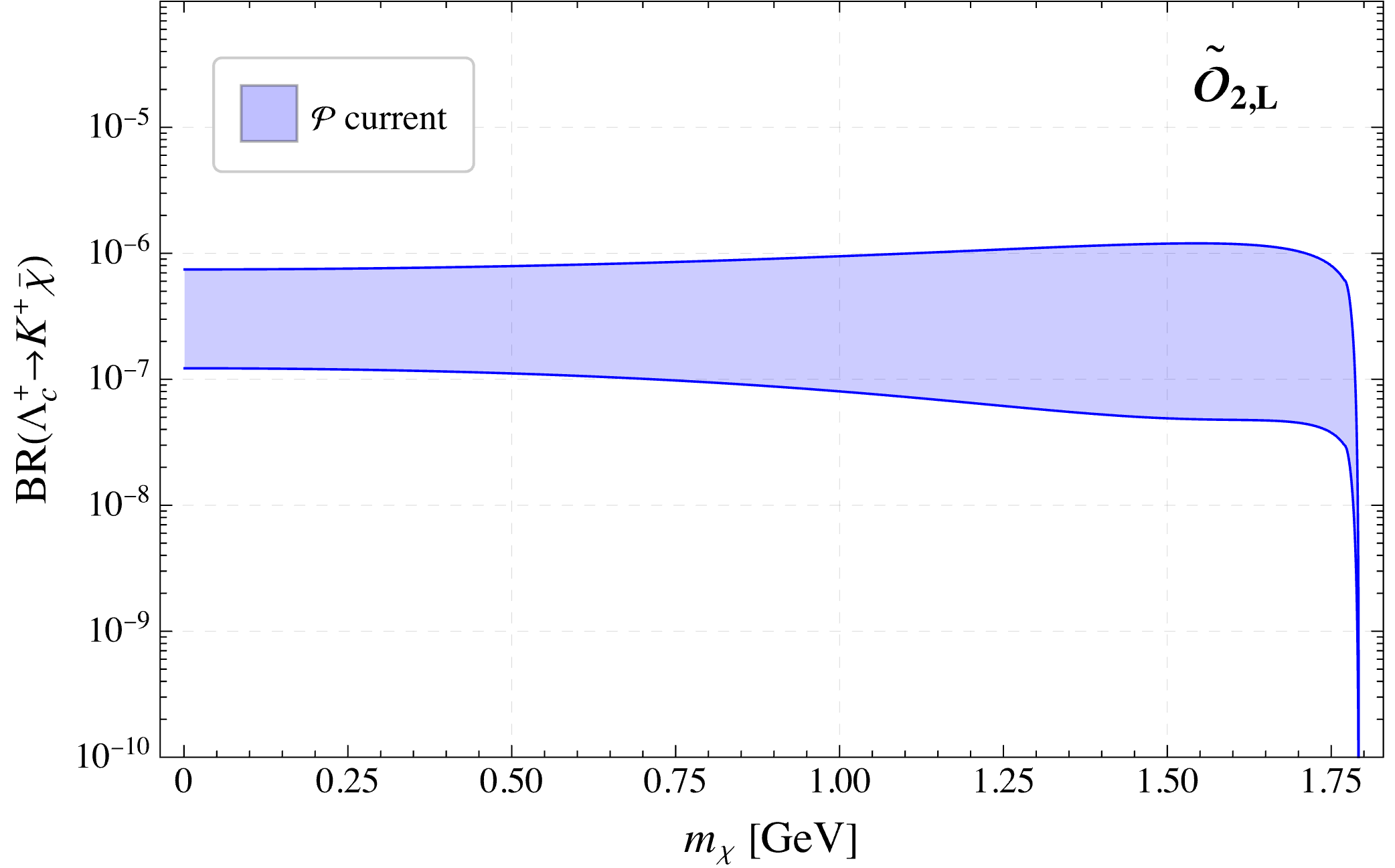}
\end{minipage}

\par\smallskip

\begin{minipage}[t]{0.48\textwidth}
\centering
\includegraphics[width=\linewidth]{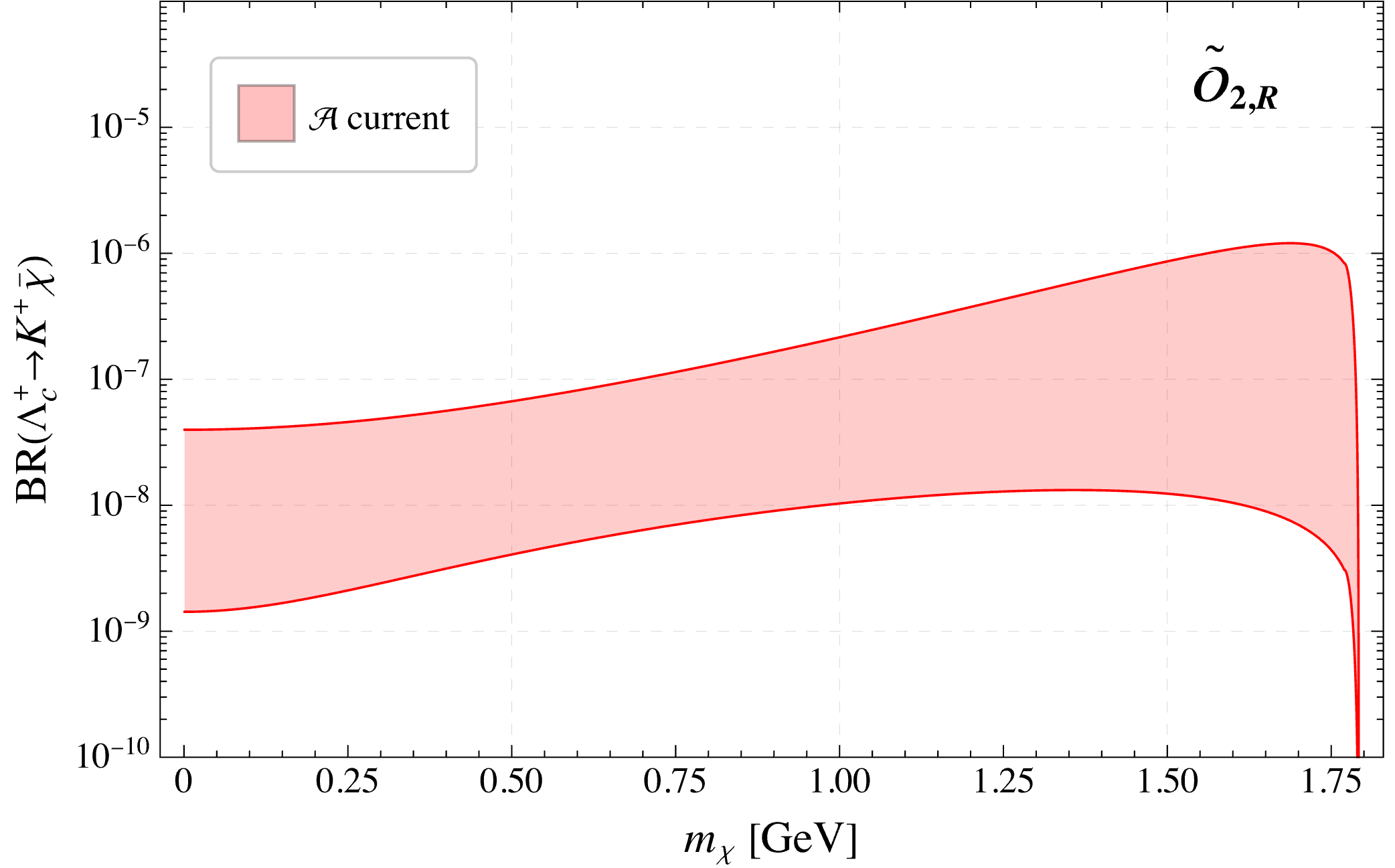}
\end{minipage}
\hfill
\begin{minipage}[t]{0.48\textwidth}
\centering
\includegraphics[width=\linewidth]{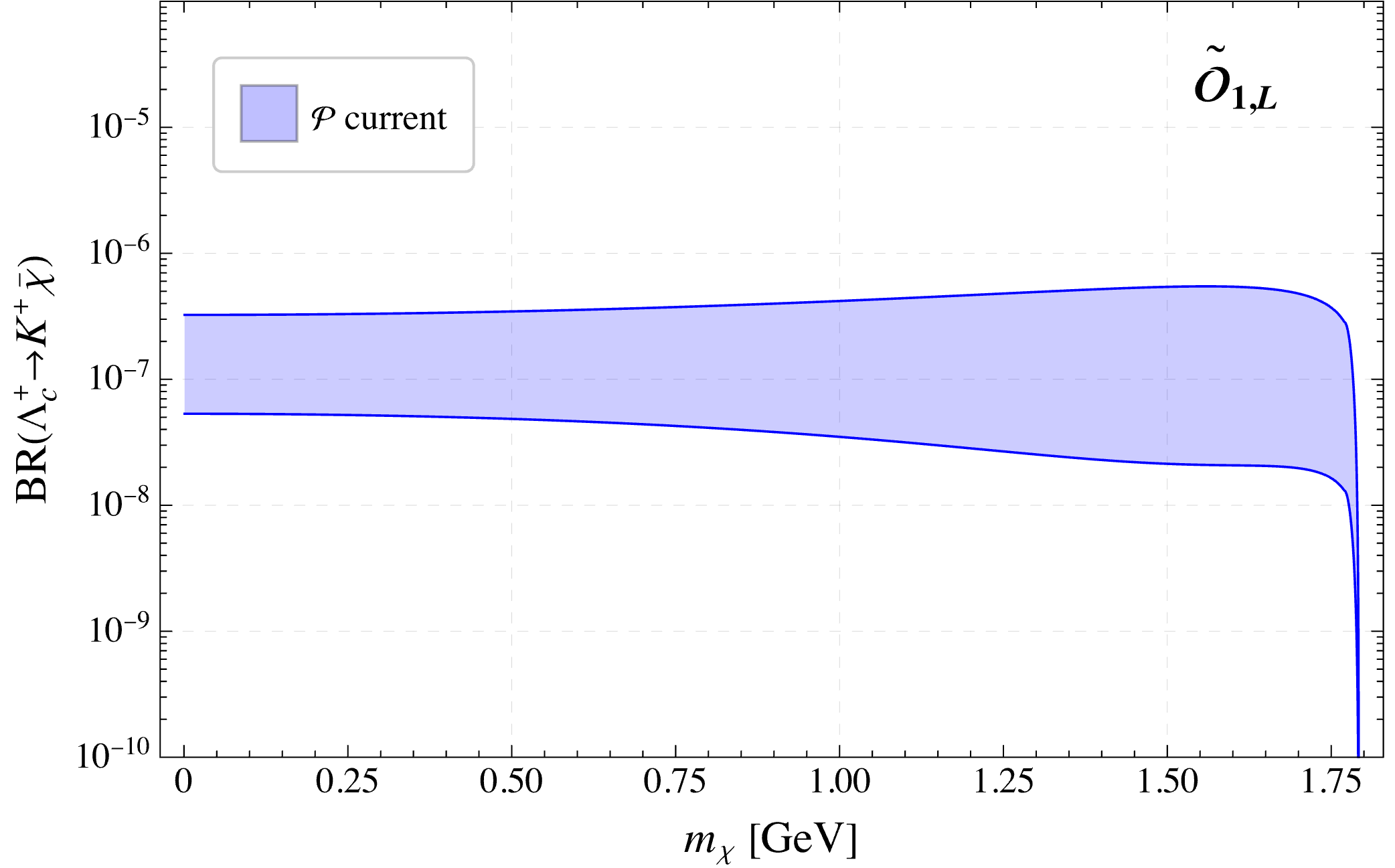}
\end{minipage}

\caption{Branching fractions for
$\Lambda_c^+\to K^+\bar\chi$ as functions of $m_\chi$
in the single-operator scenarios indicated in each panel,
using the benchmarks in \cref{eq:kaon_WC_benchmark}.
Blue and red bands denote the $\mathcal P$- and
$\mathcal A$-current uncertainty envelopes, respectively.}
\label{fig:BR_kaon_individual}
\end{figure*}

The branching fractions are shown in
\cref{fig:BR_kaon_individual}, with uncertainties propagated
using the same parameter scan as for the pion channel.
For $\widetilde{\mathcal O}_{1,R}$, the two current
determinations yield substantially overlapping uncertainty
bands. For the remaining operators, only the selected
current determination is shown.

The single-operator benchmarks yield branching fractions
of order $10^{-7}$ in several scenarios, with sizeable
form-factor uncertainties. The
$\widetilde{\mathcal O}_{2,R}$ contribution depends more
strongly on $m_\chi$, with its uncertainty envelope reaching
$10^{-6}$ towards the upper end of the physical region.

\subsection{Experimental prospects}
\label{sec:pheno}

The single-operator benchmarks in
\cref{fig:BR_pion,fig:BR_kaon_individual} motivate searches
at branching-fraction sensitivities of $10^{-7}$--$10^{-6}$.

It is instructive to compare the branching fractions obtained above
with the prospective sensitivity of present and future flavour
experiments. Following~\cite{Hiller:2026osz}, we estimate the
expected number of signal events as
\begin{equation}
N_{\rm sig}
=
N_{\Lambda_c}\,\epsilon\,
\mathrm{BR}(\Lambda_c\to M\bar\chi),
\qquad M=\pi,K,
\label{eq:Nsig_Lambdac}
\end{equation}
where $N_{\Lambda_c}$ denotes the number of produced $\Lambda_c$
baryons and $\epsilon$ the overall signal efficiency. In the
idealised zero-background limit, assuming zero observed events, the
Poisson upper limit at $90\%$ C.L. corresponds to 2.3 signal events.
The corresponding branching-fraction sensitivity is therefore
\begin{equation}
\mathrm{BR}_{90}
=
\frac{2.3}{N_{\Lambda_c}\epsilon}.
\label{eq:BR90_Lambdac}
\end{equation}

Dedicated efficiency studies for the decay modes considered here are
not yet available. 
We therefore
adopt $\epsilon=0.01$ as a benchmark efficiency, which is supported by
studies of charm decays to invisible final states in the FCC-ee
environment~\cite{Hiller:2026osz}. In addition, we consider the illustrative value
$\epsilon=0.1$ to quantify the potential improvement achievable with
a dedicated experimental analysis. Since the sensitivity in
\cref{eq:BR90_Lambdac} scales as $\epsilon^{-1}$, increasing the
efficiency from $1\%$ to $10\%$ improves the expected
branching-fraction reach by one order of magnitude.

\begin{table}[!t]
\centering
\caption{Indicative $90\%$ C.L. sensitivities to
$\mathrm{BR}(\Lambda_c\to\pi(K)\bar\chi)$, obtained from
$\mathrm{BR}_{90}=2.3/(N_{\Lambda_c}\epsilon)$ for zero
background and zero observed events. The $\Lambda_c$ yields
and benchmark efficiency $\epsilon=1\%$ follow
\cite{Hiller:2026osz}; $\epsilon=10\%$ illustrates the reach
with improved efficiency.}
\label{tab:Lambdac_exp_sensitivity}

\begingroup
\setlength{\arrayrulewidth}{0.6pt}
\setlength{\lightrulewidth}{0.6pt}
\begin{tabular}{c|c|cc}
\hline\hline
\textbf{Experiment}
&
\boldmath{$N_{\Lambda_c}$}
&
\boldmath{$\mathrm{BR}\;(\epsilon=0.01)$}
&
\boldmath{$\mathrm{BR}\;(\epsilon=0.1)$}
\\
\hline
BESIII
& \(10^{6}\)
& \(2.3\times10^{-4}\)
& \(2.3\times10^{-5}\)
\\
STCF \((0.2\,{\rm ab}^{-1})\)
& \(10^{8}\)
& \(2.3\times10^{-6}\)
& \(2.3\times10^{-7}\)
\\
STCF \((1\,{\rm ab}^{-1})\)
& \(5\times10^{8}\)
& \(4.6\times10^{-7}\)
& \(4.6\times10^{-8}\)
\\
Belle II \((10\,{\rm ab}^{-1})\)
& \(8\times10^{8}\)
& \(2.9\times10^{-7}\)
& \(2.9\times10^{-8}\)
\\
FCC-ee
& \(6\times10^{9}\)
& \(3.8\times10^{-8}\)
& \(3.8\times10^{-9}\)
\\
\hline\hline
\end{tabular}
\endgroup

\end{table}
Using the $\Lambda_c$ yields collected
in~\cite{Hiller:2026osz}, the indicative sensitivities are
summarised in \cref{tab:Lambdac_exp_sensitivity}.
At $1\%$ efficiency, STCF and Belle II could probe part
of the predicted branching-fraction range, while FCC-ee
could reach a few times $10^{-8}$. These estimates motivate
dedicated searches for $\Lambda_c\to\pi(K)\bar\chi$,
with experimental studies needed to establish realistic
efficiencies, backgrounds, and their dependence on $m_\chi$.

\section{Conclusions and outlook}
\label{sec:conclusions}

Heavy-hadron decays with missing energy offer a way to probe
apparent baryon-number violation in the visible sector,
while total baryon number may be conserved through its
transfer to an invisible fermion. We have studied
$\Lambda_c\to\pi\bar\chi$ and $\Lambda_c\to K\bar\chi$,
where $\chi$ is a light SM-singlet fermion carrying baryon
number, and determined the complete set of transition form
factors for the dimension-six operators considered here
using LCSRs with light-meson DAs.

The calculation includes LO two-particle contributions
up to twist four and factorisable four-particle terms.
The lowest negative-parity $\Lambda_c^\ast$ contribution
is explicitly eliminated from the hadronic representation.
We employ both pseudoscalar and axial-vector interpolating
currents, whose predictions overlap within the quoted
uncertainties for most form factors. Their comparison
provides a useful consistency check within the retained
approximation and guides the current selection discussed
in \cref{sec:Numerical}.

The factorisable four-particle terms can be sizeable,
with their importance depending on the operator and
interpolating current. The quoted uncertainty bands reflect
input, sum-rule, and scale variations within the retained
approximation. Including three-particle DAs and radiative
corrections, together with examining the accuracy of the
factorisation approximation, would allow a more complete
assessment of the theoretical uncertainties.

Using single-operator benchmarks based on bounds on the
Wilson coefficients from LHC missing-energy
searches~\cite{Hiller:2026osz}, we obtain
branching fractions of order $10^{-7}$ over sizeable parts
of the physical region, with uncertainty bands reaching
$10^{-6}$ in some scenarios, as shown in
\cref{fig:BR_pion,fig:BR_kaon_individual}.
These indicative estimates identify an experimentally
interesting range despite the present hadronic uncertainties.
The comparison with ChPT yields compatible branching-fraction
scales, while our larger form factors relative to the QCDF
estimates should be viewed in light of the latter's restriction
to the hard-scattering contribution.

The projected STCF and Belle II samples could probe part
of this range, while FCC-ee could extend the sensitivity
to smaller branching fractions. These prospects depend
on achievable efficiencies and backgrounds and motivate
dedicated experimental studies of
$\Lambda_c\to\pi(K)+\mathrm{invisible}$.

\begin{acknowledgments}
I am very grateful to Gudrun Hiller, whose insights and stimulating
discussions provided important motivation for this study, and to
Alexander Khodjamirian, Alexander Lenz, and Maria Laura Piscopo
for helpful discussions and valuable comments on the draft.
I gratefully acknowledge the hospitality of the CERN Theory
Department, where part of this work was carried out.
This research is supported by the Deutsche Forschungsgemeinschaft
(DFG, German Research Foundation) under Germany's Excellence Strategy,
EXC 3107 -- Project ID 533766364, and under grant
396021762 -- TRR 257.

\end{acknowledgments}
\appendix 
\numberwithin{equation}{section}

\section{Comparison with other estimates}
\label{app:comparison_previous}

 We compare our LCSR predictions for
$\Lambda_c\to\pi(K)\bar\chi$ with the ChPT
estimates~\cite{Wang:2026vxh} and the QCDF estimates based
on heavy-quark and large-recoil
expansions~\cite{Hiller:2026osz}. The operator content and
Wilson-coefficient normalisation are matched where necessary
to compare the resulting branching fractions.

\paragraph{Comparison with the ChPT estimate.}
For the pion channel, we consider the right-handed operator
$\widetilde{\mathcal O}_{3,R}$ and match the single-operator
benchmark used in~\cite{Wang:2026vxh},
\begin{equation}
\mathcal C^{udd\chi}_{211}
=
10^{-6}\,\GeV^{-2},
\end{equation}
with all other coefficients set to zero. The LCSR prediction
is evaluated using both interpolating currents.
This value serves only as a common normalisation for the
comparison, rather than a benchmark satisfying the LHC bound.

For the kaon channel, the comparison involves the coherent
contributions of the right-handed operators
$\widetilde{\mathcal O}_{1,R}$ and
$\widetilde{\mathcal O}_{2,R}$. To match the benchmark
of~\cite{Wang:2026vxh}, we set
\begin{equation}
\mathcal C^{udd\chi}_{212}
=
\mathcal C^{udd\chi}_{221}
=
10^{-6}\,\GeV^{-2},
\end{equation}
with all other coefficients set to zero. These values provide
a common normalisation for the comparison rather than an
LHC-allowed benchmark. Following our current selection,
we consider
\begin{align}
\text{Combination I}:\quad&
(\widetilde{\mathcal O}_{1,R})_{\mathcal P}
+
(\widetilde{\mathcal O}_{2,R})_{\mathcal A},
\nonumber\\*
\text{Combination II}:\quad&
(\widetilde{\mathcal O}_{1,R})_{\mathcal A}
+
(\widetilde{\mathcal O}_{2,R})_{\mathcal A}.
\label{eq:chpt_current_combinations}
\end{align}
The contributions are added coherently at the amplitude
level. The two combinations differ only in the current
used for $\widetilde{\mathcal O}_{1,R}$ and represent
the same short-distance scenario.

\begin{figure*}[t]
    \centering
    \includegraphics[width=0.49\textwidth]{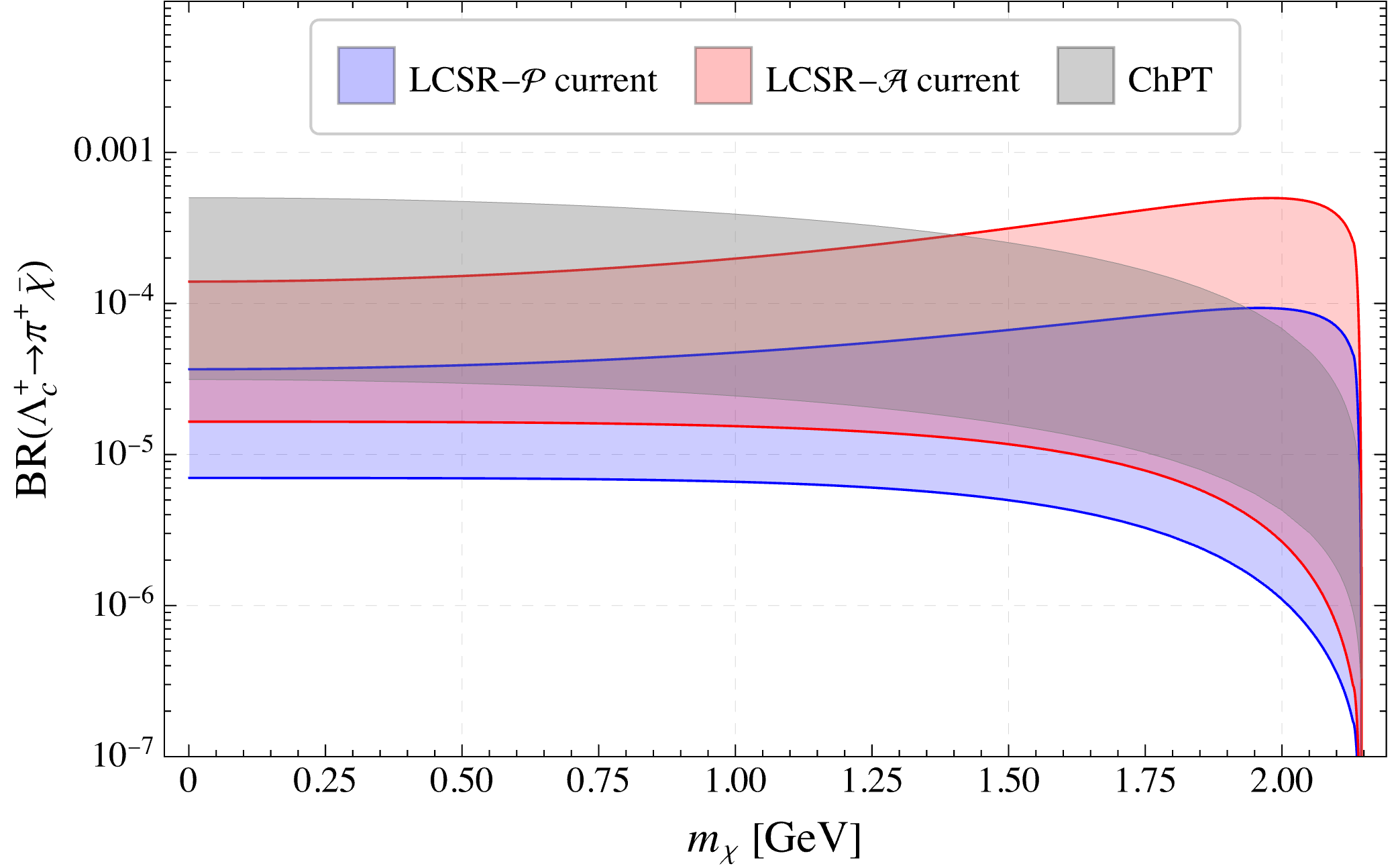}
    \hfill
    \includegraphics[width=0.49\textwidth]{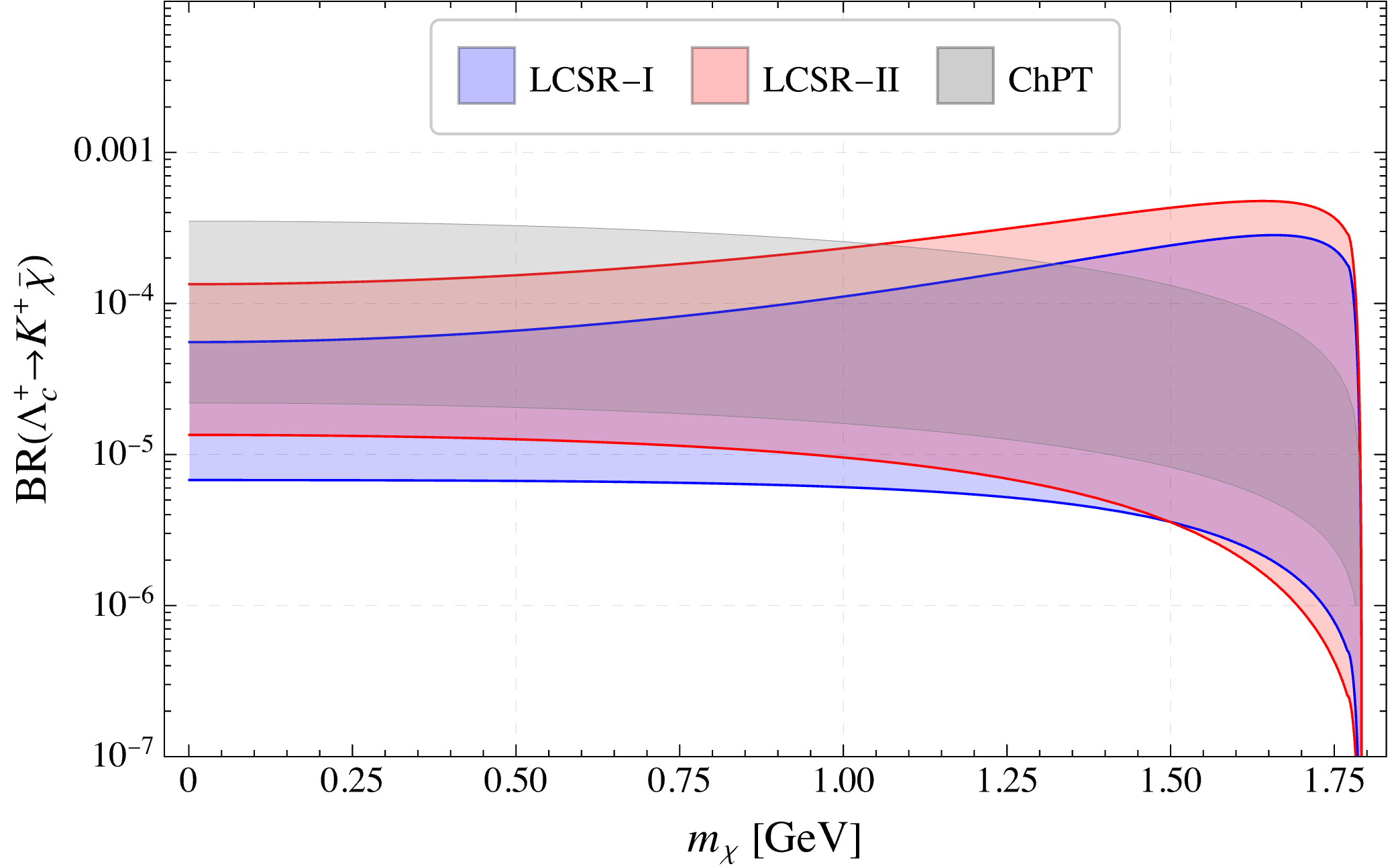}
\caption{
LCSR predictions compared with the ChPT
estimates~\cite{Wang:2026vxh} (grey bands).
Left: $\Lambda_c^+\to\pi^+\bar\chi$, with blue and red
bands denoting the $\mathcal P$- and $\mathcal A$-current
results for $\widetilde{\mathcal O}_{3,R}$.
Right: $\Lambda_c^+\to K^+\bar\chi$, with blue and red
bands denoting Combinations~I and~II, respectively, defined
in \cref{eq:chpt_current_combinations}.
The Wilson coefficients are fixed to the common comparison
benchmarks specified in the text.
}
    \label{fig:comparison-chpt}
\end{figure*}

As shown in \cref{fig:comparison-chpt}, the pion-channel
LCSR and ChPT predictions are of the same order of magnitude.
The $\mathcal A$-current result is generally larger than
the $\mathcal P$-current result, but their uncertainty bands
overlap substantially. The ChPT band also overlaps appreciably
with the LCSR bands.

For the kaon channel, Combination~II generally reaches larger
branching fractions than Combination~I, reflecting the
different current used for the
$\widetilde{\mathcal O}_{1,R}$ form factors. Their uncertainty
bands overlap substantially, and the ChPT band lies largely
within the range spanned by the LCSR predictions at small
$m_\chi$.

In both channels, differences in the mass dependence become
more pronounced at larger $m_\chi$: the upper LCSR boundaries
rise while the ChPT predictions decrease towards the endpoints.
Nevertheless, substantial overlap persists over most of the
physical region.

The agreement is non-trivial, since LCSR uses a light-cone
expansion in meson DAs, whereas ChPT employs a low-energy
chiral description. After matching the operator content
and Wilson-coefficient normalisation, the two approaches
yield compatible branching-fraction scales. This provides
a useful cross-check of the overall magnitude, rather than
a precision test, given the sizeable hadronic uncertainties
and distinct approximations.

\paragraph{Comparison with the QCDF estimate.}

We next compare our LCSR results with the QCDF estimates of~\cite{Hiller:2026osz}, obtained using a heavy-quark expansion
at large hadronic recoil, $q^2\simeq0$. Within that approximation,
the form factors satisfy
\begin{equation}
\widetilde F_{I,L}=-F_{I,L},
\qquad
\widetilde F_{I,R}=0,
\qquad
F_{1,L}=F_{2,L}.
\end{equation}
These relations are not imposed in our LCSR calculation, where
the form factors are extracted separately. In particular, we
obtain nonvanishing $\widetilde F_{I,R}$, which contribute to
the decay rates for nonzero $m_\chi$.

The right-handed form factors $F_{I,R}$ arise at higher twist
in the QCDF treatment and involve endpoint-divergent convolution
integrals. Their regularisation introduces order-one uncertainties,
leading the authors to refrain from further phenomenology for
these contributions. Our LCSR determination provides complementary
estimates, with uncertainties associated with the truncated
light-cone expansion, nonperturbative inputs, and sum-rule
parameters.

At $q^2=0$, the QCDF estimates are
$|F_{3,L}(0)|\simeq0.012\,\GeV^2$ and
$|F_{1,L}(0)|=|F_{2,L}(0)|\simeq0.0072\,\GeV^2$
\cite{Hiller:2026osz}. Our central LCSR values are roughly
one order of magnitude larger, although both approaches
carry sizeable hadronic uncertainties. We compare magnitudes, since no definite claim about the
absolute sign is made in the QCDF analysis.

These differences should be viewed in light of the scope
of the QCDF calculation, which estimates the hard-scattering
contribution but does not determine the nonperturbative
soft contribution. The authors therefore regard their
results as lower estimates of the full form factors.
The larger LCSR values need not contradict the QCDF estimates,
since the latter do not include the soft contribution to
the form factors.

\section{Expressions for the OPE invariant amplitudes}
\label{app:Pis}

We collect the nonvanishing coefficient functions
$\mathcal F_{i;I,X}^{(t,\mathcal J)}$ for the LO two-particle
contributions, defined in
\cref{eq:pion_OPE_generic,eq:pion_OPE_DA_structure},
and $\mathcal C_{i;I,X}^{(t,\mathcal J)}$ for the
factorisable four-particle contributions, defined in
\cref{eq:condensate_generic}. In both cases, the twist
labels refer to the accompanying two-particle DAs.

The expressions are given for the kaon channel. The pion
coefficients follow from the mapping specified at the end
of this appendix, including the two Wick contractions
associated with $\widetilde{\mathcal O}_{3,X}$.

\subsection{LO two-particle contributions}

For compactness, we define
\begin{align}
\mathcal D_K(u)
&=(1-u)P^2+uq^2-u(1-u)m_K^2,
\label{eq:DK}
\\
\Delta_K(u)
&=m_c^2-\mathcal D_K(u),
\label{eq:DeltaK}
\\
\mathcal R_K(u)
&=q^2-(1-u)^2m_K^2.
\label{eq:RK}
\end{align}


For the operator $\widetilde{\mathcal O}_{1,R}$ and the
$\mathcal P$ interpolating current, the independent nonvanishing
coefficient functions are
\begin{align}
\mathcal F_{1;1,R}^{(2,\mathcal P)}
&=
-\frac{f_K m_c N_c}{192\pi^2}\,
(P^2-u m_K^2)
\frac{\Delta_K^2(u)}{\mathcal D_K^2(u)},
\nonumber\\*
\mathcal F_{6;1,R}^{(2,\mathcal P)}
&=
-\frac{f_K N_c}{192\pi^2}\,
\frac{\Delta_K^2(u)}{\mathcal D_K(u)},
\\
\mathcal F_{1;1,R}^{(4,\psi,\mathcal P)}
&=
\frac{f_K m_c N_c}{96\pi^2}\,
\frac{\Delta_K(u)}{\mathcal D_K(u)},
\nonumber\\*
\mathcal F_{1;1,R}^{(4,\phi,\mathcal P)}
&=
\frac{f_K m_c^3 N_c}{384\pi^2}\,
\frac{P^2-u m_K^2}{\mathcal D_K^2(u)},
\\
\mathcal F_{6;1,R}^{(4,\psi,\mathcal P)}
&=
\frac{f_K N_c(1-u)}{192\pi^2}\,
\frac{\Delta_K(u)
[\mathcal D_K(u)+m_c^2]}{\mathcal D_K^2(u)},
\nonumber\\*
\mathcal F_{6;1,R}^{(4,\phi,\mathcal P)}
&=
\frac{f_K N_c}{384\pi^2}.
\end{align}
The remaining nonvanishing coefficients are related to the above by
\begin{align}
\mathcal F_{4;1,R}^{(2,\mathcal P)}
&=
-\frac{1}{P^2-u m_K^2}\,
\mathcal F_{1;1,R}^{(2,\mathcal P)},
\nonumber\\*
\mathcal F_{4;1,R}^{(4,\phi,\mathcal P)}
&=
-\frac{1}{P^2-u m_K^2}\,
\mathcal F_{1;1,R}^{(4,\phi,\mathcal P)},
\\
\mathcal F_{7;1,R}^{(2,\mathcal P)}
&=
-\mathcal F_{6;1,R}^{(2,\mathcal P)},
\nonumber\\*
\mathcal F_{7;1,R}^{(4,\phi,\mathcal P)}
&=
-\mathcal F_{6;1,R}^{(4,\phi,\mathcal P)},
\\
\mathcal F_{7;1,R}^{(4,\psi,\mathcal P)}
&=
\frac{u}{1-u}\,
\mathcal F_{6;1,R}^{(4,\psi,\mathcal P)}.
\end{align}
For the operator $\widetilde{\mathcal O}_{1,R}$ and the
$\mathcal A$ interpolating current, the LO two-particle contributions
involve only the twist-three distribution amplitudes
$\phi_{3K}^{p}$ and $\phi_{3K}^{\sigma}$.
The independent nonvanishing coefficient functions are

\begin{align}
\mathcal F_{1;1,R}^{(3,p,\mathcal A)}
&=
-\frac{f_K N_c\mu_K}{48\pi^2}\,
\frac{\Delta_K^2(u)}{\mathcal D_K(u)},
\nonumber\\*
\mathcal F_{6;1,R}^{(3,p,\mathcal A)}
&=
\frac{f_K m_c N_c\mu_K(1-u)}{96\pi^2}\,
\frac{\Delta_K^2(u)}{\mathcal D_K^2(u)}.
\end{align}
\begin{align}
\mathcal F_{1;1,R}^{(3,\sigma,\mathcal A)}
&=
\frac{f_K N_c\mu_K}{288\pi^2}\,\nonumber\\*
&\quad\times(P^2+q^2-m_K^2)
\frac{\Delta_K(u)[\mathcal D_K(u)+m_c^2]}
{\mathcal D_K^2(u)},
\\
\mathcal F_{6;1,R}^{(3,\sigma,\mathcal A)}
&=
-\frac{f_K m_c N_c\mu_K}{576\pi^2}\,
\frac{\Delta_K(u)}{\mathcal D_K^3(u)}\nonumber\\*
&\quad\times\Big[3\mathcal D_K^2(u)-m_c^2\mathcal D_K(u)\nonumber\\*
&\qquad+2m_c^2\big(q^2-(1-u)^2m_K^2\big)\Big],
\\
\mathcal F_{7;1,R}^{(3,\sigma,\mathcal A)}
&=
\frac{f_K m_c N_c\mu_K}{576\pi^2}\,
\frac{\Delta_K(u)}{\mathcal D_K^3(u)}\nonumber\\*
&\quad\times\Big[3\mathcal D_K^2(u)-m_c^2\mathcal D_K(u)\nonumber\\*
&\qquad+2m_c^2\big(P^2-u^2m_K^2\big)\Big].
\end{align}
The remaining nonvanishing coefficients are related to the above by
\begin{align}
\mathcal F_{4;1,R}^{(3,\sigma,\mathcal A)}
&=
\frac{-2}{P^2+q^2-m_K^2}\,
\mathcal F_{1;1,R}^{(3,\sigma,\mathcal A)},
\nonumber\\*
\mathcal F_{7;1,R}^{(3,p,\mathcal A)}
&=
\frac{u}{1-u}\,
\mathcal F_{6;1,R}^{(3,p,\mathcal A)}.
\end{align}
For the operator $\widetilde{\mathcal O}_{2,R}$ and the
$\mathcal P$ interpolating current, the independent nonvanishing
coefficient functions are
\begin{align}
\mathcal F_{1;2,R}^{(2,\mathcal P)}
&=
-\frac{f_K m_c N_c}{192\pi^2}\,
\big[q^2-(1-u)m_K^2\big]\,
\frac{\Delta_K^2(u)}{\mathcal D_K^2(u)},
\\
\mathcal F_{1;2,R}^{(4,\phi,\mathcal P)}
&=
\frac{f_K m_c^3 N_c}{384\pi^2}\,
\frac{q^2-(1-u)m_K^2}{\mathcal D_K^2(u)},
\\
\mathcal F_{1;2,R}^{(4,\psi,\mathcal P)}
&=
-\frac{f_K m_c N_c}{96\pi^2}\,
\frac{\Delta_K(u)}{\mathcal D_K(u)},
\\
\mathcal F_{6;2,R}^{(2,\mathcal P)}
&=
-\frac{f_K N_c}{576\pi^2}\,
\frac{\Delta_K^2(u)}{\mathcal D_K^3(u)}\nonumber\\*
&\quad\times\Big\{
\mathcal D_K(u)\big[\mathcal D_K(u)-m_c^2\big]
\nonumber\\*
&\quad
+\big[\mathcal D_K(u)+2m_c^2\big]
\big[q^2-(1-u)^2m_K^2\big]
\Big\},
\\
\mathcal F_{7;2,R}^{(2,\mathcal P)}
&=
\frac{f_K N_c}{576\pi^2}\,
\frac{\Delta_K^2(u)}{\mathcal D_K^3(u)}\nonumber\\*
&\quad\times\Big\{
\mathcal D_K(u)\big[\mathcal D_K(u)-m_c^2\big]
\nonumber\\*
&\quad
+\big[\mathcal D_K(u)+2m_c^2\big]
\big[P^2-u^2m_K^2\big]
\Big\},
\\
\mathcal F_{6;2,R}^{(4,\phi,\mathcal P)}
&=
\frac{f_K N_c}{768\pi^2}\,
\frac{1}{\mathcal D_K^3(u)}\nonumber\\*
&\quad\times\Big[
\mathcal D_K(u)\big[\mathcal D_K^2(u)-m_c^4\big]\nonumber\\*
&\qquad+2m_c^4\big[q^2-(1-u)^2m_K^2\big]\Big],
\\
\mathcal F_{7;2,R}^{(4,\phi,\mathcal P)}
&=
-\frac{f_K N_c}{768\pi^2}\,
\frac{1}{\mathcal D_K^3(u)}\nonumber\\*
&\quad\times\Big[
\mathcal D_K(u)\big[\mathcal D_K^2(u)-m_c^4\big]\nonumber\\*
&\qquad+2m_c^4\big[P^2-u^2m_K^2\big]\Big],
\\
\mathcal F_{6;2,R}^{(4,\psi,\mathcal P)}
&=
-\frac{f_K N_c}{192\pi^2}\,
(1-u)\,
\frac{
\Delta_K(u)\big[\mathcal D_K(u)+m_c^2\big]
}{
\mathcal D_K^2(u)
}.
\end{align}
The remaining nonvanishing coefficients are related to the above by
\begin{align}
\mathcal F_{4;2,R}^{(2,\mathcal P)}
&=
\frac{-1}{q^2-(1-u)m_K^2}\,
\mathcal F_{1;2,R}^{(2,\mathcal P)},
\nonumber\\*
\mathcal F_{4;2,R}^{(4,\phi,\mathcal P)}
&=
\frac{-1}{q^2-(1-u)m_K^2}\,
\mathcal F_{1;2,R}^{(4,\phi,\mathcal P)},
\\
\mathcal F_{7;2,R}^{(4,\psi,\mathcal P)}
&=
\frac{u}{1-u}\,
\mathcal F_{6;2,R}^{(4,\psi,\mathcal P)}.
\end{align}
For the operator $\widetilde{\mathcal O}_{2,R}$ and the
$\mathcal A$ interpolating current, only twist-three coefficient
functions contribute to the expressions considered here.
The independent nonvanishing coefficient functions are
\begin{align}
\mathcal F_{1;2,R}^{(3,p,\mathcal A)}
&=
-\frac{f_K N_c\mu_K}{96\pi^2}\,
\frac{\Delta_K^2(u)}{\mathcal D_K(u)},
\\
\mathcal F_{1;2,R}^{(3,\sigma,\mathcal A)}
&=
\frac{f_K N_c\mu_K}{576\pi^2}\,\nonumber\\*
&\quad\times\big(P^2+q^2-m_K^2\big)\,
\frac{
\Delta_K(u)\big[\mathcal D_K(u)+m_c^2\big]
}{
\mathcal D_K^2(u)
},
\\
\mathcal F_{6;2,R}^{(3,p,\mathcal A)}
&=
\frac{f_K m_c N_c\mu_K}{48\pi^2}\,
(1-u)\,
\frac{\Delta_K^2(u)}{\mathcal D_K^2(u)}.
\end{align}
The remaining nonvanishing coefficients are related to the above by
\begin{align}
\mathcal F_{4;2,R}^{(3,\sigma,\mathcal A)}
&=
\frac{-2}{P^2+q^2-m_K^2}\,
\mathcal F_{1;2,R}^{(3,\sigma,\mathcal A)},
\nonumber\\*
\mathcal F_{7;2,R}^{(3,p,\mathcal A)}
&=
\frac{u}{1-u}\,
\mathcal F_{6;2,R}^{(3,p,\mathcal A)}.
\end{align}
For the operators $\widetilde{\mathcal O}_{1,L}$ and
$\widetilde{\mathcal O}_{2,L}$ with the $\mathcal P$
interpolating current, the coefficient functions coincide.
The expressions below therefore apply to both $I=1$ and $I=2$.
Only the twist-three distribution amplitudes
$\phi_{3K}^{p}$ and $\phi_{3K}^{\sigma}$ contribute to
the terms considered here.
The independent nonvanishing coefficient functions are
\begin{align}
\mathcal F_{1;I,L}^{(3,p,\mathcal P)}
&=
\frac{f_K N_c\mu_K}{192\pi^2}\,
\frac{\Delta_K^2(u)}{\mathcal D_K(u)},
\\
\mathcal F_{1;I,L}^{(3,\sigma,\mathcal P)}
&=
-\frac{f_K N_c\mu_K}{1152\pi^2}\,\nonumber\\*
&\quad\times\big(P^2+q^2-m_K^2\big)\,
\frac{
\Delta_K(u)\big[\mathcal D_K(u)+m_c^2\big]
}{
\mathcal D_K^2(u)
},
\\
\mathcal F_{6;I,L}^{(3,p,\mathcal P)}
&=
\frac{f_K m_c N_c\mu_K}{192\pi^2}\,
(1-u)\,
\frac{\Delta_K^2(u)}{\mathcal D_K^2(u)},
\\
\mathcal F_{6;I,L}^{(3,\sigma,\mathcal P)}
&=
-\frac{f_K m_c N_c\mu_K}{1152\pi^2}\,
\frac{\Delta_K(u)}{\mathcal D_K^3(u)}\nonumber\\*
&\quad\times\Big[3\mathcal D_K^2(u)-m_c^2\mathcal D_K(u)\nonumber\\*
&\qquad+2m_c^2\big(q^2-(1-u)^2m_K^2\big)\Big],
\\
\mathcal F_{7;I,L}^{(3,\sigma,\mathcal P)}
&=
\frac{f_K m_c N_c\mu_K}{1152\pi^2}\,
\frac{\Delta_K(u)}{\mathcal D_K^3(u)}\nonumber\\*
&\quad\times\Big[3\mathcal D_K^2(u)-m_c^2\mathcal D_K(u)\nonumber\\*
&\qquad+2m_c^2\big(P^2-u^2m_K^2\big)\Big].
\end{align}
The remaining nonvanishing coefficients are related to the above by
\begin{align}
\mathcal F_{4;I,L}^{(3,\sigma,\mathcal P)}
&=
\frac{-2}{P^2+q^2-m_K^2}\,
\mathcal F_{1;I,L}^{(3,\sigma,\mathcal P)},
\nonumber\\*
\mathcal F_{7;I,L}^{(3,p,\mathcal P)}
&=
\frac{u}{1-u}\,
\mathcal F_{6;I,L}^{(3,p,\mathcal P)}.
\end{align}
For the operator $\widetilde{\mathcal O}_{1,L}$ and the
$\mathcal A$ interpolating current, the independent nonvanishing
coefficient functions are
\begin{align}
\mathcal F_{1;1,L}^{(2,\mathcal A)}
&=
-\frac{f_K m_c N_c}{96\pi^2}\,
(P^2-u m_K^2)
\frac{\Delta_K^2(u)}{\mathcal D_K^2(u)},
\nonumber\\*
\mathcal F_{6;1,L}^{(2,\mathcal A)}
&=
\frac{f_K N_c}{48\pi^2}\,
\frac{\Delta_K^2(u)}{\mathcal D_K(u)},
\\
\mathcal F_{1;1,L}^{(4,\psi,\mathcal A)}
&=
\frac{f_K m_c N_c}{48\pi^2}\,
\frac{\Delta_K(u)}{\mathcal D_K(u)},
\nonumber\\*
\mathcal F_{1;1,L}^{(4,\phi,\mathcal A)}
&=
\frac{f_K m_c^3 N_c}{192\pi^2}\,
\frac{P^2-u m_K^2}{\mathcal D_K^2(u)},
\\
\mathcal F_{6;1,L}^{(4,\psi,\mathcal A)}
&=
-\frac{f_K N_c(1-u)}{48\pi^2}\,
\frac{\Delta_K(u)
[\mathcal D_K(u)+m_c^2]}{\mathcal D_K^2(u)},
\nonumber\\*
\mathcal F_{6;1,L}^{(4,\phi,\mathcal A)}
&=
-\frac{f_K N_c}{96\pi^2}.
\end{align}
The remaining nonvanishing coefficients are related to the above by
\begin{align}
\mathcal F_{4;1,L}^{(2,\mathcal A)}
&=
-\frac{1}{P^2-u m_K^2}\,
\mathcal F_{1;1,L}^{(2,\mathcal A)},
\nonumber\\*
\mathcal F_{4;1,L}^{(4,\phi,\mathcal A)}
&=
-\frac{1}{P^2-u m_K^2}\,
\mathcal F_{1;1,L}^{(4,\phi,\mathcal A)},
\\
\mathcal F_{7;1,L}^{(2,\mathcal A)}
&=
-\mathcal F_{6;1,L}^{(2,\mathcal A)},
\nonumber\\*
\mathcal F_{7;1,L}^{(4,\phi,\mathcal A)}
&=
-\mathcal F_{6;1,L}^{(4,\phi,\mathcal A)},
\\
\mathcal F_{7;1,L}^{(4,\psi,\mathcal A)}
&=
\frac{u}{1-u}\,
\mathcal F_{6;1,L}^{(4,\psi,\mathcal A)}.
\end{align}
For the operator $\widetilde{\mathcal O}_{2,L}$ and the
$\mathcal A$ interpolating current, the independent nonvanishing
coefficient functions are
\begin{align}
\mathcal F_{1;2,L}^{(2,\mathcal A)}
&=
\frac{f_K m_c N_c}{96\pi^2}\,
\big[q^2-(1-u)m_K^2\big]\,
\frac{\Delta_K^2(u)}{\mathcal D_K^2(u)},
\\
\mathcal F_{1;2,L}^{(4,\phi,\mathcal A)}
&=
-\frac{f_K m_c^3 N_c}{192\pi^2}\,
\frac{q^2-(1-u)m_K^2}{\mathcal D_K^2(u)},
\\
\mathcal F_{1;2,L}^{(4,\psi,\mathcal A)}
&=
\frac{f_K m_c N_c}{48\pi^2}\,
\frac{\Delta_K(u)}{\mathcal D_K(u)},
\\
\mathcal F_{6;2,L}^{(2,\mathcal A)}
&=
\frac{f_K N_c}{288\pi^2}\,
\frac{\Delta_K^2(u)}{\mathcal D_K^3(u)}\nonumber\\*
&\quad\times\Big\{
2\mathcal D_K^2(u)
+m_c^2\mathcal D_K(u)\nonumber\\*
&\qquad-\mathcal D_K(u)\mathcal R_K(u)
-2m_c^2\mathcal R_K(u)
\Big\},
\\
\mathcal F_{7;2,L}^{(2,\mathcal A)}
&=
-\frac{f_K N_c}{288\pi^2}\,
\frac{\Delta_K^2(u)}
{(1-u)\mathcal D_K^3(u)}\nonumber\\*
&\quad\times\Big\{
(1-2u)\mathcal D_K^2(u)\nonumber\\*
&\qquad-(1+u)m_c^2\mathcal D_K(u)
\nonumber\\*
&\quad
+u\mathcal D_K(u)\mathcal R_K(u)
+2u m_c^2\mathcal R_K(u)
\Big\},
\\
\mathcal F_{6;2,L}^{(4,\phi,\mathcal A)}
&=
-\frac{f_K N_c}{384\pi^2}\,
\frac{1}{\mathcal D_K^3(u)}\nonumber\\*
&\quad\times\Big[
\mathcal D_K^3(u)+m_c^4\mathcal D_K(u)
-2m_c^4\mathcal R_K(u)\Big],
\\
\mathcal F_{7;2,L}^{(4,\phi,\mathcal A)}
&=
\frac{f_K N_c}{384\pi^2}\,
\frac{1}{(1-u)\mathcal D_K^3(u)}\nonumber\\*
&\quad\times\Big[
(1-u)\mathcal D_K^3(u)\nonumber\\*
&\qquad-(1+u)m_c^4\mathcal D_K(u)
+2u m_c^4\mathcal R_K(u)\Big],
\\
\mathcal F_{6;2,L}^{(4,\psi,\mathcal A)}
&=
\frac{f_K N_c}{48\pi^2}\,
(1-u)\,
\frac{
\mathcal D_K^2(u)-m_c^4
}{
\mathcal D_K^2(u)
}.
\end{align}
The remaining nonvanishing coefficients are related to the above by
\begin{align}
\mathcal F_{4;2,L}^{(2,\mathcal A)}
&=
-\frac{1}{q^2-(1-u)m_K^2}\,
\mathcal F_{1;2,L}^{(2,\mathcal A)},
\nonumber\\*
\mathcal F_{4;2,L}^{(4,\phi,\mathcal A)}
&=
-\frac{1}{q^2-(1-u)m_K^2}\,
\mathcal F_{1;2,L}^{(4,\phi,\mathcal A)},
\\
\mathcal F_{7;2,L}^{(4,\psi,\mathcal A)}
&=
\frac{u}{1-u}\,
\mathcal F_{6;2,L}^{(4,\psi,\mathcal A)}.
\end{align}
\subsection{Factorisable four-particle contributions}

For the operator $\widetilde{\mathcal O}_{1,R}$ with the
$\mathcal P$ interpolating current, the independent nonvanishing
factorisable four-particle coefficient functions are
\begin{align}
\mathcal C_{1;1,R}^{(3,p,\mathcal P)}
&=
\frac{m_c}{72\,\Delta_K(u)},
\nonumber\\*
\mathcal C_{6;1,R}^{(3,p,\mathcal P)}
&=
\frac{1-u}{72\,\Delta_K(u)}.
\end{align}
\begin{align}
\mathcal C_{1;1,R}^{(3,\sigma,\mathcal P)}
&=
\frac{m_c}
{432(1-u)\Delta_K^2(u)}\nonumber\\*
&\quad\times\Big[
m_c^2-\Delta_K(u)+\mathcal R_K(u)-2u q^2
\Big],
\\
\mathcal C_{6;1,R}^{(3,\sigma,\mathcal P)}
&=
\frac{
2\Delta_K(u)+m_c^2+\mathcal R_K(u)
}{
432\,\Delta_K^2(u)
},
\\
\mathcal C_{7;1,R}^{(3,\sigma,\mathcal P)}
&=
-\frac{
(1-2u)\Delta_K(u)+(2-u)m_c^2-u\mathcal R_K(u)
}{
432(1-u)\Delta_K^2(u)
}.
\end{align}
The remaining nonvanishing coefficients are related to the above by
\begin{align}
\mathcal C_{4;1,R}^{(3,\sigma,\mathcal P)}
&=
-\frac{1}{3\,\Delta_K(u)}\,
\mathcal C_{1;1,R}^{(3,p,\mathcal P)},
\nonumber\\*
\mathcal C_{7;1,R}^{(3,p,\mathcal P)}
&=
\frac{u}{1-u}\,
\mathcal C_{6;1,R}^{(3,p,\mathcal P)}.
\end{align}
For the operator $\widetilde{\mathcal O}_{1,R}$ with the
$\mathcal A$ interpolating current, the independent nonvanishing
coefficient functions are
\begin{align}
\mathcal C_{1;1,R}^{(2,\mathcal A)}
&=
-\frac{P^2-u m_K^2}{36\,\Delta_K(u)},
\nonumber\\*
\mathcal C_{6;1,R}^{(2,\mathcal A)}
&=
\frac{m_c}{18\,\Delta_K(u)},
\\
\mathcal C_{1;1,R}^{(4,\phi,\mathcal A)}
&=
\frac{
(P^2-u m_K^2)[\Delta_K(u)+2m_c^2]
}{
144\,\Delta_K^3(u)
},
\nonumber\\*
\mathcal C_{6;1,R}^{(4,\phi,\mathcal A)}
&=
-\frac{m_c^3}{36\,\Delta_K^3(u)},
\\
\mathcal C_{1;1,R}^{(4,\psi,\mathcal A)}
&=
-\frac{\Delta_K(u)+m_c^2}{36\,\Delta_K^2(u)},
\nonumber\\*
\mathcal C_{6;1,R}^{(4,\psi,\mathcal A)}
&=
\frac{m_c(1-u)}{18\,\Delta_K^2(u)}.
\end{align}
The remaining nonvanishing coefficients are related to the above by
\begin{align}
\mathcal C_{4;1,R}^{(2,\mathcal A)}
&=
-\frac{1}{P^2-u m_K^2}\,
\mathcal C_{1;1,R}^{(2,\mathcal A)},
\nonumber\\*
\mathcal C_{4;1,R}^{(4,\phi,\mathcal A)}
&=
-\frac{1}{P^2-u m_K^2}\,
\mathcal C_{1;1,R}^{(4,\phi,\mathcal A)},
\\
\mathcal C_{7;1,R}^{(2,\mathcal A)}
&=
-\mathcal C_{6;1,R}^{(2,\mathcal A)},
\nonumber\\*
\mathcal C_{7;1,R}^{(4,\phi,\mathcal A)}
&=
-\mathcal C_{6;1,R}^{(4,\phi,\mathcal A)},
\\
\mathcal C_{7;1,R}^{(4,\psi,\mathcal A)}
&=
\frac{u}{1-u}\,
\mathcal C_{6;1,R}^{(4,\psi,\mathcal A)}.
\end{align}
For the operator $\widetilde{\mathcal O}_{2,R}$ with the
$\mathcal P$ interpolating current, the independent nonvanishing
coefficient functions are
\begin{align}
\mathcal C_{1;2,R}^{(3,p,\mathcal P)}
&=
-\frac{m_c}{72\,\Delta_K(u)},
\\
\mathcal C_{1;2,R}^{(3,\sigma,\mathcal P)}
&=
\frac{m_c}
{432(1-u)\Delta_K^2(u)}\nonumber\\*
&\quad\times\Big[
m_c^2-\Delta_K(u)
+\mathcal R_K(u)
-2u q^2
\Big],
\\
\mathcal C_{6;2,R}^{(3,p,\mathcal P)}
&=
-\frac{1-u}{72\,\Delta_K(u)},
\\
\mathcal C_{6;2,R}^{(3,\sigma,\mathcal P)}
&=
\frac{
2\Delta_K(u)+m_c^2+\mathcal R_K(u)
}{
432\,\Delta_K^2(u)
},
\\
\mathcal C_{7;2,R}^{(3,\sigma,\mathcal P)}
&=
-\frac{
(1-2u)\Delta_K(u)
+(2-u)m_c^2
-u\mathcal R_K(u)
}{
432(1-u)\Delta_K^2(u)
}.
\end{align}
The remaining nonvanishing coefficients are related to the above by
\begin{align}
\mathcal C_{4;2,R}^{(3,\sigma,\mathcal P)}
&=
\frac{1}{3\,\Delta_K(u)}\,
\mathcal C_{1;2,R}^{(3,p,\mathcal P)},
\nonumber\\*
\mathcal C_{7;2,R}^{(3,p,\mathcal P)}
&=
\frac{u}{1-u}\,
\mathcal C_{6;2,R}^{(3,p,\mathcal P)}.
\end{align}
For the operator $\widetilde{\mathcal O}_{2,R}$ with the
$\mathcal A$ interpolating current, the independent nonvanishing
coefficient functions are
\begin{align}
\mathcal C_{1;2,R}^{(2,\mathcal A)}
&=
\frac{q^2-P^2+(2u-1)m_K^2}{36\,\Delta_K(u)},
\nonumber\\*
\mathcal C_{6;2,R}^{(2,\mathcal A)}
&=
\frac{m_c}{36\,\Delta_K(u)},
\\
\mathcal C_{1;2,R}^{(4,\phi,\mathcal A)}
&=
-\frac{
[q^2-P^2+(2u-1)m_K^2]
[\Delta_K(u)+2m_c^2]
}{
144\,\Delta_K^3(u)
},
\nonumber\\*
\mathcal C_{6;2,R}^{(4,\phi,\mathcal A)}
&=
-\frac{m_c^3}{72\,\Delta_K^3(u)},
\\
\mathcal C_{1;2,R}^{(4,\psi,\mathcal A)}
&=
-\frac{\Delta_K(u)+m_c^2}{18\,\Delta_K^2(u)},
\nonumber\\*
\mathcal C_{6;2,R}^{(4,\psi,\mathcal A)}
&=
\frac{m_c(1-u)}{36\,\Delta_K^2(u)}.
\end{align}
The remaining nonvanishing coefficients are related to the above by
\begin{align}
\mathcal C_{7;2,R}^{(2,\mathcal A)}
&=
-\mathcal C_{6;2,R}^{(2,\mathcal A)},
\nonumber\\*
\mathcal C_{7;2,R}^{(4,\phi,\mathcal A)}
&=
-\mathcal C_{6;2,R}^{(4,\phi,\mathcal A)},
\\
\mathcal C_{7;2,R}^{(4,\psi,\mathcal A)}
&=
\frac{u}{1-u}\,
\mathcal C_{6;2,R}^{(4,\psi,\mathcal A)}.
\end{align}
For the operators $\widetilde{\mathcal O}_{1,L}$ and
$\widetilde{\mathcal O}_{2,L}$ with the $\mathcal P$ interpolating
current, the coefficient functions coincide.
The expressions below therefore apply to both $I=1$ and $I=2$.
The independent nonvanishing coefficient functions are
\begin{align}
\mathcal C_{1;I,L}^{(2,\mathcal P)}
&=
-\frac{m_c^2-\Delta_K(u)-u q^2}
{72(1-u)\Delta_K(u)},
\nonumber\\*
\mathcal C_{4;I,L}^{(2,\mathcal P)}
&=
\frac{1}{72\,\Delta_K(u)},
\\
\mathcal C_{1;I,L}^{(4,\psi,\mathcal P)}
&=
-\frac{\Delta_K(u)+m_c^2}
{72\,\Delta_K^2(u)},
\nonumber\\*
\mathcal C_{4;I,L}^{(4,\phi,\mathcal P)}
&=
-\frac{\Delta_K(u)+2m_c^2}
{288\,\Delta_K^3(u)},
\\
\mathcal C_{6;I,L}^{(2,\mathcal P)}
&=
-\frac{m_c}{72\,\Delta_K(u)},
\nonumber\\*
\mathcal C_{6;I,L}^{(4,\phi,\mathcal P)}
&=
\frac{m_c^3}{144\,\Delta_K^3(u)}.
\end{align}
\begin{align}
\mathcal C_{1;I,L}^{(4,\phi,\mathcal P)}
&=
\frac{1}{288(1-u)\Delta_K^3(u)}\nonumber\\*
&\quad\times\Big[
2m_c^4-m_c^2\Delta_K(u)-\Delta_K^2(u)\nonumber\\*
&\qquad-u[\Delta_K(u)+2m_c^2]q^2\Big],
\\
\mathcal C_{6;I,L}^{(4,\psi,\mathcal P)}
&=
-\frac{m_c(1-u)}{72\,\Delta_K^2(u)}.
\end{align}
The remaining nonvanishing coefficients are related to the above by
\begin{align}
\mathcal C_{7;I,L}^{(2,\mathcal P)}
&=
-\mathcal C_{6;I,L}^{(2,\mathcal P)},
\quad
\mathcal C_{7;I,L}^{(4,\phi,\mathcal P)}
=
-\mathcal C_{6;I,L}^{(4,\phi,\mathcal P)},
\\
\mathcal C_{7;I,L}^{(4,\psi,\mathcal P)}
&=
\frac{u}{1-u}\,
\mathcal C_{6;I,L}^{(4,\psi,\mathcal P)},
\qquad I=1,2.
\end{align}
For the operator $\widetilde{\mathcal O}_{1,L}$ with the
$\mathcal A$ interpolating current, the independent nonvanishing
coefficient functions are
\begin{align}
\mathcal C_{1;1,L}^{(3,p,\mathcal A)}
&=
-\frac{m_c}{18\,\Delta_K(u)},
\\
\mathcal C_{1;1,L}^{(3,\sigma,\mathcal A)}
&=
\frac{m_c}
{108(1-u)\Delta_K^2(u)}\nonumber\\*
&\quad\times\Big[
\Delta_K(u)-m_c^2
-\mathcal R_K(u)
+2u q^2
\Big],
\\
\mathcal C_{6;1,L}^{(3,p,\mathcal A)}
&=
\frac{1-u}{36\,\Delta_K(u)},
\\
\mathcal C_{6;1,L}^{(3,\sigma,\mathcal A)}
&=
\frac{
2\Delta_K(u)+m_c^2+\mathcal R_K(u)
}{
216\,\Delta_K^2(u)
},
\\
\mathcal C_{7;1,L}^{(3,\sigma,\mathcal A)}
&=
-\frac{
(1-2u)\Delta_K(u)
+(2-u)m_c^2
-u\mathcal R_K(u)
}{
216(1-u)\Delta_K^2(u)
}.
\end{align}
The remaining nonvanishing coefficients are related to the above by
\begin{align}
\mathcal C_{4;1,L}^{(3,\sigma,\mathcal A)}
&=
-\frac{1}{3\,\Delta_K(u)}\,
\mathcal C_{1;1,L}^{(3,p,\mathcal A)},
\nonumber\\*
\mathcal C_{7;1,L}^{(3,p,\mathcal A)}
&=
\frac{u}{1-u}\,
\mathcal C_{6;1,L}^{(3,p,\mathcal A)}.
\end{align}
For the operator $\widetilde{\mathcal O}_{2,L}$ with the
$\mathcal A$ interpolating current, the independent nonvanishing
coefficient functions are
\begin{align}
\mathcal C_{1;2,L}^{(3,p,\mathcal A)}
&=
-\frac{m_c}{18\,\Delta_K(u)},
\nonumber\\*
\mathcal C_{6;2,L}^{(3,p,\mathcal A)}
&=
\frac{1-u}{36\,\Delta_K(u)}.
\end{align}
\begin{align}
\mathcal C_{6;2,L}^{(3,\sigma,\mathcal A)}
&=
-\frac{
2\Delta_K(u)+m_c^2+\mathcal R_K(u)
}{
216\,\Delta_K^2(u)
},
\\
\mathcal C_{7;2,L}^{(3,\sigma,\mathcal A)}
&=
\frac{
(1-2u)\Delta_K(u)+(2-u)m_c^2-u\mathcal R_K(u)
}{
216(1-u)\Delta_K^2(u)
}.
\end{align}
The remaining nonvanishing coefficient is related to the above by
\begin{equation}
\mathcal C_{7;2,L}^{(3,p,\mathcal A)}
=
\frac{u}{1-u}\,
\mathcal C_{6;2,L}^{(3,p,\mathcal A)}.
\end{equation}
For completeness, we now specify the mapping from the kaon coefficient
functions listed above to those of the pion channel. For
$\widetilde{\mathcal O}_{3,X}$, two distinct Wick contractions
contribute, corresponding to the outer-$d$ and diquark-$d$
contractions introduced in \cref{sec:OPE}. These are in one-to-one
correspondence with the contractions generated by
$\widetilde{\mathcal O}_{2,X}$ and
$\widetilde{\mathcal O}_{1,X}$, respectively, in the kaon channel.
Consequently, for either chirality $X=L,R$ and either interpolating
current $\mathcal J=\mathcal P,\mathcal A$, the two pion contributions
are obtained by taking the pion limit of the corresponding kaon
coefficient functions:
\begin{align}
\mathcal F_{i;3,X}^{(t,\mathcal J),\mathrm{diq}}
&=
\left.
\mathcal F_{i;1,X}^{(t,\mathcal J)}
\right|_{K\to\pi}, \qquad
\mathcal F_{i;3,X}^{(t,\mathcal J),\mathrm{out}}
=
\left.
\mathcal F_{i;2,X}^{(t,\mathcal J)}
\right|_{K\to\pi},
\\
\mathcal C_{i;3,X}^{(t,\mathcal J),\mathrm{diq}}
&=
\left.
\mathcal C_{i;1,X}^{(t,\mathcal J)}
\right|_{K\to\pi},
\qquad
\mathcal C_{i;3,X}^{(t,\mathcal J),\mathrm{out}}
=
\left.
\mathcal C_{i;2,X}^{(t,\mathcal J)}
\right|_{K\to\pi}.
\end{align}
The complete pion coefficient functions are obtained by summing the
diquark-$d$ and outer-$d$ contributions. Here, $t$ labels the twist structure of the accompanying
two-particle DA, while $K\to\pi$ denotes the replacement
of the kaon DAs and their nonperturbative parameters by their pion
counterparts, together with $m_K^2\to m_\pi^2\simeq0$ in the
kinematic functions.

\bibliography{references-prd}

@misc{Hiller:2026osz,
  author        = {Hiller, Gudrun and Rodr{\'\i}guez-S{\'a}nchez, Antonio and Wendler, Daniel},
  title         = {Probing baryon number with missing energy},
  eprint        = {2602.15936},
  archivePrefix = {arXiv},
  primaryClass  = {hep-ph},
  year          = {2026}
}

@article{Ball:2006wn,

    author = "Ball, Patricia and Braun, Vladimir M. and Lenz, Alexander",

    title = "{Higher-twist distribution amplitudes of the K meson in QCD}",

    eprint = "hep-ph/0603063",

    archivePrefix = "arXiv",

    primaryClass = "hep-ph",

    journal = "JHEP",

    volume = "05",

    pages = "004",

    year = "2006"

}

@article{Duplancic:2008ix,
    author = "Duplancic, G. and Khodjamirian, A. and Mannel, Th. and Melic, B. and Offen, N.",
    title = "{Light-cone sum rules for $B\to\pi$ form factors revisited}",
    eprint = "0801.1796",
    archivePrefix = "arXiv",
    primaryClass = "hep-ph",
    reportNumber = "SI-HEP-2007-15",
    doi = "10.1088/1126-6708/2008/04/014",
    journal = "JHEP",
    volume = "04",
    pages = "014",
    year = "2008"
}

@article{Khodjamirian:2009ys,
    author = "Khodjamirian, A. and Klein, Ch. and Mannel, Th. and Offen, N.",
    title = "{Semileptonic charm decays $D \to \pi \ell \nu_\ell$ and $D \to K \ell \nu_\ell$ from QCD Light-Cone Sum Rules}",
    eprint = "0907.2842",
    archivePrefix = "arXiv",
    primaryClass = "hep-ph",
    reportNumber = "SI-HEP-2009-06, LPT-09-60",
    doi = "10.1103/PhysRevD.80.114005",
    journal = "Phys. Rev. D",
    volume = "80",
    pages = "114005",
    year = "2009"
}

@article{Khodjamirian:2017fxg,
    author = "Khodjamirian, Alexander and Rusov, Aleksey V.",
    title = "{$B_{s}\to K \ell \nu_\ell$ and $B_{(s)} \to \pi (K) \ell^+\ell^-$ decays at large recoil and CKM matrix elements}",
    eprint = "1703.04765",
    archivePrefix = "arXiv",
    primaryClass = "hep-ph",
    reportNumber = "SI-HEP-2017-03, QFET-2017-03",
    doi = "10.1007/JHEP08(2017)112",
    journal = "JHEP",
    volume = "08",
    pages = "112",
    year = "2017"
}

@article{Cheng:2020vwr,
    author = "Cheng, Shan and Khodjamirian, Alexander and Rusov, Aleksey V.",
    title = "{Pion light-cone distribution amplitude from the pion electromagnetic form factor}",
    eprint = "2007.05550",
    archivePrefix = "arXiv",
    primaryClass = "hep-ph",
    reportNumber = "SI-HEP-2020-15, P3H-20-032, IPPP/20/27",
    doi = "10.1103/PhysRevD.102.074022",
    journal = "Phys. Rev. D",
    volume = "102",
    number = "7",
    pages = "074022",
    year = "2020"
}

@article{Chetyrkin:2007vm,
    author = "Chetyrkin, K. G. and Khodjamirian, A. and Pivovarov, A. A.",
    title = "{Towards NNLO Accuracy in the QCD Sum Rule for the Kaon Distribution Amplitude}",
    eprint = "0712.2999",
    archivePrefix = "arXiv",
    primaryClass = "hep-ph",
    reportNumber = "SI-HEP-2007-19, TTP-07-36, SFB-CPP-07-83",
    doi = "10.1016/j.physletb.2008.02.031",
    journal = "Phys. Lett. B",
    volume = "661",
    pages = "250--258",
    year = "2008"
}

@article{Braun:1999uj,
    author = "Braun, Vladimir M. and Khodjamirian, A. and Maul, M.",
    title = "{Pion form-factor in QCD at intermediate momentum transfers}",
    eprint = "hep-ph/9907495",
    archivePrefix = "arXiv",
    reportNumber = "TPR-99-10, WUE-ITP-99-014, NORDITA-1999-43-HE",
    doi = "10.1103/PhysRevD.61.073004",
    journal = "Phys. Rev. D",
    volume = "61",
    pages = "073004",
    year = "2000"
}

@article{Khodjamirian:2011jp,
    author = "Khodjamirian, A. and Klein, Ch. and Mannel, Th. and Wang, Y. -M.",
    title = "{Form Factors and Strong Couplings of Heavy Baryons from QCD Light-Cone Sum Rules}",
    eprint = "1108.2971",
    archivePrefix = "arXiv",
    primaryClass = "hep-ph",
    reportNumber = "SI-HEP-2011-05",
    doi = "10.1007/JHEP09(2011)106",
    journal = "JHEP",
    volume = "09",
    pages = "106",
    year = "2011"
}

@article{Biswas:2026oxq,
    author = "Biswas, Aritra and Khodjamirian, Alexander and Mohamed, Ali",
    title = "{Light-cone sum rules with B-meson distribution amplitudes for the B{\textrightarrow}p form factors in B-mesogenesis models}",
    eprint = "2603.01723",
    archivePrefix = "arXiv",
    primaryClass = "hep-ph",
    reportNumber = "SI-HEP-2026-03, P3H-26-012",
    doi = "10.1103/b2ll-41mf",
    journal = "Phys. Rev. D",
    volume = "113",
    number = "11",
    pages = "116035",
    year = "2026"
}

@article{FLAG:2024oxs,
    author = "Aoki, Y. and others",
    collaboration = "Flavour Lattice Averaging Group (FLAG)",
    title = "{FLAG review 2024}",
    eprint = "2411.04268",
    archivePrefix = "arXiv",
    primaryClass = "hep-lat",
    reportNumber = "CERN-TH-2024-192, FERMILAB-PUB-24-0785-T",
    doi = "10.1103/nfzp-p5dn",
    journal = "Phys. Rev. D",
    volume = "113",
    number = "1",
    pages = "014508",
    year = "2026"
}

@article{ParticleDataGroup:2026aaa,
    author = "Takahashi, F. and others",
    collaboration = "Particle Data Group",
    title = "{Review of Particle Physics}",
    doi = "10.1142/S0217751X26300115",
    journal = "Int. J. Mod. Phys. A",
    volume = "41",
    pages = "2630011",
    year = "2026"
}

@article{Gell-Mann:1968hlm,
    author = "Gell-Mann, Murray and Oakes, R. J. and Renner, B.",
    title = "{Behavior of current divergences under $SU(3) \times SU(3)$}",
    doi = "10.1103/PhysRev.175.2195",
    journal = "Phys. Rev.",
    volume = "175",
    pages = "2195--2199",
    year = "1968"
}

@article{Wang:2026vxh,
    author = "Wang, Zeren Simon and Tang, Xin-Ru and Zhang, Yu and Zhang, Yu and Zhou, Xiaorong",
    title = "{Searching for apparent baryon number violation in $\Lambda_c^+$ decays at the Super Tau-Charm Facility}",
    eprint = "2604.11329",
    archivePrefix = "arXiv",
    primaryClass = "hep-ph",
    doi = "10.1103/q8kq-jh6f",
    journal = "Phys. Rev. D",
    volume = "114",
    number = "1",
    pages = "015033",
    year = "2026"
}

@misc{Mohamed:2026hab,
    author = "Mohamed, Ali and Piscopo, Maria Laura",
    title = "{Enhanced Standard Model bound on charm CP violation from QCD penguins}",
    eprint = "2609.16219",
    archivePrefix = "arXiv",
    primaryClass = "hep-ph",
    reportNumber = "P3H-26-072, SI-HEP-2026-22, CERN-TH-2026-220",
    month = "9",
    year = "2026"
}

@article{Bourrely:2008za,
    author = "Bourrely, Claude and Caprini, Irinel and Lellouch, Laurent",
    title = "{Model-independent description of $B\to\pi\ell\nu$ decays and a determination of $|V_{ub}|$}",
    eprint = "0807.2722",
    archivePrefix = "arXiv",
    primaryClass = "hep-ph",
    reportNumber = "CPT-P36-2007",
    doi = "10.1103/PhysRevD.82.099902",
    journal = "Phys. Rev. D",
    volume = "79",
    pages = "013008",
    year = "2009",
    note = "[Erratum: Phys.Rev.D 82, 099902 (2010)]"
}

@article{Sakharov:1967dj,
    author = "Sakharov, A. D.",
    title = "{Violation of CP Invariance, C asymmetry, and baryon asymmetry of the universe}",
    doi = "10.1070/PU1991v034n05ABEH002497",
    journal = "Pisma Zh. Eksp. Teor. Fiz.",
    volume = "5",
    pages = "32--35",
    year = "1967"
}

@article{Dev:2022jbf,
    author = "Dev, P. S. B. and others",
    title = "{Searches for baryon number violation in neutrino experiments: a white paper}",
    eprint = "2203.08771",
    archivePrefix = "arXiv",
    primaryClass = "hep-ex",
    doi = "10.1088/1361-6471/ad1658",
    journal = "J. Phys. G",
    volume = "51",
    number = "3",
    pages = "033001",
    year = "2024"
}

@article{Elor:2018twp,
    author = "Elor, Gilly and Escudero, Miguel and Nelson, Ann",
    title = "{Baryogenesis and Dark Matter from $B$ Mesons}",
    eprint = "1810.00880",
    archivePrefix = "arXiv",
    primaryClass = "hep-ph",
    reportNumber = "KCL-18-53, IFIC-18-35",
    doi = "10.1103/PhysRevD.99.035031",
    journal = "Phys. Rev. D",
    volume = "99",
    number = "3",
    pages = "035031",
    year = "2019"
}

@article{Alonso-Alvarez:2021qfd,
    author = "Alonso-\'Alvarez, Gonzalo and Elor, Gilly and Escudero, Miguel",
    title = "{Collider signals of baryogenesis and dark matter from B mesons: A roadmap to discovery}",
    eprint = "2101.02706",
    archivePrefix = "arXiv",
    primaryClass = "hep-ph",
    reportNumber = "TUM-HEP 1299/20",
    doi = "10.1103/PhysRevD.104.035028",
    journal = "Phys. Rev. D",
    volume = "104",
    number = "3",
    pages = "035028",
    year = "2021"
}

@article{Elahi:2021jia,
    author = "Elahi, Fatemeh and Elor, Gilly and McGehee, Robert",
    title = "{Charged B mesogenesis}",
    eprint = "2109.09751",
    archivePrefix = "arXiv",
    primaryClass = "hep-ph",
    reportNumber = "LCTP-21-24, MITP-21-041",
    doi = "10.1103/PhysRevD.105.055024",
    journal = "Phys. Rev. D",
    volume = "105",
    number = "5",
    pages = "055024",
    year = "2022"
}

@article{Alonso-Alvarez:2021oaj,
    author = "Alonso-\'Alvarez, Gonzalo and Elor, Gilly and Escudero, Miguel and Fornal, Bartosz and Grinstein, Benjam\'\i{}n and Martin Camalich, Jorge",
    title = "{Strange physics of dark baryons}",
    eprint = "2111.12712",
    archivePrefix = "arXiv",
    primaryClass = "hep-ph",
    reportNumber = "TUM-HEP-1373/21, MITP-21-060",
    doi = "10.1103/PhysRevD.105.115005",
    journal = "Phys. Rev. D",
    volume = "105",
    number = "11",
    pages = "115005",
    year = "2022"
}

@article{Nelson:2019fln,
    author = "Nelson, Ann E. and Xiao, Huangyu",
    title = "{Baryogenesis from B Meson Oscillations}",
    eprint = "1901.08141",
    archivePrefix = "arXiv",
    primaryClass = "hep-ph",
    doi = "10.1103/PhysRevD.100.075002",
    journal = "Phys. Rev. D",
    volume = "100",
    number = "7",
    pages = "075002",
    year = "2019"
}

@article{Khodjamirian:2022vta,
    author = "Khodjamirian, Alexander and Wald, Marcel",
    title = "{B-meson decay into a proton and dark antibaryon from QCD light-cone sum rules}",
    eprint = "2206.11601",
    archivePrefix = "arXiv",
    primaryClass = "hep-ph",
    reportNumber = "SI-HEP-2022-08, P3H-22-060",
    doi = "10.1016/j.physletb.2022.137434",
    journal = "Phys. Lett. B",
    volume = "834",
    pages = "137434",
    year = "2022"
}

@article{Boushmelev:2023huu,
    author = "Boushmelev, Anastasia and Wald, Marcel",
    title = "{Higher twist corrections to B-meson decays into a proton and dark antibaryon from QCD light-cone sum rules}",
    eprint = "2311.13482",
    archivePrefix = "arXiv",
    primaryClass = "hep-ph",
    reportNumber = "SI-HEP-2023-29, P3H-23-093",
    doi = "10.1103/PhysRevD.109.055049",
    journal = "Phys. Rev. D",
    volume = "109",
    number = "5",
    pages = "055049",
    year = "2024"
}

@article{Elor:2022jxy,
    author = "Elor, Gilly and Guerrera, Alfredo Walter Mario",
    title = "{Branching fractions of B meson decays in Mesogenesis}",
    eprint = "2211.10553",
    archivePrefix = "arXiv",
    primaryClass = "hep-ph",
    doi = "10.1007/JHEP02(2023)100",
    journal = "JHEP",
    volume = "02",
    pages = "100",
    year = "2023"
}

@article{Lenz:2024rwi,
    author = {Lenz, Alexander and Mohamed, Ali and W{\"u}thrich, Zachary},
    title = "{Constraining B-mesogenesis models with inclusive and exclusive decays}",
    eprint = "2412.14947",
    archivePrefix = "arXiv",
    primaryClass = "hep-ph",
    reportNumber = "P3H-24-099,SI-HEP-2024-30",
    doi = "10.1007/JHEP08(2025)141",
    journal = "JHEP",
    volume = "08",
    pages = "141",
    year = "2025"
}

@article{Mohamed:2025zgx,
    author = "Mohamed, Ali",
    title = "{Are subleading effects really subleading? B-meson decays in mesogenesis}",
    eprint = "2511.04858",
    archivePrefix = "arXiv",
    primaryClass = "hep-ph",
    reportNumber = "P3H-25-089, SI-HEP-2025-24",
    doi = "10.1007/JHEP06(2026)219",
    journal = "JHEP",
    volume = "06",
    pages = "219",
    year = "2026"
}

@article{BaBar:2023dtq,
    author = "Lees, J. P. and others",
    collaboration = "BaBar",
    title = "{Search for Evidence of Baryogenesis and Dark Matter in $B^+ \to \psi_D+p$ Decays at BABAR}",
    eprint = "2306.08490",
    archivePrefix = "arXiv",
    primaryClass = "hep-ex",
    reportNumber = "BABAR-PUB-23/03 SLAC-PUB-17731",
    doi = "10.1103/PhysRevLett.131.201801",
    journal = "Phys. Rev. Lett.",
    volume = "131",
    number = "20",
    pages = "201801",
    year = "2023"
}

@misc{Belle-II:2026tyb,
    author = "Abumusabh, M. and others",
    collaboration = "Belle-II, Belle",
    title = "{A search for feebly-interacting particles in $B$ decays with missing energy at Belle}",
    eprint = "2601.07104",
    archivePrefix = "arXiv",
    primaryClass = "hep-ex",
    reportNumber = "Belle II Preprint: 2025-029 KEK Preprint: 2025-36",
    month = "1",
    year = "2026"
}

@article{BaBar:2023rer,
    author = "Lees, J. P. and others",
    collaboration = "BaBar",
    title = "{Search for B mesogenesis at BaBar}",
    eprint = "2302.00208",
    archivePrefix = "arXiv",
    primaryClass = "hep-ex",
    doi = "10.1103/PhysRevD.107.092001",
    journal = "Phys. Rev. D",
    volume = "107",
    number = "9",
    pages = "092001",
    year = "2023"
}

@article{BaBar:2024qqx,
    author = "Lees, J. P. and others",
    collaboration = "BaBar",
    title = "{Search for baryogenesis and dark matter in $B^+ \to \Lambda_{c}^{+}$+ invisible decays}",
    eprint = "2412.06950",
    archivePrefix = "arXiv",
    primaryClass = "hep-ex",
    reportNumber = "BABAR-PUB-24/001, SLAC-PUB-241025",
    doi = "10.1103/PhysRevD.111.L031101",
    journal = "Phys. Rev. D",
    volume = "111",
    number = "3",
    pages = "L031101",
    year = "2025"
}

@article{Belle:2021gmc,
    author = "Hadjivasiliou, C. and others",
    collaboration = "Belle",
    title = "{Search for $B^0$ meson decays into $\Lambda$ and missing energy with a hadronic tagging method at Belle}",
    eprint = "2110.14086",
    archivePrefix = "arXiv",
    primaryClass = "hep-ex",
    reportNumber = "Belle Preprint 2021-24; KEK Preprint 2021-29; PNNL-SA-167279",
    doi = "10.1103/PhysRevD.105.L051101",
    journal = "Phys. Rev. D",
    volume = "105",
    number = "5",
    pages = "L051101",
    year = "2022"
}

@article{BurgosMarcos:2026lwv,
    author = "Burgos Marcos, M. and Verheyden, A. and Vos, K. K.",
    title = "{From B$_{c}$ mesons to the baryon asymmetry: a unified B mesogenesis framework}",
    eprint = "2601.19651",
    archivePrefix = "arXiv",
    primaryClass = "hep-ph",
    reportNumber = "Nikhef 2026-001",
    doi = "10.1007/JHEP09(2026)077",
    journal = "JHEP",
    volume = "09",
    pages = "077",
    year = "2026"
}

@article{Khodjamirian:2002pk,
    author = "Khodjamirian, Alexander and Mannel, Thomas and Urban, Piotr",
    title = "{Gluonic penguins in B ---{\ensuremath{>}} pi pi from QCD light cone sum rules}",
    eprint = "hep-ph/0210378",
    archivePrefix = "arXiv",
    reportNumber = "TTP02-16",
    doi = "10.1103/PhysRevD.67.054027",
    journal = "Phys. Rev. D",
    volume = "67",
    pages = "054027",
    year = "2003"
}

@article{Rusov:2017chr,
    author = "Rusov, Aleksey V.",
    title = "{Higher-twist effects in light-cone sum rule for the $B\rightarrow \pi $ form factor}",
    eprint = "1705.01929",
    archivePrefix = "arXiv",
    primaryClass = "hep-ph",
    reportNumber = "SI-HEP-2017-07, QFET-2017-06",
    doi = "10.1140/epjc/s10052-017-5000-9",
    journal = "Eur. Phys. J. C",
    volume = "77",
    number = "7",
    pages = "442",
    year = "2017"
}

\end{document}